\documentclass[a4paper,UKenglish,oldauthorstyle,numberwithinsect]{lipics-v2021}
\hideLIPIcs
\usepackage{mathtools,booktabs,microtype,array}
\usepackage{xspace,tikz}
\usetikzlibrary{arrows.meta,positioning,shapes.geometric}
\numberwithin{equation}{section}
\newcommand{\val}{\operatorname{val}}
\newcommand{\SAS}{\operatorname{SAS}}
\newcommand{\Prb}{\mathbb P}
\newcommand{\Exp}{\mathbb E}
\newcommand{\Max}{\ensuremath{\mathrm{Max}}\xspace}
\newcommand{\Min}{\ensuremath{\mathrm{Min}}\xspace}
\title{Attaining Optimal Values under Sure Constraints}
\titlerunning{Attaining Optimal Values under Sure Constraints}
\author{Takuto Yamauchi\footnote{Corresponding author: Takuto Yamauchi.}}{[Waseda University, Institute of Science Tokyo]}{takuto.yamauchi@aoni.waseda.jp}{}{}
\author{Kenji Tei}{Institute of Science Tokyo}{tei@comp.isct.ac.jp}{}{}
\authorrunning{T. Yamauchi and K. Tei}
\ccsdesc[500]{Theory of computation~Stochastic games}
\ccsdesc[300]{Theory of computation~Automata over infinite objects}
\keywords{Stochastic games, sure constraints, finite memory, exact attainment, complexity}
\category{Preprint}
\AtBeginDocument{%
  \nolinenumbers
  \hypersetup{
    pdftitle={Attaining Optimal Values under Sure Constraints},
    pdfauthor={Takuto Yamauchi, Kenji Tei},
    pdfsubject={Preprint},
    pdfcreator={LaTeX with lipics-v2021.cls}}}
\begin{document}
\maketitle
\begin{abstract}
A game may have, for each target below its value, a finite controller
meeting that target, yet no finite controller attaining the value itself. We study finite stochastic games
with a hard specification that must hold on every supported execution,
including paths of probability zero. Max uses a pure finite controller;
Min may use any pure public-history strategy. We distinguish three
numerical questions: can a supplied target be met, can the unspecified
optimum be attained, and what is one exact value?
For explicit hard and Boolean soft Streett conditions, the strict and non-strict
supplied-threshold problems are coNP-complete, attainment of the
unsupplied supremum is $\mathrm P^{\mathrm{NP}}$-complete, and one exact
reduced fraction is $\mathrm{FP}^{\mathrm{NP}}$-complete under polynomial
metric reductions. The proofs preserve both the value and its finite
attainability when fixing a single opponent, and encode adaptive queries
without computing unknown subgame values.
The same construction yields matching value-comparison and exact-output
lower bounds for ordinary stochastic Streett games with unrestricted
randomized history strategies.

We also bound the memory needed for optimal finite implementations
for arbitrary Muller hard conditions and rational rewards in $[0,1]$
determined by infinitely recurring states. One controller is sure from
the entire original sure region and attains at every finitely attained
start, with no attainment premise on the other states. For $n$ states
in that region it uses $2^{O(n\log(n+1))}Q(1+\log R)$ modes, where $Q$
is the product of the least common denominators of full original chance rows
sourced in that sure region and $R$ clears all reward values. Its joint dependence on $Q$ and $R$ is worst-case
sharp at each fixed $n\ge6$ when $Q,R\ge2$. Optimal dependence on arena
size and efficient general synthesis are not resolved here.
\end{abstract}
\clearpage
\section{Introduction}
A supplied target, an unspecified optimum and an exact numerical output
are different requests to synthesis. Meeting a target asks for a
sufficient implementation. Attaining an optimum asks whether the best
guarantee has any finite implementation. Computing its exact value asks
for the number itself. We classify these requests and bound the memory
needed when an optimal finite implementation exists.

Hard specifications and expected quality impose different obligations.
Hard must hold on every supported execution, including executions of
probability zero; quality is an expected reward against an adversarial
environment and actual random transitions. Max uses a pure finite
controller with observed-edge updates, while Min may use the entire
public history. This finite restriction can even change the value:
a parity MDP can have soft probability zero for every finite-memory
sure strategy and probability one with infinite memory
\cite[Figure~2 and Theorem~15]{BRR17}.

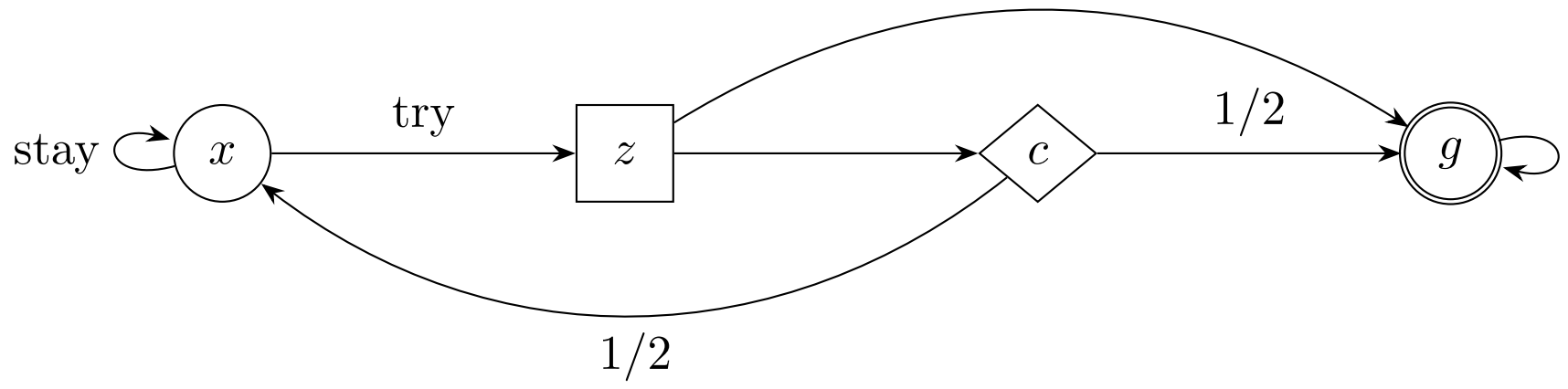
\begin{figure}[ht]
\centering
\begin{tikzpicture}[>=Stealth,node distance=22mm,st/.style={draw,circle,minimum size=7mm}]
\node[st] (x) {$x$};
\node[draw,rectangle,right=of x,minimum size=7mm] (z) {$z$};
\node[draw,diamond,aspect=1.2,right=of z,minimum size=7mm] (c) {$c$};
\node[st,double,right=of c] (g) {$g$};
\draw[->] (x)--node[above]{try}(z);
\draw[->] (x) edge[loop left] node[left]{stay} (x);
\draw[->] (z)--(c);\draw[->] (z) to[bend left=32] (g);
\draw[->] (c)--node[above]{$1/2$}(g);
\draw[->] (c) to[bend left=38] node[below]{$1/2$}(x);
\draw[->] (g) edge[loop right] (g);
\end{tikzpicture}
\caption{A value-preserving opponent need not preserve attainment. Circles
are Max vertices, squares Min vertices and diamonds chance vertices;
the double circle is absorbing and unlabelled deterministic edges have
probability one. Hard is coB\"uchi avoidance of $\{z,c\}$; soft is
B\"uchi success at $g$. Fixing Min to the coin keeps the endpoint open;
fixing Min to the direct edge $z\to g$ closes it.}
\label{fig:endpoint}
\end{figure}

If Min always chooses the coin in Figure~\ref{fig:endpoint}, trying
$N$ times and then staying at $x$ guarantees $1-2^{-N}$. Every sure
controller must stop on the supported all-failure execution, so the
supremum one is not attained. If Min instead chooses the direct edge
to $g$, one trial attains the same value. Knowing the value therefore
does not settle finite attainability. A reduction that fixes the
opponent must preserve both.

Nonexistence of an optimum is known for sure-parity expected-mean-cost
MDPs~\cite{AKV16}; finite-memory retry separates sure/limit-sure from
sure/almost-sure parity~\cite[Theorem~14]{CP19}. The additional obstacle
here is a change in admissibility: fixing Min may admit finite
controllers that were not sure in the original game. A value-preserving
policy can consequently close an originally open endpoint.

\paragraph*{Two main contributions.}
For explicit hard and Boolean soft Streett conditions, supplied strict
and non-strict thresholds are coNP-complete, attainment of the
unsupplied supremum is $\mathrm P^{\mathrm{NP}}$-complete, and one exact
reduced value is $\mathrm{FP}^{\mathrm{NP}}$-complete under polynomial
metric reductions (Theorem~\ref{thm:main-complexity}). Unless
$\mathrm{NP}=\mathrm{coNP}$, asking to implement an unspecified optimum
is outside coNP. These are classifications for pure finite sure Max
against arbitrary pure public-history Min, not unrestricted synthesis.

For arbitrary Muller hard conditions and rational rewards determined
by infinitely recurring states, one controller is sure from the whole
original sure region and attains at every finitely attained start,
without attainment assumptions elsewhere. Up to an arena-size factor,
its mode count is $Q(1+\log R)$: $Q$ is the product of original
chance-row denominators and $R$ clears the rewards. This joint
arithmetic dependence has matching fixed-size lower families
(Theorem~\ref{thm:main-memory}). Optimal state-count dependence and
efficient general synthesis are not resolved here.

These results address two obstacles to finite implementation. For the
classification, a single opponent must preserve feasibility and the
open or closed endpoint, even for newly admissible controllers. For
the memory bound, a finite cutoff must protect every supported path
without losing the exact guarantee. We explain the invariants that
make these two tasks possible; the appendices supply complete proofs.

\paragraph*{Scientific reach and closest prior settings.}
The adaptive numerical construction also applies with hard true.
For ordinary stochastic Streett games, allowing arbitrary possibly
randomized history strategies for both players, two-value comparison
is $\mathrm P^{\mathrm{NP}}$-complete and one exact value is
$\mathrm{FP}^{\mathrm{NP}}$-complete. Known value-vector and strategy
results supply the upper bounds~\cite{CDH04full,CH06,UW11}; our
construction supplies the comparison and exact-output lower bounds.
Thus the numerical distinction reaches beyond the sure-finite model.

MDPs already have finite witnesses for sure parity with reachability
thresholds~\cite[Theorem~4]{BRR17} and existing lexicographic
sink-reachability optima in clean MDPs, allowing randomization
\cite[arXiv v1, Theorem~3]{BKW24}. Parity-probability thresholds can
instead require infinite memory~\cite[Theorem~23]{BRR17}. Here Min is
adversarial and Max is pure finite. Qualitative constructions
\cite[Theorem~8]{CP19}, \cite[Theorem~2]{DG26} already cover fixed-index
finite controllers at non-strict threshold one when combined. Our
parity supplied-threshold classification treats arbitrary rational
targets in the pure-finite model, without resolving the unrestricted
question of~\cite[p.~34:3]{DG26}.

Ordinary value and strategy-transfer methods preserve expectation
\cite{DGG25,GH08,GK23,DM26}. Our additional obligation is to preserve
finite endpoints and the original sure domain. We reuse qualitative
split/pause and finite-SAS procedures; they are not new constructions
here. Appendix J compares the models and their precise scopes.
Table~\ref{tab:reading-guide} gives entry points to the complete proofs.

\section{Model and requested outputs}\label{sec:model}
A finite public arena has vertex set
$V=V_{\mathrm{Max}}\sqcup V_{\mathrm{Min}}\sqcup V_c$, partitioned
into Max, Min and chance vertices. Every vertex has an outgoing edge. A chance row lists positive rational probabilities
summing to one. Play has perfect information: both players observe every
actual edge, so a public history is the sequence of visited vertices and
edges. We call a vertex
\emph{original} to distinguish it from a product state carrying memory.

Max uses a pure finite controller
$\mathcal C=(\mathcal M,\iota,\mathrm{act},\mathrm{upd})$.
At a Max vertex $x$ and mode $m$, the chosen edge is $\mathrm{act}(m,x)$;
after observing an actual edge $e$, the mode becomes $\mathrm{upd}(m,e)$.
The initial mode is $\iota(s)$. Mode count is $|\mathcal M|$, including
all witness tags, counters, the mode that stops a retry, and fallback. Min may use any pure strategy on
the public history. A positional Min policy chooses one original edge
at each Min vertex.

A supported execution follows the chosen actions and positive-probability
chance edges. Hard $A$ must hold on every such infinite execution against
every Min strategy. Almost-sure satisfaction means probability one.
For a fixed pure controller, behavioral Min strategies are mixtures of
pure history strategies and give the same worst guarantee; each
supported execution is also consistent with a pure opponent.

For the computational classifications, hard $A$ and soft $B$ are explicit
Streett lists: a pair $(E_i,F_i)$ requires infinitely many $F_i$ visits
whenever $E_i$ recurs infinitely often. The reward is $F=\mathbf1_B$, so expectation
is soft satisfaction probability. Parity uses the maximum-even convention; its index is the number of
priorities after order- and parity-preserving normalization.
Input size includes all vertices, edges, objective occurrences and
binary rational entries.

Treating chance support adversarially for the hard condition, let $W=W_G$
be the original states with a pure finite sure-$A$ controller,
and $\mathcal F_A(s)$ the set of those controllers from $s$. For $s\in W$,
with $\tau$ ranging over pure public-history Min strategies, define
\[
 v(s)=\sup_{\sigma\in\mathcal F_A(s)}\inf_\tau
          \Pr_s^{\sigma,\tau}[B],\qquad
 I(s)=\{a:\exists\sigma\in\mathcal F_A(s)\ \forall\tau\
                    \Pr_s^{\sigma,\tau}[B]\ge a\}.
\]
The interval is $(-\infty,v(s))$ or $(-\infty,v(s)]$; its endpoint
records whether a finite controller attains the supremum. We also write
$I_G(s)=I(s)$ to identify the game.
Write $G$ for the encoded game. The supplied-target decision problems
have input $(G,s,q)$, with rational $q$, and ask whether a pure finite
sure controller guarantees $>q$, respectively $\ge q$, against every Min
strategy. \textsc{Own-Attain} has input $(G,s)$ and asks whether
$v(s)\in I(s)$. These decision problems are false when $s\notin W$.
The exact-value problem is the partial function with domain
$\{(G,s):s\in W\}$ that outputs $v(s)$ as a binary reduced fraction.
Fixing Min can enlarge the set of sure controllers; our preservation
theorem retains the original domain and every interval.
\begin{theorem}[Three numerical tasks for Streett]
\label{thm:main-complexity}\label{thm:main-own}\label{thm:main-exact}
For explicit hard and Boolean soft Streett, with pure finite sure Max
and arbitrary pure public-history Min:
\par
\noindent\begin{tabular}{@{}p{.59\linewidth}p{.36\linewidth}@{}}
\toprule
\textbf{Request} & \textbf{Classification}\\\midrule
(1) Given a rational $q$, decide whether a guarantee $>q$ or $\ge q$
can be met.
& coNP-complete.\\
(2) Without a supplied optimum, decide whether $v(s)$ is finitely attained.
& $\mathrm P^{\mathrm{NP}}$-complete.\\
(3) Return one exact designated value at a sure-feasible start as a binary
reduced fraction.
& $\mathrm{FP}^{\mathrm{NP}}$-complete under polynomial metric reductions.\\
\bottomrule
\end{tabular}\par
\end{theorem}
Own-attainment hardness already holds with one hard coB\"uchi pair and
an unbounded soft Streett list; exact-value hardness holds with hard true.

\section{Why classification needs more than values}\label{sec:classification}
\subsection{Preserving endpoints and admissibility}\label{sec:criterion}\label{sec:upper}
An explicit controller exhibits a lower guarantee. Rejecting a target
requires an upper bound for \emph{every} admissible controller. We
therefore cannot fix an arbitrary optimal opponent and then ignore
controllers that become sure only after the restriction.

The invariant has two parts: preserve the qualitative test for reaching
the endpoint, and also the stricter test that would expose an excessive
reward. Let $K_v$ retain value-preserving player edges on the original
sure region $W$ and every full original chance row. Bellman's equations
make player choices nonempty and every chance row harmonic for $v$;
this is mean preservation, not equality of individual outcomes.
For nonempty recurrent sets $S$, define
\[
 B_v(S)\iff \mathbf1_B(S)\ge\max_{x\in S}v(x),\qquad
 C_v(S)\iff \mathbf1_B(S)>\max_{x\in S}v(x).
\]
The weak test permits equality; the strict test detects an excess.
These are recurrent tests, not a demand that a fractional expectation
be achieved on each path. Write $\SAS(A,\Phi)$ for a pure finite
controller that is sure $A$ and almost-sure $\Phi$ against every
public-history Min strategy.

The endpoint test is
\begin{equation}\label{eq:main-endpoint}
 s\text{ admits an attaining controller}
 \iff s\text{ admits finite }\SAS(A,B_v)\text{ in }K_v.
\end{equation}
Why can it use the restricted arena? An attainer has zero deficit from
$v$ at its start. Under preserving Min moves, fixed-controller baseline
inequalities and full-row averaging propagate that zero deficit to
every supported successor. Conversely, when Min first leaves $K_v$,
it moves to a higher-value state. Switch there to a finite sure
continuation beating the \emph{old} value; attaining the new value is
unnecessary. This first-deviation repair~\cite{GH08} restores the
original guarantee and hard acceptance (Appendix B).

\begin{theorem}[The interval opponent]
\label{thm:main-joint}\label{thm:main-interval}
For hard and Boolean soft Streett conditions, there is a positional Min
policy $\tau^*$ on the full original arena. Let $G[\tau^*]$ fix Min to it
and keep every original chance row unchanged. Compute its sure region
and intervals using all pure finite controllers sure against $\tau^*$,
including controllers not admissible in $G$. Then
\[
 W_{G[\tau^*]}=W_G,\qquad
 I_{G[\tau^*]}(s)=I_G(s)\quad(s\in W_G).
\]
Here $W_G=W$ is the original sure region.
\end{theorem}

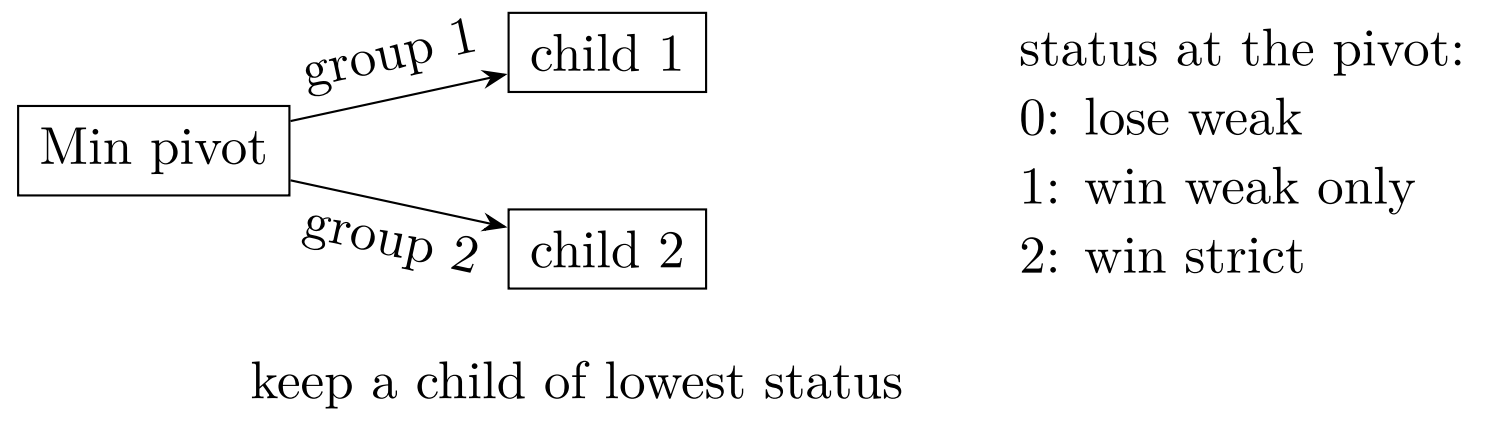
\begin{figure}[!h]\centering
\begin{tikzpicture}[>=Stealth,
box/.style={draw,rectangle,align=center,inner sep=4pt}]
\node[box] (pivot) at (0,0) {Min pivot};
\node[box] (left) at (3,.65) {child 1};
\node[box] (right) at (3,-.65) {child 2};
\draw[->] (pivot)--node[above,sloped]{group 1}(left);
\draw[->] (pivot)--node[below,sloped]{group 2}(right);
\node[align=left] at (7.2,0) {status at the pivot:\\
$0$: lose weak\\$1$: win weak only\\$2$: win strict};
\node[align=center] at (2.8,-1.55) {keep a child of lowest status};
\end{tikzpicture}
\caption{One split of Min's available edges into two nonempty groups.
Children keep the same vertices and full original chance rows; only
Min's choices at the pivot change. Status is measured at that pivot.}
\label{fig:r56-opponent-split}
\end{figure}

\paragraph*{One local choice for both tests.}
Repeat the local choice in Figure~\ref{fig:r56-opponent-split} to build
the common opponent.

\emph{Choose.} If a lowest-status child wins either test at the pivot,
both children win that test there. If it loses the test there, its
winning witnesses avoid the pivot on every supported history. This
same dichotomy serves both ordered tests.

\emph{Merge.} When both children win a fixed test, keep both finite child
memories~\cite{CP19}. Min's observed pivot edge chooses the active child;
only that memory is updated, while the other is paused, never reset.
If both projections are infinite, their shared pivot fixes the same
constant recurrent value. Streett acceptance and the comparison survive
the union. The active child is chosen before each chance outcome and
its full original row is executed, giving fairness under pausing.
Appendix E proves the simultaneous invariant for all starting states
and both tests. Max's witnesses may depend on the start and the test.

The strict region in $K_v$ is empty. Otherwise finite recurrence gives
a uniform positive strict gap, and a repair after Min's first
value-increasing move preserves a positive gain over the old value.
Combining the no-switch gap and switch gains would beat the supremum.
Now fix the opponent selected by the invariant and consider \emph{any}
controller sure against it. Its finite product makes $v$ a
supermartingale. An excessive bottom-component reward would yield a
strict witness in the restricted arena and, by preservation, in $K_v$:
a contradiction. This bounds even newly admissible controllers.

The preserved weak test then decides whether the bound is attained.
An outside-$W$ hard-losing spoiler prevents new sure starting states,
using supported paths even when they have probability zero. These
three obligations---upper bound, endpoint and original domain---give
Theorem~\ref{thm:main-joint}. The full argument preserves all original
chance rows (Theorem~\ref{thm:joint-interval} and
Corollary~\ref{cor:full-arena}); it is stronger than value preservation.

\subsection{From the invariant to a short rejecting certificate}
A hard-losing start can be rejected directly. Otherwise a certificate
guesses $U\ni s$, one positional opponent and a rational upper potential
$z$ with $z(s)\le q$; it does not receive the true $W$ or $v$.
Hard-losing spoilers for excluded vertices prove $W\subseteq U$.
The guessed domain keeps every original Min and chance successor and
an internal Max move, so all original sure controllers are covered.
Local nonincrease on internal Max and selected Min edges, full-row
nonincreasing means, and negative strict tests from \emph{every} state
bound all admitted sure controllers, including newly admitted ones.

If $z(s)<q$, both targets fail. At equality the strict target fails;
to reject the non-strict target, also exclude the weak test after
retaining zero-loss player moves and full zero-mean-loss chance rows,
then pruning states where these cannot stay. Thus equality needs an
additional test, not merely an upper bound. Lemmas~\ref{lem:local-upper}--\ref{lem:zero-slack}
give the full soundness checks. For completeness, the interval opponent
supplies $U=W,z=v$ and preserves that equality test. The verifier still
checks only guesses, not unknown true quantities.

\paragraph*{Why the guesses have short bits.}
Measure denominators on original states, rather than a large controller
product. Finite qualitative witnesses and their finite-cap realization
place $v$ at a vertex of original-row constraints. A determinant
expansion charges each original chance row once, giving a common
clearer $h_v\le Q$, where $Q$ is the product of least denominators of
full original chance rows sourced in $W$
(Theorem~\ref{thm:r48-general-denominators}). D.2--D.4 supply the complete
realization and vertex proof, and H supplies the finite-SAS certificate
oracle. The explicit quantitative mode count of C.1 is not needed for
this polynomial value-bit bound; the qualitative procedures remain shared.

The resulting complement has a polynomial NP verifier. Deterministic
Streett winning is coNP-hard already for two-parity conjunctions
\cite[Theorem~2(5)]{CHP07}. Continuation-or-fallback gates make all
states sure feasible, while fallback loses soft, giving threshold
$1/2$ hardness for both comparisons with $W=V$ and attained yes
instances~\cite{Chatterjee07}. The parity conjunction has unbounded
indices. Appendix G gives the full classifications and function upper
bounds.

\subsection{Unknown values can carry adaptive answers}\label{sec:lower}
The upper bound reconstructs the value by supplied-threshold queries
and then asks its endpoint question. A matching lower bound must put
adaptive numerical information into a single game \emph{before}
learning any answer. A later satisfiability query may depend on earlier
answers, but its constructor must use syntax, not a SAT oracle.

Every designated entry has two guarantees. A pure finite controller
sure against every original Min strategy attains its value. One common
positional Min policy bounds every original-sure continuation by that
value, uniformly over the inherited memory mode and prior public
history. Irreversible calls preserve this entry contract for older
modules (Definition~\ref{def:r52-entry-contract}). This contract is
about original-sure continuations, rather than the larger class in the
interval theorem.

Store a bit in entries whose absolute gap $\delta$ is known:
\[
 v(p_i)-v(q_i)=(2b_i-1)\delta.
\]
A reference lottery calls both old entries with the same known weight
$\lambda$. Replacing one call by the other gives test lotteries
\[
 v(r_i^\pm)=v(\mathsf{ref})\pm\lambda\bigl(v(p_i)-v(q_i)\bigr),
 \qquad \eta=\lambda\delta.
\]
Their coefficients are known although the reference value and gap sign
are not. A positive literal $b_i$ uses $r_i^-$ and a negative one uses
$r_i^+$, so a false literal gives the higher value
$v(\mathsf{hi})=v(\mathsf{ref})+\eta$, and a true one the lower value
$v(\mathsf{lo})=v(\mathsf{ref})-\eta$.

For a CNF formula $\varphi(\mathbf b,\mathbf y)$, Min chooses the free
assignment $\mathbf y$ and Max chooses a clause. The entry $a$ has value
\[
 v(a)=
 \begin{cases}
 v(\mathsf{hi})=v(\mathsf{ref})+\eta,&\varphi(\mathbf b,\cdot)\text{ is UNSAT},\\
 v(\mathsf{lo})=v(\mathsf{ref})-\eta,&\varphi(\mathbf b,\cdot)\text{ is SAT}.
 \end{cases}
\]
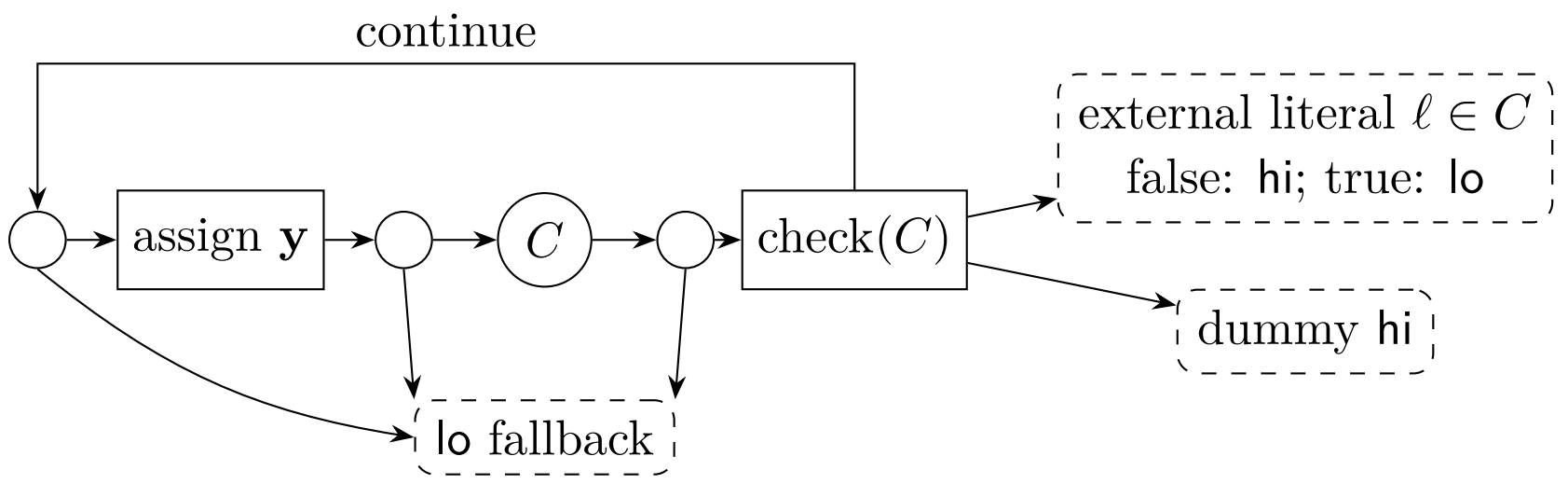
\begin{figure}[!h]\centering
\begin{tikzpicture}[>=Stealth,
gate/.style={draw,circle,minimum size=4mm,inner sep=0pt},
mn/.style={draw,rectangle,align=center,minimum height=7mm,inner sep=3pt},
mx/.style={draw,circle,align=center,inner sep=3pt},
call/.style={draw,dashed,rounded corners,align=center,inner sep=4pt}]
\node[gate] (g0) at (0,0) {};
\node[mn] (a) at (1.3,0) {assign $\mathbf y$};
\node[gate] (g1) at (2.6,0) {};
\node[mx] (cl) at (3.6,0) {$C$};
\node[gate] (g2) at (4.6,0) {};
\node[mn] (ck) at (5.8,0) {$\mathrm{check}(C)$};
\node[call] (rail) at (9.0,.65) {external literal $\ell\in C$\\false: $\mathsf{hi}$; true: $\mathsf{lo}$};
\node[call] (hi) at (9.0,-.65) {dummy $\mathsf{hi}$};
\node[call] (lo) at (3.6,-1.4) {$\mathsf{lo}$ fallback};
\draw[->] (g0)--(a);\draw[->] (a)--(g1);
\draw[->] (g1)--(cl);\draw[->] (cl)--(g2);\draw[->] (g2)--(ck);
\draw[->] (ck)--(rail);\draw[->] (ck)--(hi);
\draw[->] (ck.north)--(5.8,1.25)--node[above]{continue}(0,1.25)--(g0.north);
\draw[->] (g0.south) to[bend right=15] (lo.west);
\draw[->] (g1.south)--(lo.north west);\draw[->] (g2.south)--(lo.north east);
\end{tikzpicture}
\caption{One guard round. Squares are Min and circles Max. Each part
abbreviates source vertices and their copied markers; every actual source
vertex has a Max gate offering fallback. The check exits are specific
to $C$: its external-literal tests, dummy $\mathsf{hi}$, and continuation. Dashed
boxes call lotteries with attained values; $\mathsf{hi}/\mathsf{lo}$ identify their values $v(\mathsf{hi})/v(\mathsf{lo})$.}
\label{fig:r12-construction-map}
\end{figure}

The check exits belong to the \emph{selected} clause $C$. Min may
continue, take dummy $\mathsf{hi}$, or test an external literal of $C$.
The guard checks free literals: selecting $C$ is marked by $L_C$, and
assigning $y=b$ by $A_y^b$, including their Max-gate copies. A positive
free literal contributes $(\{L_C\},\{A_y^0\})$, and a negative one
uses $A_y^1$. Thus each free literal of a recurrently selected clause must be
assigned false infinitely often. Every source vertex also has Max's
$\mathsf{lo}$ fallback.

On UNSAT, Max can always select a false clause. Every test and dummy
exit then has value $v(\mathsf{hi})$, and an infinitely continued play
satisfies the guard and soft condition. Min matches by taking dummy
$\mathsf{hi}$. On SAT, Min repeats a satisfying assignment: it continues
when the chosen clause has a true free literal, and otherwise exits
through a true external literal at $v(\mathsf{lo})$. In this explanatory
hard-guard/B\"uchi version, infinite continuation violates hard, so a
sure controller must eventually leave. Immediate fallback matches
$v(\mathsf{lo})$.

For the hard-true core, put all guard pairs together with B\"uchi in
soft Streett. Infinite continuation under the same SAT policy then earns
zero, preserving the comparison. Exact-value hardness uses this core;
only the final own-attainment retry adds a hard coB\"uchi pair. The
constructor writes known probabilities, entry names and gaps, never an
unknown value. Lemma~\ref{lem:r47-numerical-gate} proves both cases and
the inherited-mode/history-uniform Min bound at all new entries.

Finally open only the positive reference by retry: its supremum stays
$v(\mathsf{ref})$, but no pure finite sure controller attains it.
A new Min entry chooses the retry or the attained query $a$, giving
supremum $\min\{v(a),v(\mathsf{ref})\}$. A lower query can be met by a sufficiently long finite retry on that
branch; a higher query leaves the reference open.
The two-query example in Appendix I.5 uses the same query
$v(a_2)=3/64=6/128$ in both instances:
\begin{center}
\begin{tabular}{@{}ccc@{}}
\toprule
Reference & Final Min value & Endpoint\\\midrule
$5/128$ & $5/128$ & open\\
$7/128$ & $3/64$ & closed\\
\bottomrule
\end{tabular}
\end{center}
The comparison with the unknown reference, rather than the query value
alone, carries the adaptive information.

\subsection{One exact output contains the transcript}\label{sec:exact-output}
For partial functions $f_0,f_1$, a polynomial metric reduction uses
polynomial-time $h,d$ such that $x\in\operatorname{dom}(f_0)$ implies
$h(x)\in\operatorname{dom}(f_1)$ and $f_0(x)=d(x,f_1(h(x)))$.
The decoder receives the source input and one exact target output
\cite{Krentel88}. The hard-true construction has $W=V$, so it produces
inputs in the exact-value domain.

The same numerical gate stores the whole adaptive transcript in one
value. An input-computable affine map takes the designated fraction to
an integer $c_m$, whose ternary code satisfies
\[
 c_0=0,\qquad c_{t+1}=3c_t+b_{t+1}.
\]
The known digit count includes leading zeros. Exact rational arithmetic
and repeated division by three recover all answers in polynomial time;
the source computation can then be replayed with that transcript.
No answers are computed while constructing the graph. The full
coefficient and size analysis is Theorem~\ref{thm:r47-single-value}.

\begin{corollary}[Numerical outputs in ordinary Streett games]
\label{cor:r56-main-ordinary}
For ordinary stochastic Streett games with arbitrary possibly randomized
history strategies for both players, deciding
$\val_G(s)\ge\val_G(t)$ for two specified states is
$\mathrm P^{\mathrm{NP}}$-complete, and one exact binary reduced-fraction
value is $\mathrm{FP}^{\mathrm{NP}}$-complete under polynomial metric
reductions.
\end{corollary}
Known supplied-threshold and value-vector methods give the upper bounds
\cite{CDH04full,CH06,UW11}; the gate above supplies the comparison and
single-output lower bounds (Corollary~\ref{cor:r47-value-comparison} and
Theorem~\ref{thm:r47-single-value}).

\section{Memory for exact attainment}\label{sec:memory}
The second main result concerns \emph{realizing} an attainable optimum,
not the difficulty of deciding whether it is attainable. Its scope is
broader: hard $A$ may be any Muller family of nonempty recurrent sets,
and the payoff is $F(\rho)=f(\operatorname{Inf}(\rho))\in[0,1]$ with finite
rational range. Here $v$ is the supremum of the worst expected $F$ over
pure finite sure-$A$ Max against arbitrary pure public-history Min.
The original sure region remains $W$. Put $n=|W|$; $Q$ is the product
of least common denominators of the full original chance rows sourced
in $W$, and $R$ is the least positive integer clearing all rewards.

\begin{theorem}[Exact memory for rational Inf rewards]\label{thm:main-memory}
One original pure finite controller is sure $A$ from every initialized
state of $W$ and attains $v(s)$ at every finitely attained start $s$.
It uses $2^{O(n\log(n+1))}Q(1+\log R)$ modes, with start-dependent
initialization. No attainment premise is imposed on other original states.
\end{theorem}

Modes include every witness tag, counter and permanent fallback. This
is an existence bound, not an efficient general synthesis algorithm;
transition-table bit length is a different measure, and $Q$ can be
exponential in the input's binary length.

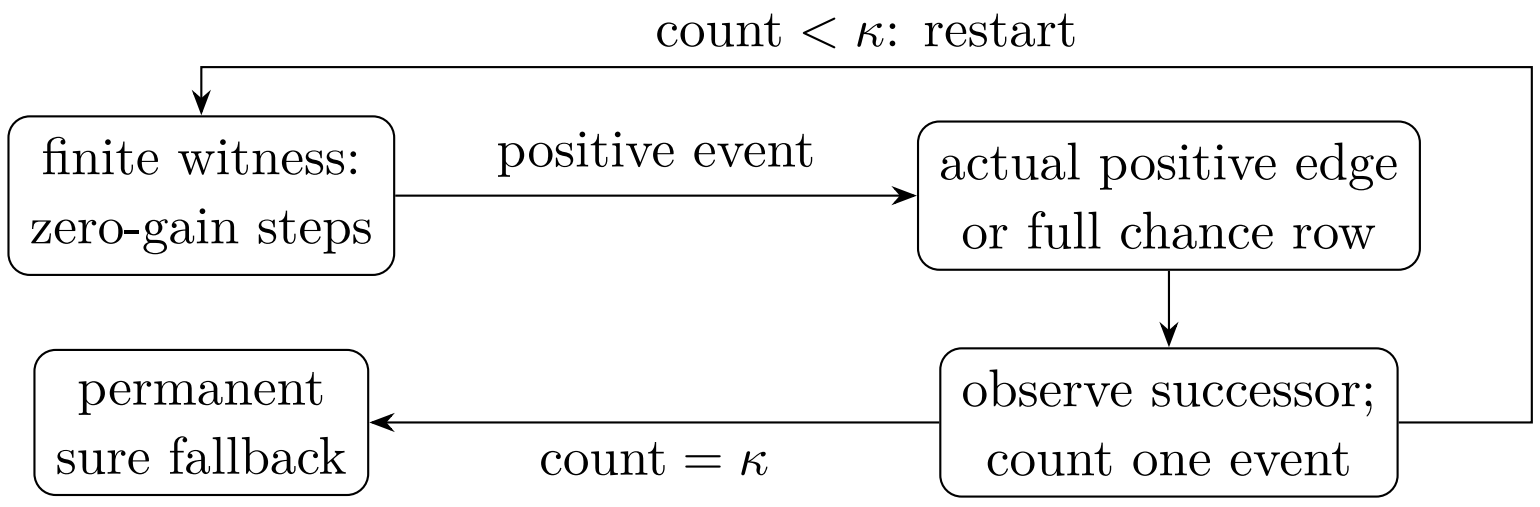
\begin{figure}[!b]\centering
\begin{tikzpicture}[>=Stealth,
box/.style={draw,rounded corners,align=center,inner sep=4pt}]
\node[box] (witness) at (0,0) {finite witness:\\zero-gain steps};
\node[box] (event) at (6.4,0) {actual positive edge\\or full chance row};
\node[box] (count) at (6.4,-1.5) {observe successor;\\count one event};
\node[box] (fallback) at (0,-1.5) {permanent\\sure fallback};
\draw[->] (witness)--node[above]{positive event}(event);
\draw[->] (event)--(count);
\draw[->] (count)--node[below]{$\text{count}=\kappa$}(fallback);
\draw[->] (count.east)--(8.8,-1.5)--(8.8,0.85)
--node[above]{$\text{count}<\kappa$: restart}(0,0.85)--(witness.north);
\end{tikzpicture}
\caption{One capped implementation run, with initial event count zero.
An event is counted after its actual edge or chance outcome. Every
chance step keeps its full original row. Restarting the witness does
not reset the event counter.}
\label{fig:r56-finite-protocol}
\end{figure}

\subsection{First obstacle: anchor the value without assuming attainment elsewhere}
Let $C$ be the finitely attained states. We seek a lower reward
potential $u$ with $u=v$ on $C$, leaving the other values free. Requiring
$u=v$ on all of $W$ would be wrong: a finite exact realization there
would attain an originally open endpoint. The controller will still
run on the whole original $W$, with every full chance row and every
original value-increasing Min departure available.

The values needed for the anchor come from the equality subgame on $C$:
keep full chance rows, preserving Min edges and all internal preserving
Max edges, with at least one at each Max state. These closure properties
follow from zero-deficit propagation along existing attainers. Yet
\emph{deleting Min edges might improve a value}, so the equality of this
subgame's values with $v|_C$ needs a further argument.

Suppose a restricted finite controller improves $v$ at its start.
Return it to the original arena with two permanent repairs. If its
worst continuation guarantee at the observed state and memory drops
below $v$, switch to an original attainer at that state of $C$.
If Min takes a removed value-increasing edge, switch instead to a
finite sure continuation beating the old value; its target need not
attain its own value. For the contradiction, mix the original value
with the restricted controller's continuation guarantee. A sufficiently
small weight on the latter and full-row averaging make both repairs
nondecreasing in expected potential while retaining the initial strict
gain. The returned original finite sure controller would beat $v$
(Lemma~\ref{lem:r51-attained-restriction}). Thus the restricted values
supply valid original-row anchors without assuming attainment outside $C$.

Existing attainers prove that an anchored lower potential and finite
qualitative witnesses exist. A new construction supplies a controlled
positive margin; it does not store the old attainers' possibly large
memories. Its potentials are analytical certificates, not real-valued
registers in the implementation.

\subsection{Second obstacle: a sure cutoff must pay for fallback}
The construction chooses $u$ nondecreasing on every Min move, on its
selected Max moves, and in conditional mean under every full chance row.
Between events, a finite qualitative witness enforces the recurrent
reward comparison. Figure~\ref{fig:r56-finite-protocol} shows how that
witness is used in one actual run.

\begin{samepage}
Min events are detected from observed edges, and the witness restarts
at the observed successor. After $\kappa$ events the switch to sure
fallback is permanent. Every supported execution, even a null path,
therefore has a sure suffix after finitely many changes; almost-sure
termination alone would not suffice.
\par
\end{samepage}

Let $J\le\kappa$ be the completed-event count on this actual capped
execution, and let $\gamma$ lower-bound its positive gains. Conditional
potential drift earns at least $\gamma\mathbb E J$. Without reaching the
cap, the qualitative witness supplies the limiting reward; at the cap,
fallback can lose at most one. Therefore
\begin{equation}\label{eq:main-event-payoff}
 \mathbb E F\ge u(s)+\gamma\mathbb E J-\Pr(J=\kappa).
\end{equation}
A chance event earns a conditional \emph{mean} gain; individual outcomes
may decrease $u$. A direct cap $\kappa\gamma\ge1$ pays for fallback via
$\mathbb E J\ge\kappa\Pr(J=\kappa)$, but its dependence on reward precision
would be linear in $R$.

\paragraph*{The innovation is a tail relative to the same earned gain.}
The second budget comes from a progress potential $w\in[0,1]^W$
with the same zero/event pattern as $u$. On this protocol, before the
cap, zero steps have zero conditional drift and each counted event
has conditional mean gain at least $1/T_{\mathrm{ev}}$, where
$T_{\mathrm{ev}}=(n+1)Q$ (Lemma~\ref{lem:r50-event-ranking}). Its constraints
use original rows and event signs, not reward thresholds, so this bound
is independent of $R$. Freeze the analysis of $w$ after the same actual
cap transition. Its range is one, so conditional telescoping bounds the
expected number of remaining events by $T_{\mathrm{ev}}$ after every
history, including inside a witness and after a restart. Both potentials
analyze this one execution; $w$ adds no alternate choices or modes.

The reward construction gives positive gains at least $\gamma$, where
we take $\gamma=1/(2RT_{\mathrm{ev}})$, charging each original row once
(Lemma~\ref{lem:r51-anchored-good-margin}).

Take $\ell_R=\lceil\log_2(2R)\rceil$ blocks and cap at
$\kappa=2T_{\mathrm{ev}}\ell_R$. For the first block, Markov gives a
probability at most $\mathbb E J/(2T_{\mathrm{ev}})$. After each completed
block, the conditional remaining-event bound makes the next-block
probability at most $1/2$. Hence
\[
 \Pr(J=\kappa)\le 2^{-\ell_R}\frac{\mathbb E J}{T_{\mathrm{ev}}}
                  \le\gamma\mathbb E J.
\]
This uses the \emph{same actual execution and same $J$} as the reward
estimate, so its loss term is cancelled without lowering $u(s)$.
The cutoff costs $T_{\mathrm{ev}}(1+\log R)$ rather than $RT_{\mathrm{ev}}$.
The reward estimate holds for every cap; the relative argument does
not assume the direct sufficient condition $\kappa\gamma\ge1$.
Combining the counter with fresh qualitative witness modes proves
Theorem~\ref{thm:main-memory} (Appendix D). C.1 supplies the explicit
arena-size factor, separately from the denominator proof used for
classification.

\paragraph*{Why the arithmetic scale is necessary.}
In the six-state lower family, Min chooses between retry and an
absorbing safe reward $1-1/R$. Retry succeeds with probability $1/Q$;
to attain the safe optimum, its all-failure cutoff must satisfy
$(1-1/Q)^j\le1/R$, hence $j=\Theta(Q\log R)$ for $Q,R\ge2$.
Pure finite sure play cannot repeat a TRY mode on that supported
all-failure path, since the resulting infinite path violates hard.
STOP consumes a mode too. This yields the matching fixed-size scale
(Theorem~\ref{thm:r50-joint-memory-lower}), not an optimal dependence on $n$.
At $R=1$, the Boolean $O_n(Q)$ boundary has matching worst-case
dependence using even $Q$ (Proposition~\ref{prop:closed-endpoint-memory}).
A seven-state serial family needs $\Theta(q^2\log R)$ modes with two
row denominators $q$, ruling out linear replacement of $Q$ by their
maximum or lcm up to an arena-only factor
(Corollary~\ref{cor:r55-serial-product-memory}).

\section{Consequences and scope}\label{sec:positive-implementation}\label{sec:related}
For hard and soft parity with at least one fixed index, supplied
thresholds and \textsc{Own-Attain} lie in
$\mathrm{NP}\cap\mathrm{coNP}$ in the same pure-finite sure model
(Appendix G.5). For one hard parity condition and one soft parity
condition, with neither index fixed in the problem definition, supplied
thresholds are coNP-complete. These results address the supplied-threshold
pure-finite direction of~\cite[p.~34:3]{DG26}, not its unrestricted-memory
question. The matching own-attainment and exact-output lower bounds
here use Streett. Randomized finite Max under sure constraints remains
separate.

The ordinary Streett results are stated in
Corollary~\ref{cor:r56-main-ordinary}.
Beyond-worst-case synthesis instead evaluates expectation against a
supplied stochastic environment~\cite{BFRR14}; here expectation is also
minimized over Min histories. The parity-MDP equality test
of~\cite[Theorem~4]{BRR17} deletes and normalizes target-dependent rows;
our construction retains full original rows. Qualitative Boolean
combinations in MDPs have another task and strategy scope~\cite{BGR20}.
Appendix J gives detailed model and transfer comparisons.

We classify numerical requests and establish implementation-size
existence bounds; we do not give an efficient general compiler.
Optimal state-count dependence is not resolved here. Appendix K records
checks and methods, including the exact human/AI verification scope.
The following map is for complete verification, rather than a substitute
for the technical ideas explained above.

\begin{table}[!htbp]\centering
\caption{Entry points for checking the retained claims. The cited proofs
give their additional dependencies.}
\label{tab:reading-guide}
\begin{tabular}{@{}p{.26\linewidth}p{\dimexpr.74\linewidth-2\tabcolsep\relax}@{}}
\toprule
Claim & Complete proof\\\midrule
Thresholds (\ref{thm:main-complexity}) & Theorem~\ref{thm:tight-complexity},
Lemmas~\ref{lem:local-upper}--\ref{lem:zero-slack} and
Lemma~\ref{lem:r36-streett-mdp}. Polynomial bit length uses finite
realization (Lemma~\ref{lem:r51-exact-event-cap}) and the original-row
bound (Theorem~\ref{thm:r48-general-denominators}).\\
Interval opponent (\ref{thm:main-joint}) &
Theorem~\ref{thm:joint-interval} on $W$ and
Corollary~\ref{cor:full-arena} on $V$.\\
Own attainment (\ref{thm:main-own}) &
Theorem~\ref{thm:r47-general-own}, using
Lemma~\ref{lem:r47-numerical-gate}.\\
Exact value (\ref{thm:main-exact}) and ordinary consequences
(\ref{thm:main-ordinary}) &
Theorem~\ref{thm:tight-complexity}'s function part,
Corollary~\ref{cor:r47-value-comparison}, and
Theorem~\ref{thm:r47-single-value}.\\
Fixed-index parity (\ref{thm:fixed-index-parity}) and SSG lower bound &
Lemmas~\ref{lem:fixed-index-exact-realization}--\ref{lem:fixed-index-envelope},
Appendix~\ref{app:fixed-index-certificates}; the separate SSG reduction is in
Appendix~\ref{app:fixed-index-ssg-own}.\\
Memory (\ref{thm:main-memory}) &
Lemma~\ref{lem:qualitative-memory-interface} supplies the explicit
mode-count factor; Theorem~\ref{thm:r51-rational-exact-memory} gives exact
rational memory;
Theorem~\ref{thm:r50-joint-memory-lower},
Corollary~\ref{cor:r55-serial-product-memory}, and Proposition~\ref{prop:closed-endpoint-memory} give its rational and
Boolean fixed-size lower families.\\
\bottomrule
\end{tabular}
\end{table}

\clearpage
\appendix
\section{Finite products and the one-player base}
\label{app:products}
These finite-controller facts provide the starting point for the
quantitative common-opponent theorem.

For the appendices, allow the normalized finite-valued Inf payoff
$F(\rho)=f(\operatorname{Inf}(\rho))$, where
$f:2^V\setminus\{\varnothing\}\to[0,1]$; the Boolean case is
$f=\mathbf1_B$. These semantic arguments allow real reward values;
effective rational statements assume rational values and effective
level sets where explicitly stated.

Write $\SAS(A,\Phi)$ for a pure finite controller that is sure $A$ and
almost-sure $\Phi$ against every public-history Min strategy.

Table~\ref{tab:notation-guide} separates global notation from symbols
declared within one proof. Table~\ref{tab:reading-guide} in the main
text gives entry points to the complete proofs.
In dependency order, A--B supply finite-product facts and attainment;
C is the qualitative memory interface and D the direct arithmetic bounds.
The common-opponent proof E is independent of D's quantitative controller
bound. F extends its domain, G--H give certificates, and I gives reductions.

\begin{table}[ht]\centering
\caption{Notation. Local symbols keep the meaning
declared in their own proof; they do not change the global model.}
\label{tab:notation-guide}
\begin{tabular}{@{}p{.24\linewidth}p{.69\linewidth}@{}}
\toprule
Global notation & Meaning\\\midrule
$V,W,n$ & Original arena, original finite sure region, and $n=|W|$.\\
$v,I,B_v,C_v$ & Constrained value, guaranteed interval, and weak/strict
recurrent comparisons. $u$ proves reward, $w$ counts progress, and $z$
is a guessed upper potential.\\
$Q,R,D,M_{\mathrm{qual}}(n)$ & $Q=\prod_{x\in W\cap V_c}q_x$; $R$ clears reward values;
$D=QR$; $M_{\mathrm{qual}}(n)$ is the qualitative witness bound.\\
Proof-local notation & $H$ is a fixed-domain arena, $U$ a certificate
domain, $S_{=}$ its zero-slack core; $\kappa$ is the event cap,
$T_{\mathrm{ev}}$ bounds the conditional expected event count, and $\ell$
counts its tail blocks.
In I, $\mathsf{lo},\mathsf{ref},\mathsf{hi}$ are lottery roots,
$r_i^\pm$ are test lotteries, $\Xi_t$ is a decoded
ratio, and $D_0$ normalizes the final lottery. Parenthesized $(U_A)$,$(U_f)$ name
the full-domain hypotheses in Appendix E.\\\bottomrule
\end{tabular}
\end{table}

\paragraph*{Randomized Min for a fixed pure controller.}
For a fixed pure controller, allowing Min
behavioral randomization changes neither its worst expected reward nor
its sure requirement: presampling choices on countably many public
histories gives a mixture of pure opponents, and each supported path is
consistent with some pure opponent. This does not purify a randomized
Max controller or classify unrestricted sure Max.

\subsection{Finite-product evaluation and the one-player base}
Fix a pure finite Max controller. Its reachable public product is an MDP
controlled by Min; all chance successors and the fixed Max edges are
retained. An end component $C$ is nonempty and strongly connected, is
closed under those fixed/chance edges, and has an internal Min choice at
each Min vertex. It is cyclic, including the required self-loop in the
singleton case. Under any history policy the recurrent states and edges
form such a component almost surely~\cite{CH07,CDH04full}.

For a set $C$ of product states, write $\operatorname{proj}_V C$ for
its original-state projection. For the general payoff above, $f$ is a function of
original-state sets: any reward-colour map has already been pulled back.
The product reward is therefore $f(\operatorname{proj}_V C)$ when $C$
is the recurrent product set; $g=1-f$ uses the same domain.

\begin{lemma}[Touring a recurrent product]\label{lem:general-tour}
Min can remain in an end component $C$ and make every product vertex
of $C$ recur almost surely, with a pure finite controller. Consequently,
for the baseline's minimizing expected reward $q$, if
$a=f(\operatorname{proj}_V C)$, then $q(x)\le a$ at every $x\in C$.
\end{lemma}
\begin{proof}
Store a cyclic target in a list of the vertices of $C$. Until reaching
that target, choose at Min vertices an edge on a shortest internal path
to it. Every other vertex has a fixed edge or a positive chance edge
on such a path. From every current state, a block of $|C|$ steps hits the
target with conditional probability at least $\eta^{|C|}>0$, where $\eta$
is the smallest positive chance probability (one if there is no chance).
Repeated conditional blocks hit it almost surely. Advance the target
and intersect the countably many probability-one hitting events. Closure
keeps the path in $C$; all vertices recur, giving exactly its original-state projection.
This proves the asserted expected payoff bound. Touring the whole
component ensures the required arbitrary-Muller reward.
\end{proof}

\begin{lemma}[Exact baseline values]\label{lem:general-q}
Every finite baseline has a pure finite minimizing reply. When the reward
values are rational and are effectively given together with their level
sets, its values $q$ are rational and effectively computable.
Its fixed Max edges preserve $q$,
chance edges average $q$, and each Min edge does not decrease $q$.
Every supported continuation of a sure baseline has $q(s,m)\le v_H(s)$.
\end{lemma}
\begin{proof}
The last inequality follows from admissibility and prefix independence.
For evaluation set $g=1-f$. Enumerate all end components $C$,
and at each vertex $x$ offer a stop reward
\[
 w(x)=\max\bigl(\{g(\operatorname{proj}_V C):x\in C\}\cup\{0\}\bigr).
\]
Maximize this ordinary stopping reward, with Min as stopping maximizer
and never-stop reward zero. A stop of reward $w$ is represented
by a chance edge to success/failure absorbers with probabilities $w,1-w$;
zero probabilities are omitted. This auxiliary finite reachability MDP
has a pure positional optimum, even when these auxiliary probabilities
are real; rationality is needed only for the effective rational conclusion.

Let $u$ be its value. Against any original history policy, $u(X_n)$ is
a bounded supermartingale and converges. Almost surely the recurrent
support is an end component $C$; every recurrent $x$ has
$u(x)\ge w(x)\ge g(\operatorname{proj}_V C)$. Hence
$\mathbb E[g]\le\mathbb E[u_\infty]\le u(X_0)$.
Conversely, follow an optimal positional stopping policy. On a component
stop, tour an end component attaining its chosen reward by
Lemma~\ref{lem:general-tour}; on an artificial zero-reward stop with no
chosen component, use any fixed pure continuation, whose actual $g$
reward is nonnegative. Never-stop paths also have nonnegative actual
$g$ reward, at least the auxiliary zero reward. Thus the resulting finite
pure policy achieves at least $u$; the upper bound forces equality.
For effectively given rational rewards, the auxiliary probabilities and
reachability values are rational and computable, so $q=1-u$ is too.
Prefix independence and splicing optimal continuations give the Bellman
identities. The positional policy is used only in the auxiliary
reachability MDP; touring supplies the finite continuation after each
component stop.
\end{proof}

Finite tagged unions of successor near-optimal controllers prove
the constrained Bellman equations on every fixed-domain subgame $H$.
Only successors in that fixed original domain are included: Max edges
leaving it have been deleted, while every original Min and chance
successor remains in the domain:

\[
 v_H(s)=\begin{cases}
 \max_{s\to t}v_H(t),&s\in V_{\mathrm{Max}},\\
 \min_{s\to t}v_H(t),&s\in V_{\mathrm{Min}},\\
 \sum_t p(s,t)v_H(t),&s\in V_c.
 \end{cases}
\]
For the upper inequalities, every supported continuation remains sure,
and the opponent may condition on the observed successor. Let the
continuation tolerances tend to zero. The equations apply also at
nonattained suprema because the construction uses near-optimal finite
continuations.

\begin{lemma}[One-player stopping characterization]\label{lem:general-mdp}
In an MDP on the fixed original $W$, let $\mathcal R_f$ be the finite reward range
with zero added, and set
\[
 R_a=\{s:\text{finite sure-}A/\text{AS}(f\ge a)\text{ is possible}\},
 \qquad w(s)=\max\{a\in \mathcal R_f:s\in R_a\}.
\]
Then the constrained value is the ordinary optimal stopping value with
stop reward $w$ and never-stop reward zero. It admits sure approximations with
$O_H(1+\log(1/\epsilon))$ memory states.
The value is rational, and the value and these witnesses are effectively
computable, when the finite reward values are rational and are effectively
given together with their level sets.
\end{lemma}
\begin{proof}
The original sure fallback makes $R_0=W$. A finite sure controller
induces a finite chain. In a bottom class $C$ its reward is
$a_C=f(\operatorname{proj}_V C)$ almost surely. At every product state in $C$, its
inherited continuation is sure $A$ and almost surely at least $a_C$.
Thus $w(s)\ge a_C$ at its projected vertex. In the auxiliary stopping
MDP follow this controller and stop upon first entry into any of its
bottom classes. This stopping time is observable in its known finite
product, almost surely finite, and gives at least its actual reward.
Ordinary optimal stopping therefore upper-bounds the constrained value.

Conversely run an optimal positional stopping policy for a finite
timeout. On a stop at $s$ enter a fixed finite SAS witness for $w(s)$;
on timeout enter the original sure fallback. A finite prefix has no
effect on $A$. The expected reward is at least the reward of stops before
timeout, by nonnegativity. These increase to the stopping value.
The fixed finite stopping policy has a geometric eventual-stop tail:
from every state with positive stop probability there is a positive path
to a stop of length at most $|W|$. Blocks of that length give a uniform
positive hitting probability; states of zero stop probability add no
missing stop mass. A timeout $O_H(1+\log(1/\epsilon))$ loses at most
$\epsilon$. Its clock, finitely many tagged SAS witnesses and the finite
fallback have the asserted state count.

For effectively given reward level sets, all SAS sets are effective:
deterministic Muller-to-parity monitors and
the finite parity-SAS procedure~\cite{CP19} preserve finite existence in
both directions. Here the monitor implements the oracle; the final Min
policy remains on original vertices. For Boolean $f=\mathbf1_B$, this stopping problem is precisely
maximal reachability to the finite SAS region $R_1$.
\end{proof}

\section{Exact attainment for finite Inf rewards}
\label{app:graded-attainment}

Fix the original sure-$A$ region $W$ and the pure finite-controller class.
The hard condition is prefix independent and has a finite sure fallback on
$W$; the payoff is $F(\rho)=f(\operatorname{Inf}(\rho))\in[0,1]$, where
$f$ is a finite-valued function of recurrent original states. Original
colour objectives are pulled back to state sets. In products restricted
to $W$, $\operatorname{proj}_W$ denotes the same original-state projection
as $\operatorname{proj}_V$, with image contained in $W$. Write
\[
 v(s)=\val_F(s)=\sup_{\sigma\text{ finite, sure }A}
                         \inf_\tau\Exp_s^{\sigma,\tau}[F].
\]
Finite tagged unions of successor near-optimal controllers give the
constrained Bellman equations without assuming attainment. Max edges
leaving the fixed original $W$ are deleted; all original Min and chance
successors remain in $W$. Thus retained Max edges do not increase $v$,
Min edges do not decrease $v$, and chance
averages $v$. Let $K_v$ keep exactly the value-preserving edges of both
players and \emph{every original chance edge with its original probability}.
Player successor sets remain nonempty. Define the Muller objective
\[
 B_v(\rho)\iff
 f(\operatorname{Inf}(\rho))\ge
            \max\{v(t):t\in\operatorname{Inf}(\rho)\}.
\]
The maximum refers to original states, including when several states have
the same reward colour. No chance conditioning or positive-value restriction
is used.

\begin{theorem}[Finite graded attainment]\label{thm:graded-attainment}
For every $s\in W$, a finite sure-$A$ controller attains $v(s)$ in the
original game if and only if $s$ admits a finite sure-$A$/almost-sure-$B_v$
controller in $K_v$.
\end{theorem}

We first record the finite-product facts needed in both directions. For a
fixed finite sure controller, let $q(x,m)$ be its minimizing expected
continuation payoff in the full original Min-controlled product. Every
supported suffix is admissible, so $q(x,m)\le v(x)$. Prefix independence
and continuation splicing give equality on its fixed Max edges, the
original averaging equation at chance, and $q(\mathrm{next})\ge q(\mathrm{now})$
on each Min edge.

\begin{lemma}[Whole-component tour]\label{lem:graded-tour}
If $D$ is a cyclic product end component of a fixed finite controller,
then
\[
 q(x,m)\le f(\operatorname{proj}_W D)\qquad((x,m)\in D).
\]
Moreover, a fixed finite controller is almost-sure $B_v$ precisely when
every reachable product end component satisfies
\[
 f(\operatorname{proj}_W D)\ge\max_{(x,m)\in D}v(x). \tag{EC}
\]
\end{lemma}
\begin{proof}
Min cycles a target list containing every vertex of $D$, choosing internal
shortest paths towards the current target. Positive finite chance support
gives a uniform positive chance of reaching each target within a bounded
block. Iterating conditional block bounds makes every target recur almost
surely, and closure prevents exits. The resulting Inf set is exactly the
projection of $D$, proving the payoff bound. A reachable violating
component can first be reached with positive probability and then toured,
proving necessity of (EC). Conversely, under any history-dependent Min
policy, the recurrent product support is almost surely an end component:
conditional chance fairness closes its recurrent sources under every
positive successor. Its reward then satisfies (EC). A single recurring
target would not suffice for a general Muller reward.
\end{proof}

\begin{proof}[Proof of Theorem~\ref{thm:graded-attainment}]
Suppose first that a finite sure controller $\sigma$ attains $v(s)$, so\\
$q(s,m_0)=v(s)$. Restrict Min to preserving edges. At every supported
history under this restriction, $q=v$ propagates as follows. At Max,
\[
 q(\mathrm{next})=q(\mathrm{now})=v(x),\qquad
 q(\mathrm{next})\le v(t)\le v(x),
\]
forcing a preserving action and equality at the successor. At chance,
each successor deficit $v(t)-q(t,m')$ is nonnegative and their weighted
sum is zero. Positivity forces every deficit to vanish, even when successor
values differ. At a preserving Min edge,
\[
 q(\mathrm{next})\ge q(\mathrm{now})=v(x)=v(t)\ge q(\mathrm{next}).
\]
Thus $\sigma$ is legal in $K_v$ on all reachable histories and remains
sure $A$. Unreachable entries can be completed by arbitrary preserving
actions without changing reachability. Any reachable product end component
in this restricted game is also a legal component in the original fixed
controller product: fixed Max and all original chance successors remain,
and its internal Min choices are original choices. The tour bound and
$q=v$ at every component vertex give (EC). Hence $\sigma$ is AS $B_v$.

Conversely let $\pi$ be a finite sure-$A$/AS-$B_v$ controller in $K_v$.
For each original Min edge $e=(x,y)$ with $v(y)>v(x)$, choose a finite
original sure controller $\beta_e$ from $y$ with guarantee strictly above
the \emph{old} value $v(x)$. For example use tolerance
$(v(y)-v(x))/2$ at $y$. This does not assume that $v(y)$ is attained.
Run $\pi$ in a main mode and, on the first strict Min improvement,
switch permanently to its tagged $\beta_e$. Finitely many tags suffice.

Every supported path without a switch is a main path in $K_v$ and
satisfies $A$. Every switched path has a finite prefix and a sure suffix.
Thus the resulting finite controller $\sigma$ is sure $A$, including on
probability-zero paths.

Fix an arbitrary original Min history strategy. Use potential $v$ in main
mode and the exact baseline $q_e$ in a fallback mode. Preserving player
steps and original chance rows have zero main drift. The switching edge
has positive drift because $q_e(y,m_e)>v(x)$; inside a fallback the
baseline equations give nonnegative drift. This bounded submartingale
$Y_n$ converges and
\[
                  \Exp[Y_\infty]\ge Y_0=v(s).
\]
Almost surely its finite-controller recurrent support $D$ lies in one
mode. In fallback mode the whole-component tour bounds every recurrent
$q_e$ by the actual reward, so $Y_\infty\le F$. In main mode all preceding
Min edges were preserving, and full original chance fairness makes $D$ a
reachable end component of $\pi$ in $K_v$. By (EC),
\[
 Y_\infty\le\max_{(x,m)\in D}v(x)
             \le f(\operatorname{proj}_W D)=F.
\]
This argument uses the unconditioned process; it does not condition chance
on the event of never switching. Therefore $\Exp_s^{\sigma,\tau}[F]\ge v(s)$
against the arbitrary opponent. Admissibility gives exact attainment.
\end{proof}

The restriction of Min in the main game is essential. If Min can choose
between an absorber paying $\alpha<1$ and a retry state of unattained
value $1$, the choice state can attain $\alpha$ by a finite retry-and-stop
continuation. Retaining the improving edge in the main test would wrongly
demand attainment of $1$ downstream. Similarly, a chance split between
rewards $0$ and $1$ attains its expectation $1/2$, but fails AS$(F\ge1/2)$.
The recurrent threshold in $B_v$ and the old-value repair address these two
different obstructions. Ordinary value-class threshold theory~\cite{DGG25}
does not impose the separate sure admissibility requirement.

\begin{corollary}[Effective boundary decisions]\label{cor:graded-thresholds}
For finite Muller $A$ and a supplied effective rational vector $v$, finite
attainment is decidable. A non-strict expected threshold $\theta$ is feasible
iff $v(s)>\theta$, or $v(s)=\theta$ and the test of
Theorem~\ref{thm:graded-attainment} succeeds. A strict expected threshold is
feasible iff $v(s)>\theta$.
\end{corollary}
\begin{proof}
The objective $B_v$ is finite Muller, without any required union closure.
Deterministic Muller monitors~\cite{CCFL24} and finite parity
$\SAS$~\cite{CP19} decide the test and preserve finite strategy existence.
Below the supremum a finite near-optimal controller suffices. For the
strict boundary, a fixed finite-controller Muller-reward MDP has a finite
minimizing reply: evaluate $1-f$ by optimal stopping at its finitely many
end-component rewards and then tour the selected component. Hence one
controller cannot have every reply strictly above $\theta$ while its
infimum equals $\theta$.
\end{proof}

\section{Qualitative memory for the finite implementation}
\label{app:r44-attained-memory}
\subsection{The qualitative interface}
The quantitative construction needs a finite sure/almost-sure witness
whose memory depends only on the original arena size. The reductions
below use one interface. The forward translation is used only for
existence: a chance start maps to its gadget source, and gadget vertices
copy hard labels. The reverse translation retains the same memory set
and every original chance edge and probability. Between the translations,
a deterministic parity monitor supplies the memory bound.
Appendix~\ref{app:fixed-index-certificates} uses this same interface with
a monitor tailored to the fixed indices.

\begin{lemma}[Qualitative memory interface]
\label{lem:qualitative-memory-interface}
Let $A,B\subseteq 2^V$ be arbitrary Muller conditions on an $n$-vertex
finite stochastic arena, where $n\ge1$. If a pure finite controller is
sure $A$ and almost-sure $B$ from a specified start against arbitrary
pure public-history \Min, such a controller exists with at most
\[
 M_{\mathrm{qual}}(n)=n\,n!\,(2n+1)^{n+1}
\]
original controller modes. The bound counts monitor modes; the final
controller retains every original chance edge with its original probability.
\end{lemma}
\begin{proof}
\paragraph*{Turn both Muller conditions into parity.}
Use one move-to-front latest-appearance record for both conditions.
Its mode is $(\pi,k)$, where $\pi$ is a permutation of $V$ and $k$ is
the old position of the last vertex read. Reading a vertex moves it to
the front. Let $S$ be the first $k$ vertices of the updated permutation.
For $J\in\{A,B\}$ emit the max-even priority $2k$ if $S\in J$, and
$2k-1$ otherwise. There are $n\,n!$ modes and both priorities are at most
$2n$. If the recurrent set is $I$, then eventually the first $|I|$
positions are exactly $I$, and the largest position read infinitely often
is $|I|$. Thus each emitted parity condition is equivalent to its Muller
condition. A shared record suffices even when $A$ and $B$ use different
colour maps, since both are pulled back to original state sets.
The record is determined by the public history and updated on each
observed original edge; the product preserves all chance probabilities.

\paragraph*{Use finite SAS only as an existence interface.}
Write $G_{\rm L}$ for this stochastic record product. Apply the
finite-memory equivalence of \cite[Theorem~8]{CP19}, using the even
order reversal $p\mapsto2n-p$ to match its min-even convention.
A pure finite SAS winner in $G_{\rm L}$ exists if and only if a winner
exists in the corresponding deterministic conjunction game $G_{\rm D}$.
A player start is unchanged, while a chance start $x$ maps to its
gadget source $\bar x$~\cite[p.~6:8]{CP19}.
The forward proof is \cite[Appendix, Lemma~4.8]{CP18full}.
Gadget vertices copy their source's hard priority, and their soft
priorities remain in $\{0,\ldots,2n\}$. Only existence is needed in this
direction: the input controller's modes and its fixed-product ranks
will not enter the new memory bound.

\paragraph*{Bound the new memory on the deterministic arena.}
Convert the two gadget priorities back to max-even by the same even
order reversal. The intersection monitor of
\cite[Section~3, Theorem~2]{DG26} reduces their conjunction to one
parity condition on every path. It has one register in
$\{0,\ldots,2n\}$ for each even first priority, hence at most
$(2n+1)^{n+1}$ register states. The resulting finite parity game has a
pure positional winner. Viewed on $G_{\rm D}$, this winner therefore
uses only the registers as memory; the number of gadget vertices is
not an additional mode factor.

\paragraph*{Fold back without adding gadget modes.}
CP19's reverse construction retains that same memory set
\cite[pp.~6:9--6:11]{CP19}. At a chance source, its current memory and
the observed original successor determine a compatible virtual gadget
path and its terminal memory update. These updates are compiled into
the controller's public-edge update function; no intermediate gadget
location is stored. Every original supported chance successor is kept
with its original probability. The cited construction preserves hard
parity on every supported path and soft parity almost surely against
arbitrary public-history \Min. Finally project $G_{\rm L}$ back to the
original arena, storing the shared record together with the registers.
The resulting number of modes is at most
$n\,n!\,(2n+1)^{n+1}=M_{\mathrm{qual}}(n)$, as required.
\end{proof}

The explicit formula gives $2^{O(n\log(n+1))}$ dependence on arena
size. For the fixed-$n$ dependence on $Q$ and $R$, only a qualitative witness bound depending on $n$ is needed.
This is a qualitative mode-existence statement for arbitrary Muller
conditions, not a complexity classification or a quantitative repair.
The latter is constructed separately in
Appendix~\ref{app:r48-general-subpotential}.

\section{General original-controller memory from subpotentials}
\label{app:r48-general-subpotential}

This section needs neither $(U_A)$ nor $(U_f)$. Let $A$ be arbitrary Muller and
$f$ an arbitrary rational finite-Inf reward in $[0,1]$. Retain the
original pure-finite sure region $W$, all original Min edges and full
chance rows, and put $n=|W|\ge1$, $Q=\prod_{x\in W\cap V_c}q_x$, $D=QR$, where $q_x$
clears an original chance row and $R$ clears every reward. Max remains
pure finite, Min arbitrary pure public history, and hard is required on
every supported path, including null paths. Write
$M_{\mathrm{qual}}(m)=m\,m!\,(2m+1)^{m+1}$ for the arbitrary-Muller qualitative
sure/AS bound of Appendix~\ref{app:r44-attained-memory}.

\paragraph*{The exact-memory proof.}
We first define good lower potentials and give them finite exact
realizations. This makes the value-denominator argument independent
of approximation. The denominator and event-ranker estimates then
lead to the attained-state restriction, an anchored reward margin,
and the relative tail estimate. These ingredients are consecutive,
ending with Theorem~\ref{thm:r51-rational-exact-memory} and its
fixed-size rational and Boolean lower families.

The final controller is sure from every initialized state of the
original $W$ and attains the value at every finitely attained start;
no attainment premise is imposed on the other states.

\subsection{Good subpotentials and their realization}
Call $u\in[0,1]^W$ a subpotential if all Min edges are nondecreasing,
every Max source has a nondecreasing action, and each original chance
row has average at least its source value. Its qualitative exit game
$Z(u)$ has $W$ and one accepting absorbing marker. Delete negative Max
edges; retain zero player differences and full zero-slack chance rows;
redirect positive player edges and positive chance \emph{sources} to
the marker. On marker-free Inf sets the predicates are $A$ and
$B_u(S)\iff f(S)\ge\max_{x\in S}u(x)$; all marker-containing sets
accept. Call $u$ good if every original state has a finite sure/AS
witness in $Z(u)$. The marker is a qualitative device, not an original
chance-row modification.

\begin{lemma}[Max-mode extraction]
\label{lem:r48-sharp-extraction}
For every $\eta>0$ there is a good $u_\eta$ with
$v-\eta\le u_\eta\le v$, without assuming rationality of $v$.
\end{lemma}
\begin{proof}
Tag finite sure $\eta$-approximations at all original starts. In their
reachable initialized products take $u_\eta(x)=\max_mq(x,m)$.
Every mode in this finite maximum is a sure admissible residual, and
each original state has its initialized mode. The finite-product
equations give the claimed bounds and subpotential inequalities.

Initialize at a maximizing mode. On marker-free histories $q=u_\eta$
propagates: at zero player edges by preservation/nondecrease of $q$
and equal endpoint upper bounds; at zero chance slack because all
nonnegative successor deficits have zero weighted average under the
full row. Every selected Max edge is therefore nonnegative. A
marker-free product end component remains an original component;
whole-component touring proves $f(S)\ge\max_{x\in S}u_\eta(x)$.
Original sure acceptance proves the hard part on every marker-free
path, and finite-product recurrence proves the AS part. Marker paths
accept both predicates.
\end{proof}

Implement good-$u$ exit witnesses as follows. Zero moves continue the
current witness. A positive Max marker choice executes a fixed
corresponding original positive action; a positive Min edge is observed
normally; a positive chance source executes its \emph{entire original
row}. After the actual edge/outcome initialize the witness assigned to
the actual successor. Thus each positive step is one event, with no
persistent marker or extra phase mode.

The potential proving reward need not also bound the number of events.
For a fixed $u$, call $w\in[0,1]^W$ an \emph{event ranker of margin $1/T_{\mathrm{ev}}$}
if every zero player difference and chance-source slack of $u$ remains
zero for $w$, and every positive such difference or slack is at least
$1/T_{\mathrm{ev}}$ for $w$. Negative Max edges need no constraint: the witnesses
never execute them. No reward comparison or goodness is required of $w$.

\subsection{Exact caps for the reward subpotential}
\label{app:r51-exact-event-cap}

We first realize a good lower potential with a finite cap. The positive
increments of the reward subpotential $u$ itself provide accumulated
expected gain that pays for a permanent sure fallback. The event-ranker
and value-denominator estimates enter later.

Throughout this lemma $A$ is Muller and $F=f(\operatorname{Inf})\in[0,1]$
as defined above. The lemma also applies on a finite original-state domain
$\mathcal U$ closed under all original Min edges and supported chance
outcomes, with an internal Max successor at every Max state. Keep all
internal Max edges and every full original chance row, and assume a
finite sure-$A$ fallback staying in $\mathcal U$ from every state.
Goodness and $Z(u)$ are defined on this domain as above. Write
$d=|\mathcal U|$ and retain the qualitative bound $M_{\mathrm{qual}}(d)$.

\begin{lemma}[Exact event cap]
\label{lem:r51-exact-event-cap}
Let $u\in[0,1]^{\mathcal U}$ be good. Suppose every positive player
difference and positive chance-source slack of $u$ is at least
$\gamma>0$. For every integer $\kappa\ge1$, the witness/reset protocol
defined in the proof, capped after $\kappa$ completed actual events,
is a pure finite original controller sure $A$ from every initialized
$z\in\mathcal U$. A chance event samples its full original row before
the reset or fallback switch. The protocol uses at most
\[
                         \kappa dM_{\mathrm{qual}}(d+1)+dM_{\mathrm{qual}}(d)
\]
modes. Against every pure public-history Min strategy, let $J\le\kappa$
be its completed-event count. Its reward satisfies
$\mathbb E F\ge u(z)+\gamma\mathbb E J-\Pr(J=\kappa)$.
In particular, if $\kappa\gamma\ge1$, it guarantees at least $u(z)$.
If there is no positive form, the tagged auxiliary witnesses give
this guarantee without a cap.
\end{lemma}
\begin{proof}
Choose a sure-$A$/AS-$B_u$ auxiliary witness for each original start.
Implement its zero moves unchanged. At a positive Max marker choice
execute a corresponding original positive edge; observe a positive
Min edge normally; at a positive chance source sample its entire
original row. After each such actual transition reinitialize the
witness assigned to its observed successor. Count these completed
positive events, and after event $\kappa$ initialize original sure fallback
there permanently. If the $\kappa$-th event is at a chance source,
sample its full row before this fallback switch.

Fix an arbitrary pure public-history Min strategy. Let $\mathcal F_t$
be the original history after $t$ transitions, let $I_t$ indicate a
main-mode positive event on transition $t\to t+1$, and put
\[
 J_t=\sum_{r<t}I_r,\qquad
 \theta=\inf\{t:J_t=\kappa\},\qquad
 Y_t=u(X_{t\wedge\theta}),
\]
where $\theta=\infty$ if the cap is never reached and $I_t=0$ after
the cap. A positive chance source is known before its outcome is
sampled. At player states, the fixed pure controller or fixed pure
history opponent determines the action from $\mathcal F_t$. Hence
$I_t$ is $\mathcal F_t$-measurable in this analysis; the implemented
controller detects a Min event only on its observed edge.

Zero steps have zero conditional $u$ drift. Positive player steps
have increment at least $\gamma$, and positive chance sources have
conditional mean increment at least $\gamma$ under the full original
row. This is a conditional mean statement: individual outcomes may
decrease $u$. Thus
\[
 \mathbb E[Y_{t+1}-Y_t\mid\mathcal F_t]\ge\gamma I_t,
 \qquad
 \mathbb E Y_t\ge u(z)+\gamma\mathbb E J_t.
\]
The bounded submartingale $Y_t$ converges almost surely and in $L^1$;
also $J_t\uparrow J\le \kappa$. Finite telescoping at deterministic
horizons, followed by bounded and monotone convergence, gives
\begin{equation}\label{eq:r51-exact-cap-drift}
 \mathbb E Y_\infty\ge u(z)+\gamma\mathbb E J.
\end{equation}

On a noncap path there are fewer than $\kappa$ events, hence a final
marker-free auxiliary-witness suffix. For each finite initialization
history, its original continuation and the auxiliary witness have
the same law until the next event or marker. The auxiliary event
``no later marker and $B_u$ fails'' is null. Excluding these events
for the countably many initialization histories proves, almost
surely on noncap paths,
\[
             F\ge\max_{x\in\operatorname{Inf}X}u(x)\ge Y_\infty.
\]
This is a null-event inclusion in the original unconditioned law,
not conditioning a row on the final restart. On cap paths the reward
is nonnegative and $Y_\infty\le1$. Therefore, with
$p_{\rm cap}=\Pr(\theta<\infty)=\Pr(J=\kappa)$, \eqref{eq:r51-exact-cap-drift} implies
\begin{equation}\label{eq:r51-exact-cap-payoff}
 \mathbb E F\ge u(z)+\gamma\mathbb E J-p_{\rm cap}.
\end{equation}
This proves the reward estimate for every integer cap $\kappa\ge1$.
If $\kappa\gamma\ge1$, then $\mathbb E J\ge\kappa p_{\rm cap}$ gives
$\mathbb E F\ge u(z)+(\gamma\kappa-1)p_{\rm cap}\ge u(z)$.
The reward potential $u$ supplies this estimate. The event ranker $w$
will instead bound the number of events.

Sure acceptance and mode counting need only a finite cap. Take any
supported path, including a null path.
It either enters sure fallback after a finite prefix or has finitely
many restarts and a final marker-free sure auxiliary suffix.
Prefix independence gives $A$ in both cases. There are $\kappa$ pre-cap
counter values, at most $dM_{\mathrm{qual}}(d+1)$ tagged main modes, and at most
$dM_{\mathrm{qual}}(d)$ tagged fallback modes. Every update uses the actual original
edge, giving the stated original-controller count. Initialization
depends on the original start; the same construction works against
every pure public-history opponent. If no positive form exists, no
restart or cap occurs and the same no-event argument applies.
\end{proof}

\subsection{Cell closure and a common value denominator}
\begin{lemma}[Good-cell closure]
\label{lem:r48-sharp-cell-closure}
Partition the subpotential cube by player-edge signs, chance
zero/positive slack, and comparisons $u(x)\le f(S)$ versus
$u(x)>f(S)$ for $x\in S\ne\varnothing$. Within a nonempty cell
goodness is constant. Its closure is its bounded polytope of weakened
strict inequalities. Every good-cell closure is coordinatewise bounded
above by $v$, and one contains $v$ as its unique coordinate-sum
maximizer, hence a vertex.
\end{lemma}
\begin{proof}
Within a cell the graph and both predicates of $Z(u)$ are fixed, so
goodness is constant. Let $P_{\mathcal C}$ be the weak relaxation of a nonempty
cell $\mathcal C$, and fix a seed $c\in \mathcal C$. For any $x\in P_{\mathcal C}$ and $0<t\le1$,
the point $(1-t)x+tc$ satisfies every strict row: the weak margin at
$x$ is nonnegative and the corresponding seed margin is positive.
Weak and equality rows also remain valid. These points lie in $\mathcal C$ and
approach $x$, proving $P_{\mathcal C}\subseteq\overline{\mathcal C}$; the reverse inclusion
follows from closedness of the relaxed inequalities.

If $\mathcal C$ is good, every point of $\mathcal C$ is at most $v$ by
Lemma~\ref{lem:r51-exact-event-cap}: when positive forms exist, take
their finite minimum $\gamma>0$ and $\kappa=\lceil1/\gamma\rceil$; with no
positive form, use the lemma's no-cap case. Coordinatewise order is closed,
so $P_{\mathcal C}\subseteq\{z:z\le v\}$ as well. The boundary upper
bound follows by closedness, including at points whose exit game is
not good.

The good $u_\eta\to v$ from Lemma~\ref{lem:r48-sharp-extraction}
have an infinite subsequence in one of finitely many cells. Hence
$v\in P_{\mathcal C}$ for one good $\mathcal C$. Every other point has no larger coordinate
and at least one smaller coordinate, so it has a strictly smaller
coordinate sum. Thus $v$ is the unique maximizer of this linear
functional, and hence a vertex even if $P_{\mathcal C}$ is lower dimensional.
\end{proof}

\begin{theorem}[General original-state denominators]
\label{thm:r48-general-denominators}
The original pure-finite value vector has a \emph{common} clearing
denominator $h_v\le D=QR$. Every reduced denominator is therefore at
most $D$; divisibility by $D$ is not asserted.
\end{theorem}
\begin{proof}
Choose $n$ independent active rows at the vertex $v$ of the good-cell
closure in Lemma~\ref{lem:r48-sharp-cell-closure}, including
equality rank. In their unscaled matrix $B$, each row up to sign is a
unit $e_i$, an incidence $e_i-e_j$, or an original chance row
$e_i-\sum_jp(i,j)e_j$. There is at most one independent chance row
per source. RHS entries are $0,\pm1,\pm f(S)$, hence cleared by $R$.

A square matrix of signed unit/incidence rows has determinant
$0,\pm1$. Induct by expanding a column with at most one nonzero entry.
Otherwise every column has at least two; as every row has at most two,
all rows have two and sum to zero, making the matrix singular.
Each chance row is a convex combination of incidence rows, so
multilinear expansion gives $|\det B|\le1$.

Multiply only selected chance rows by their $q_x$ and call the integer
matrix $A=CB$. Its nonzero integer determinant $\Delta$ satisfies
$|\Delta|\le\prod_{\rm selected}q_x\le Q$. If $b$ is the original
RHS, $A(Rv)=C(Rb)$ is an integer system. Cramer's rule gives the common
clearer $h_v=R|\Delta|\le RQ$. The construction uses only original chance rows and the reward clearer.
\end{proof}

For ordinary Boolean Muller, the bound $h\le Q$ also follows by a
determinant estimate from the original-state chain of~\cite[Lemma~9]{CH07};
its printed Lemma~12 is coarser.

\subsection{An event ranker independent of reward precision}
\begin{lemma}[Original-row event ranking]
\label{lem:r50-event-ranking}
Every subpotential $u\in[0,1]^W$ has an event ranker of margin at least
$1/((n+1)Q)$, so one may take $T_{\mathrm{ev}}=(n+1)Q$.
The bound is independent of $u$, the reward denominator,
and the requested approximation tolerance.
\end{lemma}
\begin{proof}
Use variables $w\in[0,1]^W$ and $\gamma\in[0,1]$. Preserve as
equalities every zero player difference and zero chance-source slack
of $u$. Require every positive player difference and positive
chance-source slack to be at least $\gamma$. Negative Max edges are
unused and impose no condition. These are precisely the event-ranking
constraints. Their coefficients use the event pattern and original
chance rows alone.

The substitution $w=u$ is feasible for some $\gamma>0$, because the
finite positive list has a positive minimum; if that list is empty,
take $\gamma=1$. Maximize $\gamma$ on this nonempty bounded rational
polytope and choose an optimal vertex. The vertex has positive
$\gamma$, although the feasible seed $u$ need not be rational.

Choose $n+1$ independent active rows, including equality rank. Their
unscaled $\gamma$ coefficients lie in $\{0,1,-1\}$. Deleting that
column leaves signed unit, incidence or original chance rows, or a zero
row from a $\gamma$-only bound. By the determinant argument in
Theorem~\ref{thm:r48-general-denominators}, each $n$-square minor has
absolute determinant at most one. Expansion along the $\gamma$ column
therefore bounds the unscaled determinant by $n+1$.

At most one independent chance row occurs per original source: a zero
source has one equality form, and a positive source has one margin
form. Signed copies of an equality are dependent. Multiply each selected
chance row, including its $\gamma$ coefficient, by $q_x$. The resulting
integer invertible matrix $A$ satisfies
\[
                         1\le|\det A|\le(n+1)Q.
\]
Every right-hand side is zero or one, and each chance form has right-hand
side zero. Thus the scaled right-hand side is already integer. Cramer's
rule expresses the positive $\gamma$ as a nonzero integer divided by
$\det A$, proving the bound. Original row clearers suffice for these
integral right-hand sides.
\end{proof}

\subsection{The restriction to finitely attained states}
\label{app:r51-attained-restriction}

Continue with the normalized finite rational Inf reward $F=f(\operatorname{Inf})$
and its chosen positive integer reward clearer $R$. Put
\[
 C=\{x\in W:v(x)\text{ is finitely attained in the original game}\},
 \qquad N=W\setminus C,
\]
and let $Q_C$ and $Q_N$ be the products of the original chance-row
clearers sourced in $C$ and $N$, respectively. Thus $Q=Q_CQ_N$.
The following value restriction supplies rational constants on $C$
without assuming small original attaining controllers.

\begin{lemma}[Attained-state restriction]
\label{lem:r51-attained-restriction}
Every chance source in $C$ has all its supported successors in $C$.
Every value-preserving Min successor of a state in $C$ belongs to
$C$, and each Max state in $C$ has a value-preserving successor in
$C$. Let $G_C$ have vertex set $C$. It keeps every full original
chance row sourced in $C$ and exactly the original value-preserving
player edges with both endpoints in $C$. Its
pure-finite sure values are $v|_C$. If $C\ne\varnothing$, the
original-row value-vertex construction supplies an integer nonsingular
matrix $A_C$ with $1\le|\det A_C|\le Q_C$ and a common clearer
\[
 H_C=R|\det A_C|,\qquad R\mid H_C,\qquad H_C\le RQ_C.
\]
This same $H_C$ clears all values $v|_C$ and all original rewards $f(S)$.
\end{lemma}
\begin{proof}
\emph{Closure and restriction.}
For an original attaining controller, every admissible residual
baseline $b(x,m)$ is at most $v(x)$. Starting with $b=v$, a full
chance row has a zero weighted sum of nonnegative successor
deficits, so every supported successor residual attains its value.
On a preserving Min edge, baseline nondecrease and the equal
endpoint values likewise force equality. On the selected Max edge,
baseline preservation and the original Max Bellman inequality force
a preserving attained successor. These are the claimed closure
properties. Min has a preserving successor by its finite Bellman
minimum, so every player row of $G_C$ is nonempty.

Restrict an original attainer to preserving Min play. The same
zero-deficit propagation makes every reached original state belong
to $C$ and every selected Max edge preserving. It is a legal finite
sure controller in $G_C$ with its original guarantee; unreachable
entries can be completed by legal preserving actions. Thus the
restricted value is at least $v|_C$, and the finite-sure region of
$G_C$ is all of $C$.

\emph{Return an improved controller to the original game.}
For the reverse inequality, suppose a finite restricted sure
controller $\sigma$ has baseline $b(s,m_0)>v(s)$. Use its true
minimizing residual baseline $b(x,m)$ in its finite restricted
Min-MDP. It preserves its fixed Max moves, averages every full
chance row, and is nondecreasing on restricted Min edges. Choose
an original finite attainer $\alpha_x$ for every $x\in C$; no size
bound on these controllers is assumed.

If strict original Min improvements from $C$ exist, let $\delta>0$
be the smallest such value gap. Choose
$0<\mu\le\min\{1/2,\delta/4\}$. For each target $y$ of such
an edge choose an original finite sure controller $\beta_y$ whose
initial guarantee is at least $v(y)-\delta/4$. Such a controller
exists by the definition of supremum, whether or not $y\in C$.
If there is no strict improvement from $C$, take $\mu=1/2$
and omit this repair family.

The next potential combines the original value with the restricted
baseline. The chosen coefficient ensures that both permanent repairs
have nonnegative drift.
Run $\sigma$ in main mode while $b(x,m)\ge v(x)$, using potential
\[
              H(x,m)=(1-\mu)v(x)+\mu b(x,m).
\]
After a preserving actual step, perform $\sigma$'s memory update.
If its observed successor mode has $b(x,m)<v(x)$, permanently
initialize $\alpha_x$ there. If original Min chooses a strict
improvement $x\to y$, execute that edge and permanently initialize
$\beta_y$ instead. Main states are in $C$, so these rules cover
every possible original departure from the restricted play. After
a permanent switch use the chosen original controller's true
residual baseline as potential.

Fix any original pure public-history Min strategy. On preserving
main steps, the conditional drift of $H$ is nonnegative, using
both full-row averaging equations at chance. Switching at a
deficient successor increases potential because
$H(x,m)<v(x)$ there. At a strict Min improvement, the repair starts
at least at
\[
 v(y)-\delta/4\ge v(x)+3\delta/4,
 \qquad H(x,m)\le v(x)+\mu\le v(x)+\delta/4.
\]
This switch also increases potential. Inside a permanent suffix,
its baseline has nonnegative conditional drift. The combined
process, say $Y_t$, is therefore a bounded submartingale starting
at $Y_0=H(s,m_0)>v(s)$, and
$\mathbb E Y_\infty\ge H(s,m_0)$.

Consider its recurrent support in the combined finite controller,
under this arbitrary history opponent. Almost surely it is an end
component. If it lies in a permanent suffix, the original
whole-component tour of Lemma~\ref{lem:graded-tour} bounds its
limiting baseline by $F$. If it
lies in main, full chance fairness closes every recurrent main
chance source under all its original successors; all recurrent
Min choices are preserving. Its support is consequently a legal
end component of the restricted $\sigma$ product. Main modes have
$b\ge v$, hence $H\le b$, and the restricted whole-component tour
again bounds its limiting potential by $F$. This uses the
unconditioned recurrent-support fact, not a chance law conditional
on no switch. Thus $Y_\infty\le F$ almost surely and the combined
controller guarantees strictly more than $v(s)$.

Every supported no-switch path is a path of the restricted sure
controller, and every switched path has a finite prefix followed
by an original sure suffix. Hard therefore holds also on null
paths. The combined controller is an original pure finite sure
controller, contradicting the definition of $v(s)$. The restricted
values are exactly $v|_C$.

\emph{Clear the restricted values.}
Apply the actual value-vertex construction in
Theorem~\ref{thm:r48-general-denominators} to $G_C$. Its full state
set is its finite-sure region, its finite rational reward is cleared
by the same original integer $R$, and its chance rows are the full
original rows sourced in $C$. Select independent active value rows
and multiply each selected chance row once by its original $q_x$.
The resulting nonsingular integer matrix $A_C$ satisfies
$1\le|\det A_C|\le Q_C$, and $A_C(Rv|_C)$ has an integer right-hand
side. Cramer's rule therefore gives the particular clearer
$H_C=R|\det A_C|\le RQ_C$ of $v|_C$.
Its divisibility by $R$ follows from this construction, so it also
clears every original reward $f(S)$, including those outside the
restricted game. The explicit factor of the reward clearer ensures
this divisibility. The restriction/value-cap proof applies to the full
finite-Inf reward range $0\le F\le1$ by the tour argument above.
The original attainers and repairs establish the restricted value
identity; the bounded construction below uses new witnesses.
\end{proof}

\subsection{Anchoring all finitely attained values}

Write $m=|N|$. If $C\ne\varnothing$, choose the constructed,
$R$-divisible common clearer $H_C$ of
Lemma~\ref{lem:r51-attained-restriction}; otherwise put $H_C=R$.
The attained-start conclusion is vacuous when $C$ is empty.

\begin{lemma}[Anchored good margin]
\label{lem:r51-anchored-good-margin}
There is a good subpotential $u$ on $W$ with $u|_C=v|_C$ such that
every positive player difference and positive chance-source slack
of $u$ is at least
\[
                    \frac{1}{2H_C(m+1)Q_N}.
\]
\end{lemma}
\begin{proof}
\emph{Choose a seed that attains the fixed values.}
Tag one original finite attainer for each state in $C$ together
with finite sure controllers initialized at every state of $W$.
The max-mode extraction of Lemma~\ref{lem:r48-sharp-extraction}
applies to this finite tagged family: all retained residuals are
admissible, all their actual updates are retained, and every
original state has an initialized mode. It supplies a good seed
$u^0\le v$ with $u^0|_C=v|_C$. These seed controllers are used
only for this existence argument.

Let $\mathcal O$ be its good cell and let $P$ be the weak closure
from Lemma~\ref{lem:r48-sharp-cell-closure}. In particular $P\le v$
coordinatewise. Intersect $P$ with the equations $u|_C=v|_C$.
This slice contains $u^0$. Every chance row sourced in $C$ stays
in $C$ and is harmonic for the fixed values there, so its slack
vanishes identically on the slice. No positive chance form is
sourced in $C$. Positive Min forms sourced in $C$ remain ordinary
player-incidence forms and must still be included below.

\emph{Maximize a common positive margin.}
Use variables $u_N\in[0,1]^m$ and $\gamma\in[0,1]$ after
substituting the fixed $C$ values.
For a positive chance row sourced at $x\in N$, its margin inequality is
\[
 u(x)-\sum_{y\in N}p(x,y)u(y)+\gamma
       \le\sum_{y\in C}p(x,y)v(y).
\]
The right side contains only the fixed $C$ values; multiplying by $q_x$
makes their coefficients integers, and the one common $H_C$ clears it.
 Retain every weak cell row
of $P$, and require every seed-positive player difference or
chance-source slack $L_e(u)$ to satisfy $L_e(u)\ge\gamma$.
Zero forms remain equalities and negative Max forms retain their
weak cell inequalities. The seed is feasible for some
$\gamma>0$; if the positive list is empty take $\gamma=1$.
Maximize $\gamma$ on this bounded polytope and choose an optimal
vertex $(z,\gamma_*)$, with $\gamma_*>0$.

Choose $m+1$ independent active rows, including equality rank.
After removing the $\gamma$ column, their coefficient parts are
signed units, player incidences restricted to $N$, or original
N-source chance rows restricted to $N$; a $\gamma$-only bound
leaves a zero row. The $\gamma$ coefficients belong to
$\{0,1,-1\}$. A restricted chance row is a convex combination
of restricted incidences, which are units, incidences or zero rows.
As in Theorem~\ref{thm:r48-general-denominators}, every
$m$-square minor has absolute determinant at most one.
Expansion along the $\gamma$ column therefore bounds the
unscaled augmented determinant by $m+1$.

There is at most one independent selected chance row per source
in $N$. A zero source has one equality form; signed copies
are dependent. For a positive source its weak row $L_e\ge0$
is inactive because $L_e\ge\gamma_*>0$, leaving only its
margin row. Chance rows sourced in $C$ have vanished. Scale each
selected chance row, including its $\gamma$ coefficient, by
its original $q_x$. The resulting integer invertible matrix
$M$ satisfies
\[
                 1\le|\det M|\le(m+1)Q_N.
\]
Every scaled right side is cleared by the same $H_C$. Player
rows have only fixed $C$-value contributions, while unit and reward
comparison rows can also have constants $0,1$ and $\pm f(S)$.
These rational reward constants are cleared because $R\mid H_C$.
For a chance row sourced in $N$, multiplication by $q_x$ makes
every coefficient in its fixed-value sum on $C$ integral; one common
$H_C$ then clears that right side as well. Cramer's rule uses one
replaced right-hand-side column, and therefore gives
\[
                  \gamma_*\ge\frac{1}{H_C(m+1)Q_N}.
\]
The empty-minor convention gives the same argument when $m=0$.
The unknown values $v(y)$, $y\in N$, appear only in the conclusion
$P\le v$ of cell closure; the LP itself uses the specified cell rows.

\emph{Return to the good cell.}
Mix the boundary point $z$ with the seed by setting
$u=(z+u^0)/2$. Mixing with the seed restores every strict
cell row, so $u\in\mathcal O$ and is good. It still equals
$v$ on $C$, and every positive form is at least
$\gamma_*/2$. This proves the claimed margin. The potential's
possibly nonrational coordinates are used only in the analysis.
The implemented controller stores finite auxiliary witnesses and their
actual-edge updates; the seed attainers contribute no modes.
\end{proof}

\subsection{Relative event tails and exact rational memory}
\label{app:r51-rational-exact-memory}
Continue with the normalized finite rational Inf payoff and its
integer reward clearer $R$. The sets $C,N$ and the constructed $H_C$
now refer to that payoff. Keep the two cap arguments distinct.
Lemma~\ref{lem:r51-exact-event-cap} proves the reward inequality for
every cap and pays for fallback directly when $\kappa\gamma\ge1$.
The next lemma adds an event-ranker estimate on the same execution,
using
$\kappa=2T_{\mathrm{ev}}\ell$ and $2^{-\ell}\le\gamma T_{\mathrm{ev}}$ to pay for fallback by a relative tail.
This relative bound is proved directly for the capped execution below.
Both analytic potentials follow the \emph{same} original execution and
completed-event count $J$, stopped after the same actual cap transition.
The domain remains $W$, with every original chance row unchanged.

\begin{lemma}[Arbitrary caps and a relative event tail]
\label{lem:r51-relative-event-cap}
Let $u$ be good on the original sure region $W$, with every positive
player difference and chance-source slack at least $\gamma>0$.
Let $w\in[0,1]^W$ be an event ranker for this $u$ of margin $1/T_{\mathrm{ev}}$,
where $T_{\mathrm{ev}}\ge1$ is an integer. For every integer $\kappa\ge1$, the
witness/reset protocol capped after $\kappa$ completed actual events is
sure $A$ from every initialized start and has at most
\[
                         \kappa nM_{\mathrm{qual}}(n+1)+nM_{\mathrm{qual}}(n)
\]
original modes. Against any pure public-history Min, let $J\le \kappa$
be its own completed-event count. Its reward satisfies
\begin{equation}\label{eq:r51-arbitrary-cap-payoff}
             \mathbb E F\ge u(s)+\gamma\mathbb E J-\Pr(J=\kappa).
\end{equation}
If $\kappa=2T_{\mathrm{ev}}\ell$ for an integer $\ell\ge1$, then the same capped count obeys
\begin{equation}\label{eq:r51-relative-cap-tail}
             \Pr(J=\kappa)\le 2^{-\ell}\frac{\mathbb E J}{T_{\mathrm{ev}}}.
\end{equation}
In particular it guarantees $u(s)$ when $2^{-\ell}\le\gamma T_{\mathrm{ev}}$.
No condition $\kappa\gamma\ge1$ is imposed in this lemma.
\end{lemma}
\begin{proof}
\emph{Keep the same capped execution.}
Apply Lemma~\ref{lem:r51-exact-event-cap} on $\mathcal U=W$, with $d=n$;
the original sure region has the required closure and sure fallbacks.
Its witness/reset protocol is sure $A$, has the displayed mode count,
and satisfies~\eqref{eq:r51-arbitrary-cap-payoff} for its own completed-event
count $J$, for every integer cap $\kappa\ge1$.
It remains to prove~\eqref{eq:r51-relative-cap-tail} for this same
execution and this same $J$.

Fix a start $s\in W$ and an arbitrary pure public-history opponent.
Let $\mathcal F_t$ be the original history after $t$ transitions,
$I_t$ indicate a main-mode positive-$u$ event on $t\to t+1$, and put
\[
 J_t=\sum_{r<t}I_r,\qquad
 \theta=\inf\{t:J_t=\kappa\},\qquad J=\lim_tJ_t\le \kappa.
\]
Set $I_t=0$ after cap. The time $\theta$ is after the actual $\kappa$th
transition, including its full chance outcome. Under the fixed pure
controller/opponent, a player action is determined by the finite
history and a positive chance event is known at its source.
Thus $I_t$ is $\mathcal F_t$-measurable for this analysis, although
the implementation detects a Min event only on its observed edge.
For the additional event-count estimate, stop $w$ at this same time:
\[
                         Z_t=w(X_{t\wedge\theta}).
\]
Witness resets change only finite memory, leaving the original state
fixed. The protocol retains zero player moves and full zero-slack chance
rows; a positive $u$ event executes its original edge or full chance row.
The ranker preserves the zero forms and gives every positive form gain at least
$1/T_{\mathrm{ev}}$, in mean at a chance source. Thus, at every finite history,
\[
 \mathbb E[Z_{t+1}-Z_t\mid\mathcal F_t]\ge I_t/T_{\mathrm{ev}}.
\]
The bounded process $Z_t$ is a submartingale.

\emph{Bound the conditional number of remaining events.}
For finite $b>a$, conditional telescoping of the rank inequality gives
\[
 \mathbb E[J_b-J_a\mid\mathcal F_a]
 \le T_{\mathrm{ev}}\bigl(\mathbb E[Z_b\mid\mathcal F_a]-Z_a\bigr)\le T_{\mathrm{ev}}.
\]
Conditional monotone convergence yields
$\mathbb E[J-J_a\mid\mathcal F_a]\le T_{\mathrm{ev}}$. This concerns the capped
execution and holds inside a witness as well as after a reset.
Histories after cap have remaining count zero.

Set $b_0=2T_{\mathrm{ev}}$ and $\kappa=b_0\ell$. Let $\tau_r$ be the transition
time after completed event $rb_0$, or infinity.

\emph{First block.} Markov uses this same terminal capped $J$:
$\Pr(\tau_1<\infty)=\Pr(J\ge b_0)\le\mathbb E J/(2T_{\mathrm{ev}})$.

\emph{Later blocks.} For $1\le r<\ell$, every finite history at
$\tau_r$ is pre-cap, with $J_{\tau_r}=rb_0$.
Conditional Markov bounds the probability of at least $b_0$ further
events by $T_{\mathrm{ev}}/b_0=1/2$. Summing over the countably many finite
histories realizing that stopping time gives
\[
 \Pr(\tau_{r+1}<\infty)\le\tfrac12\Pr(\tau_r<\infty).
\]
\emph{Combine the blocks.} Iterating only the remaining $\ell-1$ blocks proves
\[
 \Pr(J=\kappa)\le 2^{-(\ell-1)}\frac{\mathbb E J}{2T_{\mathrm{ev}}}
                  =2^{-\ell}\frac{\mathbb E J}{T_{\mathrm{ev}}}.
\]
For $\ell=1$ the iteration is empty. Throughout, $\mathbb E J$ is the
expectation of this same capped count. The successive conditional bounds
allow dependence between blocks, and all telescoping is at finite
horizons. Combining~\eqref{eq:r51-arbitrary-cap-payoff}
and~\eqref{eq:r51-relative-cap-tail} proves the claimed lower guarantee
when $2^{-\ell}\le\gamma T_{\mathrm{ev}}$.

The same protocol stores only the tagged witnesses and event counter;
the block indices and analytic potentials add no modes.
\end{proof}

\begin{samepage}
\begin{theorem}[Exact memory for finite rational Inf rewards]
\label{thm:r51-rational-exact-memory}
For $n=|W|\ge1$, finite rational $F=f(\operatorname{Inf})\in[0,1]$
with integer common reward clearer $R\ge1$, and arbitrary Muller
hard $A$, one original pure finite controller is sure from every
initialized state of $W$ and attains $v(s)$ at all $s\in C$.
Set $T_{\mathrm{ev}}=(n+1)Q$ and $\ell_R=\lceil\log_2(2R)\rceil$.
With $\kappa_R=2T_{\mathrm{ev}}\ell_R$, it uses at most
\[
                  \kappa_R nM_{\mathrm{qual}}(n+1)+nM_{\mathrm{qual}}(n)
\]
modes. Thus the bound is $2^{O(n\log(n+1))}Q(1+\log R)$.
No attainment assumption is imposed on $N=W\setminus C$.
\end{theorem}
\end{samepage}
\begin{proof}
If $C$ is empty, the $nM_{\mathrm{qual}}(n)$ tagged sure fallbacks suffice.
Otherwise Lemmas~\ref{lem:r51-attained-restriction} and
\ref{lem:r51-anchored-good-margin} give a good anchored $u$ with
positive event margin
\[
 \gamma\ge\frac{1}{2H_C(|N|+1)Q_N},\qquad
 H_C\le RQ_C,
\]
and consequently
\[
 \gamma T_{\mathrm{ev}}\ge
 \frac{(n+1)Q_C}{2H_C(|N|+1)}\ge\frac1{2R}.
\]
Lemma~\ref{lem:r50-event-ranking} supplies a progress potential
for these same $u$ events with the integer $T_{\mathrm{ev}}=(n+1)Q$.
Since $2^{-\ell_R}\le1/(2R)\le\gamma T_{\mathrm{ev}}$, apply
Lemma~\ref{lem:r51-relative-event-cap} with $\kappa_R=2T_{\mathrm{ev}}\ell_R$.
The resulting controller is original, pure finite, sure from every
initialized state of $W$, and guarantees $u$ there. At $s\in C$
we have $u(s)=v(s)$; original sure admissibility bounds its guarantee
above by $v(s)$, giving exact attainment with the displayed count.
If no form is positive, the auxiliary witnesses already run in the
original game and give this conclusion without a cap; their tagged
count is smaller than the displayed bound. For $R=1$, $\ell_R=1$. The construction uses only the
bounded witnesses, one event counter and original sure fallbacks,
not the arbitrary seed attainers or analytic potential coordinates.
\end{proof}

When $W$ is empty there is no attained start and the assertion is
vacuous; no zero-mode all-start controller is asserted. The upper
allows $Q=1$, including an empty chance-row product. For each fixed
$n\ge6$, Theorem~\ref{thm:r50-joint-memory-lower}'s six-state family
and unreachable deterministic padding match the joint $Q,R$
dependence for every $Q,R\ge2$: its exact minimum is
$1+\lceil\ln R/(-\ln(1-1/Q))\rceil=\Theta(Q\log R)$.
The $R=1$ Boolean lower family is given separately below. These are worst-case
arithmetic comparisons at fixed $n$, not tight dependence on $n$,
per-instance constants, identification of $C$, or a synthesis-time bound.
\subsection{Sharp arithmetic dependence at fixed arena size}
\begin{theorem}[Rational retry frontier]
\label{thm:r50-joint-memory-lower}
For every pair of integers $q,r\ge2$, a six-state rational-reward arena
with $Q=q$ and $R=r$ has, at a designated Min entry $s$, a finitely
attained optimum and exact least attaining memory
\[
 1+\min\{j\ge0:(1-1/q)^j\le1/r\}
 =1+\left\lceil\frac{\ln r}{-\ln(1-1/q)}\right\rceil .
\]
The count includes STOP and applies to arbitrary pure finite controllers.
\end{theorem}
\begin{proof}
The retry arena has a Max state $t$, a chance state $c$, and absorbing
$g,b$. TRY goes from $t$ to $c$, STOP from $t$ to $b$; $c$ goes to $g$
with probability $1/q$ and to $t$ with probability $1-1/q$. Hard accepts
eventual absorption in $g$ or $b$, a coB\"uchi condition on this arena.
The Boolean soft reward is one at $g$ and zero at $b$. Its original
pure-finite supremum is one.

For any $m$-mode pure finite sure controller, follow the supported
all-failure path. Before STOP, its successive $t$ modes cannot repeat:
the deterministic action and updates would then repeat that failure
cycle forever, contradicting hard even if the path has probability zero.
The eventual STOP mode is distinct from all preceding TRY modes.
Consequently there are at most $m-1$ attempts. The success probability
is at most $1-(1-1/q)^{m-1}$. Conversely, modes $0,\ldots,m-1$ TRY below
$m-1$, increment on failure, and STOP at $m-1$, attaining this probability
and satisfying hard on every supported path. No additional phase mode
is needed at $c$ or the absorbers. The count includes STOP and all equality cases.

For the second arena add a Min entry $s$ choosing between $t$ and a new
absorber $a$ of reward $1-1/r$. Hard now accepts eventual absorption in
$a,g,b$; assign zero reward to other Inf sets. The entry has constrained
value $1-1/r$. It is attained precisely when the retry continuation
guarantees at least that amount. The same all-failure argument gives
the displayed formula. Conversely initialize the same counter mode at
$s$ and $t$; the action at $s$ is Min's, so no separate entry mode is
needed. Min can choose either branch, and the guarantee is exactly the
smaller of their two rewards. There is one nontrivial chance row with
least clearer $q$, and the only nonintegral reward has reduced denominator
$r$, so $Q=q$ and $R=r$.
\end{proof}

For $Q,R\ge2$, the exact count is $\Theta(Q\log R)$ because
$1/Q\le-\ln(1-1/Q)\le1/(Q-1)$. It matches
Theorem~\ref{thm:r51-rational-exact-memory} at each fixed arena size
at least six. For $n>6$, add $n-6$ new forced Max absorbers with
deterministic self-loops, without changing any original edge.
Extend the hard absorption target to include them and assign reward zero
to every recurrent set containing a new state. The new states are
unreachable from the original entry and belong to $W$. They add no
chance row, so $Q$ and $R$ and the minimum memory at that entry are
unchanged. A controller reuses an old mode on each forced self-loop.

\begin{corollary}[Serial-trial product lower bound]
\label{cor:r55-serial-product-memory}
For $k\ge1$ and integers $q_1,\ldots,q_k,r\ge2$, a $(k+5)$-state
arena with $W=V$, $Q=\prod_iq_i$ and $R=r$ has a finitely attained
optimum at a Min entry $s$ with exact least pure finite sure memory
\[
 1+N,\qquad N=\min\{j\ge0:(1-1/Q)^j\le1/r\}
             =\left\lceil\frac{\ln r}{-\ln(1-1/Q)}\right\rceil .
\]
STOP is counted. For $k=2$ and $q_1=q_2=q$, seven states require
$\Theta(q^2\log r)$ modes although the maximum and lcm of the original
row denominators are both $q$.
\end{corollary}
\begin{proof}
Replace $c$ in the preceding retry arena by chance states
$c_1,\ldots,c_k$. TRY at $t$ enters $c_1$; $c_i$ succeeds with
probability $1/q_i$, moving to $c_{i+1}$ or, at the last stage, $g$;
failure returns to $t$. Keep the Min entry $s$ choosing $t$ or the safe
absorber $a$ of reward $1-1/r$, STOP to $b$, and hard eventual absorption
in $a,g,b$. This gives $k+5$ states and $W=V$ by STOP.
Each full row has least clearer $q_i$ and $1-1/r$ is reduced, so the
stated $Q,R$ are exact. Put $p=1/Q$ and $\beta=1-p$.

Fix an $m$-mode pure finite sure controller after Min chooses $t$.
For a TRY mode $\mu$ at $t$, let $F_i(\mu)$ be its return mode when
stage $i$ fails after all earlier stages succeed, composing \emph{every}
actual-edge update. That branch has probability
$a_i=(\prod_{j<i}1/q_j)(1-1/q_i)>0$, and $\sum_i a_i=\beta$.
In the graph of reachable modes at $t$, include \emph{all} failure
edges $\mu\to F_i(\mu)$; STOP modes are terminal. A reachable cycle
would repeat supported failed-trial paths forever and violate hard,
even if that infinite path is null. The graph is therefore acyclic.
Every maximal path ends in a STOP mode distinct from its TRY modes,
so its height $L(\mu)$, counting further TRY actions, is at most $m-1$.

Let $P(\mu)$ be the success probability. At STOP, $P=0$ and $L=0$.
At TRY, every child has height at most $L(\mu)-1$, and backward induction gives
\[
 P(\mu)=p+\sum_i a_iP(F_i(\mu))
 \le p+\beta\bigl(1-\beta^{L(\mu)-1}\bigr)
 =1-\beta^{L(\mu)}\le1-\beta^{m-1}.
\]
This handles arbitrary failure-stage-dependent updates, not one chosen
failure history. The inherited mode after $s\to t$ is covered too.

Conversely use modes $0,\ldots,N$: TRY below $N$, STOP at $N$, keep the
mode on entry and success edges, and on \emph{every} stage failure
update $j\mapsto\min\{j+1,N\}$. Initialize in mode $0$.
No phase or entry mode is needed. The controller is sure
and succeeds with probability $1-\beta^N$. Min's two branches therefore
give guarantee $\min\{1-1/r,1-\beta^N\}=1-1/r$. The lower bound above
forces $\beta^{m-1}\le1/r$ for any attainer, proving the exact count.

The logarithmic inequalities used above give $\Theta(Q\log r)$.
With two equal row clearers, $Q=q^2$ whereas their maximum and lcm are
$q$. Fixing $r=2$ rules out either linear replacement of $Q$ in the
general memory bound by a finite factor depending only on arena size.
This does not rule out higher powers with state-dependent exponents or
assert optimal dependence on the number of states.
\end{proof}

\newpage
\subsection{The Boolean boundary case}
\begin{proposition}[A Boolean lower family]
\label{prop:closed-endpoint-memory}
For every integer $q\ge2$, a six-state Boolean arena with $Q=2q$
and $R=1$, all of whose original values are attained, has exact least
all-start attaining memory
\[
             1+\left\lceil\frac{\ln2}{-\ln(1-1/q)}\right\rceil
                          =\Theta(Q).
\]
The hard condition is coB\"uchi and the soft condition is B\"uchi.
The count includes STOP and start-dependent initialization, and the lower
bound applies to arbitrary pure finite sure controllers.
\end{proposition}
\begin{proof}
Use original states $x,t,c,d,g,b$. Max at $x$ chooses
STOP$\to b$ or TRY$\to t$; Min at $t$ chooses BASE$\to d$ or
IMPROVE$\to c$. The two full chance rows are
\[
 c\to\begin{cases}g,&1/q,\\x,&1-1/q,\end{cases}
 \qquad
 d\to\begin{cases}g,&1/2,\\b,&1/2.\end{cases}
\]
The states $g,b$ have unique deterministic self-loops. Hard is
coB\"uchi avoidance of $\{x,t,c\}$, equivalently eventual absorption
in $\{g,b\}$ here; soft is B\"uchi $g$. Always STOP proves that all
six states lie in $W$, and the least original row clearers give $Q=2q$.
Min's always-BASE reply gives the upper bounds in the claimed value vector
\[
 v(g)=1,\quad v(b)=0,\quad v(d)=v(x)=v(t)=\tfrac12,\quad
 v(c)=\tfrac1q+(1-\tfrac1q)\tfrac12.
                                                        \tag{**}
\]
The fixed values at $d,g,b$ are immediate; the stated bounds at
$x,t$ follow from BASE, and the one at $c$ uses its full initial row.

Write $s_q=1-1/q$ and
$N=\lceil\ln2/(-\ln s_q)\rceil$, so $s_q^N\le1/2<s_q^{N-1}$.
An attaining controller has modes $0,\ldots,N$. At $x$ it uses TRY
below $N$ and STOP at $N$. All edges retain the mode except a failure
$c\to x$: in mode $j<N$ increase to $j+1$, and in mode $N$ change
to $0$. Initialize $x,t,d,g,b$ in mode $0$ and $c$ in mode $N$.
The exceptional $c,N\to x,0$ update reuses the counted STOP mode
as the initial tag for $c$; it adds no phase. An $x,0$ or $t,0$ run
never reaches $c,N$, because $x,N$ stops first. Initial $c,N$ makes
one actual chance trial and, on failure, starts a fresh $x,0$ run.
Thus every supported initialized path absorbs after at most $N+1$
chance trials, even if all trials fail.

For an arbitrary pure history Min, its first BASE on the unique
continuing failure history from $x,0$ occurs after some $k<N$
failures or is absent before STOP. In the former case success has
probability
$1-s_q^k+s_q^k/2=1-s_q^k/2\ge1/2$; in the latter it is
$1-s_q^N\ge1/2$. This covers all history replies because success
and BASE absorb, leaving only one continuing failure history.
The same argument applies from $t,0$, including immediate BASE.
From initial $c,N$, its full row and the uniform $x,0$ guarantee
give $1/q+s_q/2$. Hence (**) is the actual value vector and every
original state attains, simultaneously with these $N+1$ modes.

Conversely, fix any $m$-mode pure finite sure controller from $x$
and let Min always IMPROVE. Along the supported all-failure path,
each TRY traverses $x\to t\to c\to x$, including every deterministic
memory update at all three owners. If an $x$ mode repeats before
STOP, purity repeats the same TRY and failure-return updates forever,
creating a supported hard-bad path. Sure semantics forbids it even
though that path is null. The eventual STOP mode is distinct from all
TRY modes, so there are at most $m-1$ attempts. Against this reply
success is at most $1-s_q^{m-1}$. Attaining $1/2$ requires
$m\ge N+1$, proving the exact least count from $x$ and therefore
also for all-start attainment. Since
$1/q\le-\ln(1-1/q)\le1/(q-1)$, this count is $\Theta(q)=\Theta(Q)$.

\end{proof}

For $R=1$, Theorem~\ref{thm:r51-rational-exact-memory} gives an
$O_n(Q)$ upper bound without assuming attainment outside $C$.
The family above gives matching worst-case arithmetic dependence at
fixed arena size at least six, using the same absorbing padding.

These lower families use Min vertices; no matching MDP memory lower bound
is claimed here.

\section{A common opponent for the complete lower-guarantee interval}
\label{app:joint}

For the general structural extension, let hard be a Muller family $A$ and reward be $f(\operatorname{Inf})$. The common-opponent theorem uses the following hypotheses, on
\emph{all nonempty original-state sets} $S,T\subseteq V$:
\begin{align}
 S,T\in A,\ S\cap T\ne\varnothing
       &\ \Longrightarrow\ S\cup T\in A,\tag{$U_A$}\label{eq:hard-union}\\
 S\cap T\ne\varnothing
       &\ \Longrightarrow\ f(S\cup T)\ge\min\{f(S),f(T)\}.
          \tag{$U_f$}\label{eq:reward-union}
\end{align}
Streett acceptance is closed under unions, so Streett hard and Boolean
Streett reward satisfy both. Hypothesis $(U_f)$ applies also to sets
rejected by the hard condition. The general memory and denominator
bounds do not assume $(U_A)$ or $(U_f)$.

Apply these hypotheses on $V$ after any colour pullback. Their
restrictions to $W$ are used here; the full-domain hard hypothesis is
also used in Appendix F.
Throughout this section the original sure-$A$ region $W$ is fixed.
Every induced MDP has exactly this domain; its own larger sure region
is never substituted for $W$. The induced arena deletes Max edges leaving
$W$ and retains every original Min and chance successor in $W$. Objectives
are prefix independent,
\Max uses pure finite controllers, and \Min may use arbitrary pure
public-history strategies. Write $F(\rho)=f(\operatorname{Inf}(\rho))$,
pulling colour objectives back to original state sets, and put
\[
 v(s)=\sup_{\sigma\text{ finite, sure }A}\inf_\tau
                         \Exp_s^{\sigma,\tau}F.
\]
Potential values below always label original states, even when different
states have the same reward colour. Finite sure fallbacks exist on $W$.
Tagged finite continuations give the constrained Bellman equations:
\Max edges weakly decrease $v$, \Min edges weakly increase $v$, and
every original chance row averages $v$. No attainment is assumed.

For an arena $H$ define its feasible lower-guarantee set by
\[
 \mathcal L_H(s)=\{a\in\mathbb R:\exists\sigma\text{ finite, sure }A,
                  \ \forall\tau\ \Exp_s^{\sigma,\tau}F\geq a\}.
\]
This is $(-\infty,v_H(s))$, with the endpoint adjoined precisely when
the finite supremum is attained. It is an interval of guaranteed
\emph{lower bounds}; no two-sided constraint on the opponent's possible
expectations is intended.

\begin{theorem}[Joint lower-guarantee intervals]
\label{thm:joint-interval}
Assume the hard intersecting-union hypothesis $(U_A)$ and reward
hypothesis $(U_f)$. There is one pure positional \Min policy
$\tau^*$ on the original $W$ such that, simultaneously for every $s\in W$,
\[
 v_{G[\tau^*]}(s)=v(s),\qquad
 \mathcal L_{G[\tau^*]}(s)=\mathcal L_G(s).
\]
In particular it preserves the whole finite-attainment vector, including
controllers that need only be sure against $\tau^*$ in the induced MDP.
For rational rewards these values and such a policy are effective, and
the values are rational.
\end{theorem}

The proof chooses one opponent for both local thresholds, weak and strict. Lemma~\ref{lem:nested-opponent} chooses the same minimum-tier
child for both; its pause-and-merge step uses
Lemmas~\ref{lem:predictable-fairness} and~\ref{lem:paused-support} for
chance fairness, and Lemma~\ref{lem:harmonic-component} for the common
recurrent potential. Lemma~\ref{lem:strict-empty} next excludes strict
local improvement. The final proof bounds every finite controller
admitted after fixing the opponent, then uses graded attainment to
preserve the endpoint flags as well as the values.

The inputs are Bellman, finite baseline evaluation and touring
(Lemmas~\ref{lem:general-q} and~\ref{lem:general-tour}), the finite-MDP
stopping characterization, and Theorem~\ref{thm:graded-attainment}.
The construction starts from these qualitative and finite-product inputs.
The split-and-pause architecture is the established qualitative shuffle
method~\cite{CP19}; first-deviation repair has its usual role~\cite{GH08}.
The simultaneous local thresholds are the additional assertion proved here.

\subsection{Two nested local thresholds}

Let $u$ be a real potential on a fixed finite arena $H$. Suppose that
every player edge preserves $u$ and every chance row is harmonic for $u$.
Define
\[
 B_u(\rho)\iff f(\operatorname{Inf}(\rho))\geq
                      \max_{x\in\operatorname{Inf}(\rho)}u(x),\qquad
 C_u(\rho)\iff f(\operatorname{Inf}(\rho))>
                      \max_{x\in\operatorname{Inf}(\rho)}u(x).
\]
These are Muller conditions and $C_u\subseteq B_u$. Their union property
will be used only on intersecting recurrent supports with a common
potential value.

\begin{lemma}[Constant recurrent potential]
\label{lem:harmonic-component}
Every product end component of a fixed finite controller in $H$ has a
constant projected $u$ value. This remains true after deleting \Min edges.
\end{lemma}
\begin{proof}
At a component vertex maximizing $u$, preserving player edges stay at
that maximum. All positive chance successors belong to the component;
harmonicity makes every one maximal as well. This propagates along its
strongly connected internal graph, proving constancy. Deleting \Min
edges changes neither these properties nor any chance row.
\end{proof}

\begin{lemma}[Fairness under predictable source selection]
\label{lem:predictable-fairness}
Consider an arbitrary history-dependent execution of a finite stochastic
arena. Attach one of finitely many labels to each executed chance step,
before observing its outcome. A label fixes the original chance source;
the remaining label data may include a child index and finite memory.
Almost surely, for every label executed infinitely often, each positive
successor of its source occurs infinitely often at steps carrying that
label.
\end{lemma}
\begin{proof}
Fix a label and one of its successors, of probability $p>0$. The label
has already been chosen when the chance outcome is drawn. Conditional
on the entire finite past at any occurrence of this label, the outcome
therefore still has probability $p$. The pauses can depend on that past;
they cannot change this conditional probability.

Fix a global time. The probability that at least $k$ further labelled
steps occur and their first $k$ outcomes all miss this successor is at
most $(1-p)^k$. This follows by successive conditioning at those finite
opportunities, without independence of their occurrence times. Let
$k$ tend to infinity. Infinite labelled execution with no subsequent
occurrence of this successor has probability zero. A countable union
over the last occurrence time, followed by a finite union over labels
and successors, proves the assertion. We never condition the chance
law on the event that a projection is infinite.
\end{proof}

\begin{lemma}[Recurrent supports under finite pauses]
\label{lem:paused-support}
Fix finitely many pure finite child controllers. A paused execution must
preserve each child's legal projected transitions and memory updates;
every executed chance step uses its original row, and before its outcome
the active child is already selected.
Almost surely, the recurrent product support of every infinite child
projection is a reachable child end component. If a child is AS winning
for an Inf-set condition against every Min history, each such recurrent
component satisfies that condition.
\end{lemma}
\begin{proof}
Label each executed chance step by child, finite memory and original
source. On the probability-one event of
Lemma~\ref{lem:predictable-fairness}, every infinite child projection
has recurrent support closed under all positive chance successors.
Fixed controller moves are retained. In a finite graph, the recurring
vertices and edges of an infinite projected path form a strongly
connected support; each recurrent Min vertex has an internal successor.
The recurrent product support is therefore a child end component.

Acceptance follows from reachability and the tour argument. The finite
projected prefix is legal in the child and has positive probability under a
suitable prefix opponent. If the component's Inf-set violated the
child's AS condition, Min could follow that prefix and then tour the
whole component as in Lemma~\ref{lem:general-tour}. The resulting
positive probability of violation would contradict the child's guarantee.
This argument uses the ordinary child game for the finite prefix and
the tour; it assigns no autonomous probability law to a projection
conditioned on being infinite.
\end{proof}

\begin{lemma}[A simultaneous nested qualitative opponent]
\label{lem:nested-opponent}
Suppose every player edge in $H$ preserves $u$ and every original chance
row is harmonic for $u$. Under $(U_A)$ and $(U_f)$, one original positional \Min
policy in $H$ preserves both whole-arena finite winning regions
$\SAS_H(A,B_u)$ and $\SAS_H(A,C_u)$ simultaneously. No sure-winning
region is preselected in this assertion.
\end{lemma}
\begin{proof}
Induct on the number of surplus \Min edges. With one outgoing edge at
every \Min vertex the assertion is immediate. Otherwise split a pivot
$z$'s outgoing edges into two nonempty groups, obtaining children $H_1,H_2$
on the same vertices and chance rows. Give $z$ tier $0$ in a child if it
loses $B_u$, tier $1$ if it wins $B_u$ but loses $C_u$, and tier $2$ if
it wins $C_u$. Choose a child of minimum tier, say $H_1$.

This order has a simple consequence for either fixed target
$J\in\{B_u,C_u\}$. If the selected child wins $B_u$, both children
have tier at least one and win $B_u$. If it wins $C_u$, both children
have tier two and win $C_u$. Thus there are only the following two cases.

\emph{The selected child loses $J$ at $z$.} Any finite
SAS controller from a winning state of $H_1$ avoids $z$ on \emph{every}
supported history. Indeed a supported finite arrival has positive
probability against a suitable prefix opponent; its inherited memory
would be sure $A$ and AS $J$ from $z$. Any positive suffix failure could
be pasted after that prefix. This contradicts loss at $z$. Its reachable
behaviour therefore remains legal in $H$. Conversely, deleting Min
choices preserves every parent witness. Thus
$\SAS_H(A,J)=\SAS_{H_1}(A,J)$.

\emph{Both children win $J$ at $z$.} Store
$(m_1,m_2,a)\in M_1\times M_2\times\{1,2\}$, where $M_i$ is the mode
set of a chosen child-$i$ witness for $J$ from $z$; initialize both
witnesses at $z$ and set $a=1$.
Away from $z$, execute and update only child $a$; on observing Min's
edge from $z$ in group $i$, set $a=i$ and update only $m_i$ along that
actual edge, leaving the other memory paused at $z$.
Each child projection concatenates its executed edge segments,
identifying their common boundary occurrence of $z$: all nonpivot
edges agree, a self-loop at $z$ remains an actual child edge, and
neither paused memory is reset.

\emph{Hard acceptance.} Every infinite child projection follows its
sure child controller. With two infinite
projections their hard Inf-colour sets contain the common pivot colour,
so $(U_A)$ accepts their global union on every supported path.
With one infinite projection its Inf set is the global one.

\emph{Soft acceptance.} Label each chance opportunity by its child,
finite memory and original source. The active child is selected before
the outcome, so Lemma~\ref{lem:paused-support} applies to these adaptive
pauses. Almost surely, every infinite projection has a reachable child
end component as its recurrent product support, and its reward satisfies $J$.

Lemma~\ref{lem:harmonic-component} makes each such recurrent potential
constant. If both projections are infinite, each returns to $z$
infinitely, so both constants equal $u(z)$. Hypothesis $(U_f)$
then bounds their union reward by the smaller child reward: it is at
least $u(z)$ for $B_u$ and strictly greater for $C_u$. With only one
infinite projection no union is needed. Thus the merged witness is SAS
from $z$ in $H$.

\emph{Other starting states.} From a winning state of either child,
follow its witness until the first visit to $z$, then use the merged
witness. A path that never visits $z$ remains a child play and satisfies
hard. Extend the pre-hit Min choices to a complete opponent in that
child. Its AS guarantee makes the joint event ``no visit to $z$ and
soft failure'' null; no law is conditioned on avoiding the pivot.
Every finite first-hit history has the merged sure/AS winning suffix.
There are countably many such histories, so soft failure has probability
zero, and prefix independence gives sure hard also after a hit.
Both child $J$ regions therefore equal the region in $H$.
For each of the two conditions, the \emph{same} minimum-tier child has
now been proved exact. Apply induction there; fixing its simultaneous
policy gives the same MDP in that child and in $H$.
\end{proof}

\subsection{Strict local improvement is impossible}

Let $K_v$ keep exactly the value-preserving edges of both players and
\emph{every original chance edge with its original probability}. Bellman
makes all its player successor sets nonempty.

\begin{lemma}[Empty strict-local region]
\label{lem:strict-empty}
No state of $K_v$ has a finite sure-$A$/AS-$C_v$ controller.
\end{lemma}
\begin{proof}
Suppose $\pi$ were such a controller from $s$. Finitely many original
state sets give a gap $\delta>0$ such that every $C_v$-accepted set $S$
satisfies $f(S)\geq\max_{x\in S}v(x)+\delta$. For each original strict
\Min improvement $e=(x,y)$, choose a finite original sure controller
$\beta_e$ from $y$ whose exact baseline satisfies $q_e(y)>v(x)$.
Tolerance $(v(y)-v(x))/2$ suffices, without attainment of $v(y)$.
Take $\gamma>0$ below all these finitely many improvement gaps; if there
are no such edges take $\gamma=1$, an unused bound.

Run $\pi$ until the first strict \Min improvement, then permanently enter
its tagged $\beta_e$. This is a finite original sure controller: no-switch
paths follow $\pi$, and switched paths have a finite prefix and a sure
suffix. Against an arbitrary history opponent let $T$ be the switch's
arrival time. Use potential $Y=v$ in main mode and the exact $q_e$ in
fallback mode. Ordinary drift is nonnegative; a switch contributes at
least $\gamma$. For each finite horizon $N$, summing conditional increments
gives
\[
 \Exp Y_N-Y_0
   =\sum_{n<N}\Exp(Y_{n+1}-Y_n)
   \geq\gamma\Prb(T\leq N).
\]
The bounded submartingale converges, so dominated convergence yields
\[
 \Exp Y_\infty\geq v(s)+\gamma\Prb(T<\infty).
\]

In fallback, the finite-product tour bound gives $F\geq Y_\infty$
almost surely. On no-switch paths, unconditional source fairness makes
the recurrent support a reachable component of $\pi$ in $K_v$; its
constant potential and AS-$C_v$ guarantee give
$F\geq Y_\infty+\delta$ almost surely. We use this unconditional fairness
event on the no-switch event, not chance laws conditioned on that event.
Consequently, uniformly over the original opponent,
\[
 \Exp F\geq v(s)+\gamma\Prb(T<\infty)+\delta\Prb(T=\infty)
            \geq v(s)+\min\{\gamma,\delta\},
\]
contradicting the definition of $v(s)$.
\end{proof}

We finish by bounding every finite controller sure against the selected
opponent, including the extra controllers admitted by this restriction.
The bottom-component bound gives equality of values; comparison of the
equality arenas then transfers attainment. Both steps keep the original $W$.

\begin{proof}[Proof of Theorem~\ref{thm:joint-interval}]
Apply Lemma~\ref{lem:nested-opponent} to $K_v$ with $u=v$. Obtain one
original positional $\tau^*$ preserving its $B_v$ and $C_v$ regions.
Lemma~\ref{lem:strict-empty} makes $C_v$ losing at \emph{every} state
of $K_v[\tau^*]$. Use the same selected \Min edges in $G$ on $W$;
each preserves $v$.

\emph{Bound the enlarged controller class.}
Fix any finite controller $\sigma$ sure $A$ against this fixed opponent,
including one that is not sure against all original \Min choices.
Its product is a finite Markov chain and $v(X_n)$ is a bounded
\emph{super}martingale: transient \Max steps may decrease $v$.
In every reachable bottom SCC $D$, $v$ is constant, say $r_D$.
At a minimum over $D$, all successors have value at least that minimum
while their average is at most it. Every positive successor is therefore
minimal, and strong connectivity propagates the minimum throughout $D$.
In particular every recurrent \Max edge preserves $v$.

The inherited controller initialized inside $D$ is thus legal in
$K_v[\tau^*]$, sure $A$, and has reward $f(\operatorname{proj}D)$
almost surely. If that reward exceeded $r_D$, it would witness SAS-$C_v$
from a component vertex, a contradiction. Unreachable controller entries
may be completed with arbitrary preserving choices. Hence every bottom
reward is at most its recurrent value, and
\[
                  \Exp_s^{\sigma,\tau^*}F
                      \leq\Exp_s\!\left[\lim_{n\to\infty}v(X_n)\right]\leq v(s).
\]
All original sure controllers remain admissible after fixing \Min, so
restriction gives the reverse supremum inequality. Therefore
$v_{G[\tau^*]}=v$ on the entire original $W$.

\emph{Recover finite endpoints.}
Apply graded attainment to $G$ and to this induced MDP, on that same
domain. Because $\tau^*$ preserves $v$, the MDP's equality arena is
exactly $K_v[\tau^*]$, with all original chance rows. Its attainment
region is its SAS-$B_v$ region, equal by construction to that of $K_v$.
Thus all attainment flags, and hence all lower-guarantee intervals,
agree. Strict lower-threshold feasibility also agrees: the exact
minimizing reply to each finite controller exists by finite baseline
evaluation, so it is equivalent to $v(s)>q$.

\emph{Compute values and a joint policy.}
Finally, for every original positional policy solve its MDP on $W$
using Lemma~\ref{lem:general-mdp}. Its stopping values are rational and
effective for rational rewards; qualitative levels use finite Muller
monitors and finite SAS~\cite{CCFL24,CP19}. Their componentwise minimum
is $v$, since restriction never lowers value and the joint policy attains
this minimum vector. Compute $K_v$, then enumerate its policies and test
the two qualitative regions to find a joint policy. This proves
rationality and gives an effective joint-policy construction.
\end{proof}

\section{The opponent on the full original arena}
\label{app:domain}
The fixed-original-$W$ convention is needed during restrictions. The final
joint opponent can nevertheless be chosen without enlarging the sure region.

\begin{corollary}\label{cor:full-arena}
Under $(U_A)$,$(U_f)$, the joint policy has a positional extension to the whole
original arena whose induced MDP has exactly the original finite sure-hard
region $W$. Its feasible lower-threshold intervals equal the original
intervals at every state, with empty intervals outside $W$.
\end{corollary}
\begin{proof}
Treat original chance vertices as additional Min vertices in the
deterministic hard game. The original sure region is $W$. Apply the
qualitative whole-region common-opponent argument to hard $A$ and soft
true: its paused merge needs only $(U_A)$, and the split/pause proof is the
classical qualitative argument~\cite[full proof of Theorem 4.2]{CP18full}.
There is one positional coalition policy $\nu$ on original Min and chance
vertices whose fixed-policy deterministic game has exactly winning region
$W$. This qualitative conclusion is independent of the quantitative
interval theorem.

An outside-$W$ Max vertex has no successor in $W$: otherwise that move
followed by a finite original sure fallback would win. An outside coalition
vertex's $\nu$-selected successor also lies outside $W$. A selected edge
into $W$, followed by the same fallback, would make that outside vertex
winning against $\nu$, contradicting its fixed-policy winning region.
Thus every $\nu$-consistent path starting outside remains outside.

Use the joint policy on original Min vertices in $W$ and $\nu$ on original
Min vertices outside $W$. Keep every original chance row. Against any
finite Max controller outside $W$, resolve chance by the $\nu$-selected
support edges. This unique deterministic play is hard losing; otherwise
that controller would win against $\nu$. It stays outside $W$, so changing
$\nu$ inside $W$ has no effect on it. The play is supported under the actual
chance rows, even if null, and therefore prevents a sure controller outside.

Original sure controllers keep $W$ feasible. A finite controller sure
against the full fixed opponent from $W$ cannot have a supported finite
arrival outside $W$: its inherited continuation there would be sure by
prefix independence, contradicting the preceding paragraph. Thus its
entire supported product stays in $W$. The fixed-domain interval theorem
applies exactly to this controller family, proving the claim.
\end{proof}

The qualitative whole-region claim used above can also be obtained directly
by splitting a coalition pivot. If it is hard losing in a child, every
finite hard-winning child controller avoids that pivot on every supported
path, so that child's whole region carries to the parent. If both children
win at the pivot, pause their finite memories on excursions; $(U_A)$ combines
their hard Inf sets at the shared pivot. Both child regions then equal
the parent region. Selecting the appropriate child at each split yields
one positional policy for the whole deterministic winning region. This
argument requires neither probabilities nor an attained quantitative value.

The extension does not license recomputing $W$ after an arbitrary Min
restriction. An arbitrary policy can enlarge the hard region; the
constructed final policy uses a particular outside spoiler to prevent it.

\clearpage
\paragraph*{Domain ledger.}
The proofs use the following distinct domains.
\begin{center}
\begin{tabular}{@{}p{.22\linewidth}p{.69\linewidth}@{}}
\toprule
Object & Required interpretation\\
\midrule
Original $W$ & Computed in the original hard game with chance adversarial;
fixed through every Min restriction.\\
$G[\tau]$ on $W$ & May allow additional controllers sure only against
$\tau$; every original chance edge and probability remains.\\
Actual $K_v$ & Equal-value player edges, all original chance rows;
its derived sure region may shrink.\\
Positive certificate $U\subseteq W$ & An original Min/chance-closed
domain with internal Max moves and sure fallbacks at every state;
those fallbacks prove inclusion in $W$.\\
Negative certificate $U\supseteq W$ & A guessed overdomain containing
the original sure region, used with an upper potential and fixed opponent.\\
Upper $H,z$ & Arbitrary superpotential, not an exact-value guess;
strict-local exclusion is required at every state.\\
Zero-slack $S_{=}$ & The certificate core in Appendix G: original chance
average is equal and all supported
successors remain; no selected-support renormalization.\\
\bottomrule
\end{tabular}
\end{center}

\section{Local certificates and tight threshold complexity}
\label{app:certificates}

All probabilities, rewards and thresholds are binary rationals. The hard
condition $A$ is an explicit Streett pair list. The soft input is a
Boolean Streett pair list $B$, with $F=\mathbf1_B$. The hard condition
is evaluated on every supported path and the controller is pure finite.
We retain the fixed original $W$ throughout.
The following uses classical qualitative Streett certificates and local
superpotentials, rather than claiming these ingredients as new
\cite{Chatterjee07,CP19,DGG25}. The joint endpoint theorem supplies
completeness at the non-strict boundary.

\begin{theorem}[Tight threshold complexity and joint policy extraction]
\label{thm:tight-complexity}
For this Boolean Streett representation, finite sure-hard synthesis with expected
reward strictly greater than $q$, and with expected reward at least $q$,
is $\mathrm{coNP}$-complete. Hardness holds with $W=V$, no chance vertices,
and a two-colour coB\"uchi soft condition. Exact values, infeasible states,
all finite-attainment flags, and one original positional joint policy of
Theorem~\ref{thm:joint-interval} are computable in
$\mathrm{FP}^{\mathrm{NP}}$.
\end{theorem}

\subsection{An arbitrary upper potential and its equality core}

Let $H$ be a finite MDP, with fixed original \Min edges treated as
deterministic edges, on an arbitrary domain $U$. No sure-winning
assumption on $U$ is imposed. For $z:U\to[0,1]$ require
\begin{align}
 z(x)&\geq z(y)
     &&\text{on every \Max edge and every fixed-\Min edge},\label{eq:local-upper}\\
 z(x)&\geq\sum_y p(x,y)z(y)
     &&\text{at every chance vertex}.\nonumber
\end{align}
Use the local predicates $B_z$ and $C_z$ of Appendix~\ref{app:joint},
comparing reward with the maximum recurrent \emph{original-state}
potential. The guessed $z$ need not be an exact value vector.

\begin{lemma}[Local upper certificate]
\label{lem:local-upper}
If no vertex of $H$ admits finite $\SAS(A,C_z)$, every finite sure-$A$
controller from $s$ has expected reward at most $z(s)$.
\end{lemma}
\begin{proof}
Fix such a controller. Its product is a finite Markov chain and
$Z_n=z(X_n)$ is a bounded supermartingale. On a reachable bottom SCC $D$
it is constant, say $r_D$: at a minimum of $z$ all positive successors
remain in $D$ and are at least that minimum, while
\eqref{eq:local-upper} bounds their average above by it. All are minimal;
strong connectivity propagates constancy. This permits transient strict
loss on either \Max or fixed-\Min moves.

If $f(\operatorname{proj}D)>r_D$, the inherited controller initialized
with its memory inside $D$ is sure $A$ and AS $C_z$. Bottom closure keeps
all supported continuations inside; every product vertex recurs almost
surely. This contradicts the negative premise at that projected vertex.
Thus $F\leq Z_\infty$ almost surely and
$\Exp F\leq\Exp Z_\infty\leq z(s)$. Negativity at \emph{every} vertex,
not just the designated initial state, is essential.
\end{proof}

Compute the greatest set $S_{=}\subseteq U$ satisfying these safety conditions:
a retained \Max vertex has a retained successor with equal $z$; a
retained fixed-\Min vertex has its fixed equal-$z$ successor in $S_{=}$;
a retained chance vertex has original average exactly $z(x)$ and
\emph{all} original supported successors in $S_{=}$. Iterated deletion is
polynomial. The MDP $H_{=}$ on $S_{=}$ keeps all equality \Max edges, its
fixed-\Min edges, and every original chance edge and probability.
There is no chance pruning or renormalization.

\begin{lemma}[Zero slack at an attained upper bound]
\label{lem:zero-slack}
Under Lemma~\ref{lem:local-upper}'s hypotheses, a finite sure-$A$
controller achieves expectation at least $z(s)$ in $H$ iff $s\in S_{=}$ and
finite $\SAS(A,B_z)$ is possible from $s$ in $H_{=}$.
\end{lemma}
\begin{proof}
For a controller reaching the upper bound, all upper inequalities are
tight: $\Exp F=\Exp Z_\infty=z(s)$ and $F=Z_\infty$ almost surely.
At a reachable product state $p$ let
$d(p)=z(p)-\Exp[z(P_{n+1})\mid P_n=p]\geq0$. For each $N$,
\[
 z(s)-\Exp Z_N=\Exp\sum_{n<N}d(P_n).
\]
Monotone and bounded convergence make the nonnegative infinite sum have
expectation zero. Every reachable product state has a positive-probability
finite prefix, so its local loss is zero. Thus every chosen \Max move
and fixed-\Min move preserves $z$, and each reachable chance row has
its original average equal to $z$. Every supported product successor
is reachable. The reachable product's projection is a post-fixed set
for the safety conditions and is contained in the greatest $S_{=}$.
This proves legality in $H_{=}$ on \emph{all} supported histories,
including prefixes of null infinite paths. Each bottom SCC has constant
$z$ and reward equal to it, giving AS $B_z$.

Conversely a finite SAS witness in $H_{=}$ makes $Z_n$ a bounded
martingale. Its bottom components satisfy $F\geq Z_\infty$, hence
$\Exp F\geq z(s)$. This direction needs no strict-local negative premise.
\end{proof}

\subsection{Polynomial negative qualitative certificates}

For Boolean Streett $B$ with pairs $(E_j,F_j)$, on any arena and for any
$z\in[0,1]^U$ the pointwise identity
\[
 C_z=B\ \cap\ \operatorname{coBuchi}(U_1),
             \qquad U_1=\{x:z(x)=1\},
\]
adds just the Streett pair $(U_1,\varnothing)$. In $H_{=}$, zero is
support closed: equality player moves preserve zero and a nonnegative
chance row of mean zero has only zero successors. Consequently $B_z$
equals the Streett list
\[
                     (E_j\cap\{x:z(x)>0\},F_j).
\]
A path hitting zero makes $B_z$ automatic and every modified request
finite. Otherwise its recurrent potential maximum is positive and both
conditions equal $B$. These are supported-path identities, including
null paths, rather than arbitrary-set closure assertions.

With Streett hard labels, apply the hard-label-preserving MDP gadget of
Lemma~\ref{lem:r36-streett-mdp}. The resulting deterministic Streett game has polynomial
size. A negative answer has a positional Rabin opponent, and fixing it
leaves a one-player Streett graph checked by SCC request pruning
\cite{Chatterjee07}. Thus each local SAS \emph{nonmembership} has an NP
certificate. The classical qualitative gadgets and finite-component
commitment are the ingredients used here; no positive NP certificate
for Streett SAS or two-player commitment oracle is assumed.

\subsection{Certificate soundness and completeness}

\begin{proof}[Threshold part of Theorem~\ref{thm:tight-complexity}]
Consider $q\in[0,1]$. An infeasibility certificate may first certify that
$s$ is outside the original sure-Streett region: make chance adversarial
and give a positional Rabin spoiler. Otherwise guess $U\ni s$ and
certify each excluded original vertex hard losing, so $W\subseteq U$.
Require $U$ closed under all original \Min and positive chance edges,
with an internal successor at each \Max vertex; retain \emph{every}
internal original edge. Guess an original positional \Min policy
$\tau$ on $U$, obtaining an MDP $H$, and polynomial-bit rationals $z$
satisfying \eqref{eq:local-upper}, $0\leq z\leq1$, and $z(s)\leq q$.
Supply a negative finite SAS-$C_z$ certificate from \emph{every} $u\in U$.
If $z(s)<q$, stop. At equality compute $S_{=},H_{=}$; if $s\notin S_{=}$, stop;
otherwise supply a negative SAS-$B_z$ certificate in $H_{=}$ from $s$.
The preceding encodings, rational comparisons and safety deletion make
this a polynomial NP verifier.

For soundness, an original finite sure controller stays inside original
$W$ on every supported history: an inherited finite sure suffix at an
outside vertex would prove that vertex belongs to $W$. It is therefore
legal in $U$ against the guessed $\tau$. Lemma~\ref{lem:local-upper}
bounds its reward by $z(s)$, and at equality Lemma~\ref{lem:zero-slack}
forces exactly the SAS witness excluded by the final certificate.
No hard fallback from $U\setminus W$ is used, and $z$ is never asserted
to be an exact value vector.

For completeness take $U=W$, the joint policy of
Theorem~\ref{thm:joint-interval}, and $z=v$. Its selected \Min edges
preserve $v$, so the full superpotential inequalities hold. No state
of the \emph{whole} induced MDP $H$, not merely of $K_v[\tau]$, admits
finite SAS-$C_v$. Otherwise a winning controller has a reachable bottom
SCC of constant $v$ and larger reward; initializing its inherited memory
inside that SCC beats $v_H$ there, contradicting $v_H=v$ at every state.
All requested strict-local negatives are therefore true.

Theorem~\ref{thm:r48-general-denominators} supplies polynomial value
bit size independently of any quantitative controller-size bound. Let
$Q$ be the product of the original chance-row clearing denominators and
$R$ clear the reward values (here $R=1$). Each reduced value denominator
is at most $QR$, and both $\log Q$ and $\log R$ are polynomial in the
written input. Values lie in $[0,1]$, so their numerators satisfy the
same bit bound. This is a bound, not divisibility by $QR$.

If $v(s)<q$, the strict branch succeeds. If $v(s)=q$ and the endpoint
is infeasible, the joint policy makes $H$ nonattaining too. Bellman and
the selected preserving \Min edges give $S_{=}=W$ and $H_{=}=K_v[\tau]$.
Graded attainment supplies the true negative $B_v$ test there. Thus the
complement of non-strict synthesis is in NP. Without the final equality
exclusion the same data certify $v(s)\leq q$, so the complement of strict
synthesis is in NP as well; the finite baseline minimizing reply justifies
equivalence of strict synthesis and $v(s)>q$. For $q>1$ infeasibility is
immediate, while for $q<0$ feasibility is exactly hard winning.

For hardness, start with deterministic Streett winning; coNP-hardness
already holds for conjunctions of two parity objectives
\cite[Theorem~2(5)]{CHP07}. Put a \Max gate
before every original vertex, offering continuation or an absorbing
fallback $b$, and copy original hard labels at gates. Put $b$ in no hard
request set. Every transformed vertex is hard winning by taking fallback
at the next gate. Let the soft condition be coB\"uchi avoidance of $b$.
An original winning strategy gives reward one. Otherwise the original
positional Rabin spoiler makes every infinite play avoiding fallback
hard losing, so any transformed sure controller must reach $b$ against
it and has reward zero. The threshold $1/2$ is therefore hard for both
comparisons, with $W=V$ and attained yes instances. The finite-winner and positional Rabin-spoiler properties are
classical~\cite[Theorem~1]{CHP07}; see also~\cite{Chatterjee07}.
\end{proof}

\paragraph*{The supplied-threshold parity specialization.}
Both supplied-threshold questions are also coNP-complete when hard and
soft are each one parity condition, without a sure-feasibility promise.
For the upper bound, under the maximum-even convention translate each
occurring odd priority $j$ into the Streett pair
$\bigl(\{x:p(x)=j\},\{x:p(x)>j\}\bigr)$.
This preserves the path event and the arena, using at most $|V|$ pairs
per objective even for binary-encoded priorities. Apply the Streett upper bound.

For hardness use the chance-free conjunction of two parity conditions
\cite[Theorems~1 and~2(5)]{CHP07}, which has finite pure winning witnesses.
Keep its arena unchanged, use the first parity as hard and the second
as Boolean soft, and supply $q=1/2$. Each pure strategy profile determines
one play, so soft probability is zero or one. Meeting either $>1/2$ or
$\ge1/2$ while sure hard is therefore exactly winning the conjunction.
This is an independent parity lower bound, rather than hardness inherited
from the larger Streett class. No fixed-priority-index or $W=V$ restriction
is asserted. Non-strict threshold one is the prior finite SAS result
\cite[Corollary~12]{CP19}; the classification with $q$ in the input does not
claim hardness at every degenerate fixed endpoint.

\subsection{Values, flags, and an actual joint policy}

\begin{proof}[Function part of Theorem~\ref{thm:tight-complexity}]
Compute original $W$ with an NP oracle. The upper certificate without
the equality exclusion gives NP cuts $v(s)\leq q$, and imposing
$z(s)<q$ gives NP cuts $v(s)<q$. Completeness uses the same joint policy
and $z=v$. A polynomial denominator-bit bound from Theorem~\ref{thm:r48-general-denominators} allows binary search and rational
reconstruction of every coordinate with polynomially many oracle queries:
distinct rationals of denominator at most $D=QR$ differ by at least
$D^{-2}$. Query non-strict synthesis at the now-known $v(s)$ to compute
each attainment flag.

Here is an NP search predicate that outputs the joint policy rather than
an arbitrary value-common policy. With these computed $W,v$ and flags,
guess an original positional $\tau$ on $W$, explicitly require
\[
       v(\tau(x))=v(x)\quad(x\in V_{\Min}\cap W),
\]
check all inequalities \eqref{eq:local-upper} with $z=v$, and supply
negative SAS-$C_v$ certificates at every state of $G[\tau]$ on $W$.
Lemma~\ref{lem:local-upper} gives $v_{G[\tau]}\leq v$; restriction of
the original opponent gives $v_{G[\tau]}\geq v$. Require additionally
a negative SAS-$B_v$ certificate in $K_v[\tau]$ for every globally
nonattaining state. The explicitly preserving \Min edges make this
the induced MDP's exact equality arena. Graded attainment therefore
excludes precisely those endpoints. Globally attained endpoints remain
attained by their original controllers after fixing \Min. Every accepted
policy is thus joint, and Theorem~\ref{thm:joint-interval} supplies a
successful witness.

Encode the original edge choices with polynomially many bits. Ask the
NP oracle whether a successful witness extends the current prefix with
zero, retain zero if possible, otherwise one, and continue. This
search-to-decision self-reduction extracts the policy in
$\mathrm{FP}^{\mathrm{NP}}$. It does not extract a small optimal
\Max controller.

To output the full-arena extension of Theorem~\ref{thm:main-interval},
also extract its qualitative outside-$W$ coalition spoiler. Guess one
edge at each outside original Min or chance vertex, require that its
successor remains outside $W$, and retain every outside Max edge.
There is no outside Max edge into $W$, by the definition of $W$.
Check that this fixed-coalition one-player graph has no accepting
Streett path from any outside state, using polynomial SCC request
pruning. This is an NP search predicate, and
Corollary~\ref{cor:full-arena} supplies a simultaneous witness.
The same bit-prefix self-reduction computes it in
$\mathrm{FP}^{\mathrm{NP}}$. Use only its original Min choices outside
$W$ and the joint policy inside, restoring every original chance row.
The corollary's supported losing-path argument proves that the resulting
full Min policy preserves exactly the original sure region. Thus the
whole-arena output, as well as the fixed-$W$ policy, has the stated
function upper bound. If the complement of $W$ is empty this step is
vacuous.
\end{proof}

Attainment at an \emph{unsupplied} optimum has the
$\mathrm P^{\mathrm{NP}}$ upper bound: compute the optimum, then ask its
non-strict query. Appendix~\ref{app:r47-general-boolean} supplies the
matching decision lower bound and the exact-value function lower bound.
The coNP classification concerns a supplied rational threshold. These
certificates do not extend that classification to arbitrary Muller hard
conditions or constitute an efficient controller compiler.


\subsection{Positive certificates for fixed-index parity}
\label{app:fixed-index-certificates}

We retain the original controller model and sure region $W$.
In this section $A$ and $B$ are parity conditions and $F=\mathbf1_B$.
The index bound is an external constant: the polynomial degree below may
depend on that constant. Only occurring priorities are normalized, in
order and with their parity preserved; a large unused binary priority
is never expanded into that many levels.

For $\bowtie\in\{>,\ge\}$, the supplied question asks whether a pure
finite controller is sure $A$ and satisfies
$\Exp_s^{\sigma,\tau}F\bowtie q$ for every pure public-history $\tau$.
It is false outside $W$. A fixed finite controller has an attained
minimum in its finite Min-MDP, by the stopping-and-tour argument in
Appendix~\ref{app:graded-attainment}. Consequently strict feasibility is
equivalent to $v(s)>q$. Non-strict feasibility also holds when
$v(s)=q$ and that endpoint is attained. Thus the proof below must retain
the equality case; it cannot replace $\ge q$ by $v(s)>q$.

\paragraph*{Positive events pay for an exact finite cap.}

We first prove a statement that does not require the parity restriction.
Let $U$ be a finite original-state domain closed under every original
Min edge and every supported chance outcome, with at least one internal
Max successor per Max state. Retain every internal Max edge and every
full original chance row. Assume that every state of $U$ has a finite
sure-$A$ fallback staying in $U$. Define subpotentials, $Z(u)$ and
goodness on $U$ exactly as in Appendix~\ref{app:r48-general-subpotential}.

\begin{lemma}[Exact lower realization]
\label{lem:fixed-index-exact-realization}
Every good $u\in[0,1]^U$ has a pure finite original sure-$A$ controller
guaranteeing at least $u(z)$ from every $z\in U$.
\end{lemma}
\begin{proof}
If a positive form exists, let $\gamma>0$ be the least one and apply
Lemma~\ref{lem:r51-exact-event-cap} on $U$ with
$\kappa=\lceil1/\gamma\rceil$, so $\kappa\gamma\ge1$.
If no positive form exists, use its no-event case.
The domain and fallback hypotheses are exactly those of that lemma.
This is a lower guarantee at least $u$; the actual minimum may be larger.
\end{proof}

\paragraph*{Every feasible controller supplies a good envelope.}

\begin{lemma}[Exact feasibility envelope]
\label{lem:fixed-index-envelope}
For $s\in W$, the non-strict supplied question is feasible exactly when
some good $u$ on $W$ satisfies $u(s)\ge q$. The strict question is
feasible exactly when some good $u$ satisfies $u(s)>q$.
\end{lemma}
\begin{proof}
Let $\sigma$ be a finite sure controller for the supplied comparison.
Tag it together with sure fallbacks initialized at every state of $W$.
In their finite initialized products retain the supported reachable
residual modes. Every such residual is sure admissible, the family is
closed under its actual updates and all Min/chance outcomes, and every
original state has at least one retained mode. If $b(x,m)$ is its true
minimizing residual baseline, put $u(x)=\max_m b(x,m)$.

At a maximizing mode, the fixed Max edge preserves $b$, every Min
edge is nondecreasing in $b$, and the full chance row averages $b$.
The successor envelope dominates each successor baseline. Thus $u$ is
a subpotential. Initialize at a maximizing mode. Along a zero player
difference, equal endpoint upper bounds force $b=u$ to persist; at a
zero chance-source slack, all nonnegative successor deficits have zero
positive-weighted sum, so each deficit vanishes. A selected positive
event goes to the auxiliary marker. Marker-free histories therefore
remain original histories of the selected tagged controller and retain
sure hard. Every marker-free recurrent product component is also a
legal original component: fixed Max moves and all chance successors
remain. Its tour bound gives $f(S)\ge\max_{x\in S}u(x)$. Hence the
initialized auxiliary controller is almost-sure $B_u$, proving goodness
from every state.

At $s$ the envelope dominates $\sigma$'s baseline. For the strict case,
that baseline is strictly above $q$, because the minimizing reply of a
fixed finite controller is attained. This proves the necessary envelope
comparison. The converse follows from
Lemma~\ref{lem:fixed-index-exact-realization}. No claim that $u=v$ is
needed. In particular an envelope at an arbitrary feasible $q$ does not
by itself say that the supremum endpoint is closed.
\end{proof}

The certificate guesses a lower potential, not the value vector. At a
closed endpoint $q=v(s)$, exact lower realization and admissibility force
only $u(s)=v(s)$. Other coordinates may remain below their unattained
suprema. Thus a good certificate at one closed start assumes neither
global equality $u=v$ nor all-state attainment. The sharp rational attained-start memory bound uses the separate
anchored-margin and relative-tail arguments of
Theorem~\ref{thm:r51-rational-exact-memory}.

\paragraph*{Choose a polynomial-bit potential.}

For Boolean reward, put $P_u=\{x:u(x)>0\}$. On every nonempty recurrent
set,
\[
                       B_u=B\ \lor\ \operatorname{coBuchi}(P_u).
\]
A reward-one set always passes, while a reward-zero set passes exactly
when all its recurrent coordinates are zero. Consequently goodness
depends only on player-edge signs, chance-source zero/positive slacks,
and the coordinate zero set.

Fix a real feasible envelope and preserve its pattern with variables
$r_x\in[0,1]$ and $\zeta\in[0,1]$. Keep every zero form equal to zero,
require positive forms to be at least $\zeta$, and negative Max
differences to be at most $-\zeta$. Set $r_x=0$ on the zero set and
$r_x\ge\zeta$ on its complement. For the non-strict task impose
$r_s\ge q$; for the strict task impose $r_s-q\ge\zeta$. The original
envelope makes this rational linear system feasible for some
$\zeta>0$, by the finite minimum of its strict margins. Empty lists
of strict pattern rows cause no problem. Maximizing $\zeta$ gives a
positive optimum at a rational vertex. The numbers of rows and
variables and all coefficient bit lengths are polynomial in the input.
The determinant bound for rational LP vertices therefore gives
polynomial-bit $r$. Its positive optimum preserves the entire pattern,
so $r$ is good and keeps the supplied comparison. This is independent
of the denominator of the original controller's product.

\paragraph*{Monitor the soft disjunction.}
Give the absorbing auxiliary marker even hard priority zero, so it
accepts hard. Extend its soft priority by even zero, and write $p:Z(u)\to\{0,\ldots,b\}$ in max-even convention.
The marker lies outside $P_u$. To monitor
$B\lor\operatorname{coBuchi}(P_u)$, keep the maximum priority $h$ since
the preceding $P_u$ checkpoint, excluding the current vertex $x$;
initially $h=0$. Emit $c$ and update by
\[
 (c,h')=\begin{cases}
   (\max\{h,p(x)\},0),&x\in P_u,\\
   (0,\max\{h,p(x)\}),&x\notin P_u.
 \end{cases}
\]
On the arena product $(x,h)$, $c$ is a vertex priority and $h'$ is
updated on the actual outgoing edge. This is a deterministic public-edge
monitor with $b+1$ modes. If $P_u$ is visited only finitely often,
eventually only even zero is emitted. Otherwise the intervals between
its visits are finite and partition the tail. Let $j$ be the largest
original priority recurring infinitely often. Priorities above $j$
affect only finitely many intervals, while $j$ occurs in infinitely
many intervals because each interval is finite. The largest emitted
priority recurring infinitely often is therefore $j$, proving the
claimed language. The output range remains $\{0,\ldots,b\}$; copying
the hard labels also preserves their range.

\paragraph*{Count the arenas and the controller modes separately.}
Let $N=|Z(u)|$, including the marker, and let $a$ be the normalized
hard maximum priority. The checkpoint arena has at most $N(b+1)$
vertices. Put
\[
 D_A=2\lceil a/2\rceil,\qquad D_B=2\lceil b/2\rceil.
\]
These even bounds permit min/max conversion without changing parity.
Use the forward-existence and same-memory reverse interface in the proof
of Lemma~\ref{lem:qualitative-memory-interface}, with the present priority
ranges: \cite[Theorem~8]{CP19}, with forward proof
\cite[Appendix, Lemma~4.8]{CP18full}. A finite auxiliary
SAS winner exists exactly when its deterministic conjunction game has
a winner. The gadget copies the hard priority and uses soft priorities
at most $D_B$. Its vertex definition gives the safe bound
\[
             N_{\rm D}\le 2(D_B+2)N(b+1).
\]
Indeed a chance source of soft min-priority $p$ has one source vertex,
at most $\lfloor(p+1)/2\rfloor+1$ tilde vertices and $p+1$ hat vertices;
each hat retains every supported exit. Only existence is used in this
forward step, not the old controller's memory or rank size.

After converting back to max-even, apply the register monitor of
\cite[Section~3, Theorem~2]{DG26}. Interchanging the two parity
conditions if beneficial gives at most
\[
 M_{\rm reg}=\min\bigl\{(D_B+1)^{D_A/2+1},
              (D_A+1)^{D_B/2+1}\bigr\}
\]
register states. The resulting single-parity game has a positional
winner, so $M_{\rm reg}$ bounds its memory on the deterministic gadget arena.
CP19's reverse construction retains those same $M_{\rm reg}$ modes on the
checkpoint arena \cite[pp.~6:9--6:11]{CP19}. At an actual chance outcome,
the source, current memory and observed successor determine the
terminal register update of a selected virtual gadget path. No gadget
location is stored, and no chance row of the checkpoint arena is
conditioned or changed.

Projection to $Z(u)$ stores the checkpoint as well as the registers.
One initialized start therefore needs at most $(b+1)M_{\rm reg}$ modes. Tagging
such witnesses for every initialized start needs at most $N(b+1)M_{\rm reg}$.
This is polynomial in the explicit input when either original parity
index is fixed. The arena size $N_{\rm D}$ is not an additional factor
in controller memory. Both monitors are determined by the public
history, and the witness remains pure finite against arbitrary
public-history \Min. This uses the finite-memory equivalence, not the
unrestricted SAS winning-region theorem. The quantitative realization
is separate: at a positive chance source it restores and executes the
full original game row. Its possibly large repair counter is not part
of the qualitative witness.

\paragraph*{Verify the certificate.}
Guess $U\ni s$, closed
under every original Min edge and supported chance successor, with an
internal Max successor at every Max state; retain every internal Max
edge. Supply a positional sure-parity fallback winning from all of $U$,
a polynomial-bit potential $u$, and a polynomial auxiliary controller
with initial modes for every state of $U$. The fallback is checked with
chance adversarial and proves $U\subseteq W$. This check is essential:
the auxiliary marker can hide an original sure-losing infinite retry.
Completeness takes $U=W$ and uses a positional sure-parity fallback.

All verification is polynomial. Exact rational arithmetic checks the
potential inequalities and builds $Z(u)$. Fix the guessed auxiliary
controller and compute the product states reachable from its designated
initializations. For each odd hard priority $j$, inspect the graph
induced by hard priorities at most $j$. A cyclic strongly connected
component containing priority $j$ gives a supported hard-losing path;
conversely every hard-losing path gives such a component. This test uses
every supported chance edge adversarially, whether or not remaining on
the cycle has positive probability.

For the almost-sure test one may check the original formula
$B\lor\operatorname{coBuchi}(P_u)$ directly. For each odd soft priority
$j$, restrict the reachable product to priorities at most $j$ and
compute its maximal end components: every fixed Max edge and every
supported chance successor must stay inside, and each Min state must
have an internal choice. Reject if one contains
both a priority-$j$ state and a state whose original vertex is in $P_u$.
Min can reach such a component with positive probability and then tour
it, making its full support recurrent almost surely; soft parity is
then false and $P_u$ recurs. Conversely the recurrent support under any
history opponent is almost surely an end component. If $B_u$ fails,
its largest soft priority is some odd $j$ and it meets $P_u$; it is
contained in one of the maximal components detected by this test.
Singleton components count only when cyclic. Standard end-component
pruning and graph searches implement all these tests in polynomial time.
They retain all auxiliary supported outcomes; the quantitative
realization separately executes every full original chance row.
These tests quantify over every original pure history Min strategy. Lemma
\ref{lem:fixed-index-exact-realization} gives quantitative soundness;
the envelope and rationalization arguments give completeness. The
possibly large quantitative repair is not part of the certificate.
This proves NP membership with at least one index fixed. The existing
supplied-threshold coNP bound applies after the polynomial parity-to-
Streett encoding, proving the decision bound in
Corollary~\ref{thm:fixed-index-parity} for both comparisons.

\paragraph*{Own attainment and one exact value.}

First query non-strict threshold zero to test original sure feasibility.
It is exactly that test because $F\ge0$. If it fails, return false for
\textsc{Own-Attain}. Otherwise choose the computable polynomial-bit
bound $Q_{\rm all}$ given by the product of the least full-row clearers
over all original chance vertices. The original value-denominator
theorem, with Boolean reward, bounds every reduced denominator by
$Q\le Q_{\rm all}$; this is a size bound, not a divisibility claim.

Strict threshold queries compare $v(s)$ with binary rationals. Binary
search to width less than $1/(2Q_{\rm all}^2)$ and rational
reconstruction recover its unique reduced fraction in polynomial time
with polynomially many queries. Every query retains the original parity
functions and hence the fixed index. One non-strict query at the recovered
$v(s)$ decides whether its endpoint is closed. At a supplied $q=v(s)$
the strict query is false, while the non-strict query is exactly this
attainment test. At $q=0$, non-strict feasibility is hard feasibility;
at $q=1$, strict feasibility is always false.

The oracle closure needed for the decision bound is elementary.
For $L\in\mathrm{NP}\cap\mathrm{coNP}$, guess a deterministic
polynomial oracle computation together with certificates for every yes
answer in $L$ and every no answer in its complement. Simulate its
adaptive queries and check all certificates. No incorrect transcript
passes. Flipping the final output gives the analogous complement
verifier. The tagged union of the two threshold languages remains in
$\mathrm{NP}\cap\mathrm{coNP}$, so
$\mathrm P^{\mathrm{NP}\cap\mathrm{coNP}}\subseteq
\mathrm{NP}\cap\mathrm{coNP}$. Thus \textsc{Own-Attain} has the stated
decision bound. The same reconstruction gives one exact feasible value
in $\mathrm{FP}^{\mathrm{NP}\cap\mathrm{coNP}}$. Neither conclusion
asserts a polynomial-size original controller.

\section{The inherited Streett MDP certificate oracle}
\label{app:streett-oracle}
The following hard-label preservation proof makes the classical
qualitative Streett reduction usable by the negative certificates.
It is not a new qualitative gadget.

\begin{lemma}
\label{lem:r36-streett-mdp}
Let \(H\) be an MDP, let \(A\) be any hard Inf-set objective, and let
\(B\) be a soft Streett condition with \(d\) pairs. Finite pure
sure-\(A\)/almost-sure-\(B\) existence reduces to sure winning in a
deterministic game with \(O((d+1)(|V|+|E|))\) vertices and edges,
with objective \(A\cap B'\), where \(B'\) is Streett with \(d+1\) pairs.
This is a graph-size bound. Hard labels are copied to gadget vertices;
their occurrence encoding, and that of the new Streett pairs, have
polynomial size in the full input. In particular, if \(A\) has \(k\)
explicit Streett pairs, finite-SAS MDP winning is in coNP.
\end{lemma}

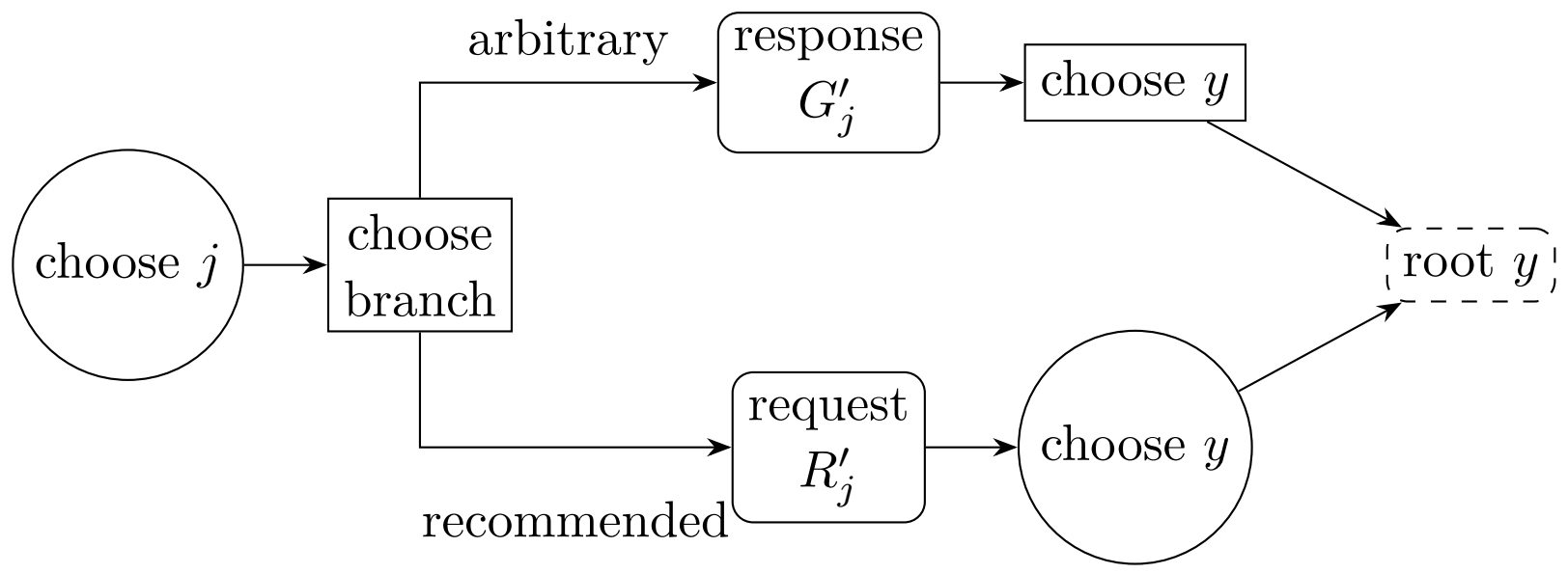
\begin{figure}[ht]\centering
\begin{tikzpicture}[>=Stealth,
 mx/.style={draw,circle,align=center,inner sep=3pt},
 mn/.style={draw,rectangle,align=center,inner sep=3pt},
 markerbox/.style={draw,rounded corners,align=center,inner sep=3pt}]
\node[mx] (j) at (0,0) {choose $j$};
\node[mn] (branch) at (2,0) {choose\\branch};
\node[markerbox] (response) at (4.8,1.25) {response\\$G'_j$};
\node[mn] (any) at (6.9,1.25) {choose $y$};
\node[markerbox] (request) at (4.8,-1.25) {request\\$R'_j$};
\node[mx] (rec) at (6.9,-1.25) {choose $y$};
\node[draw,dashed,rounded corners,align=center,inner sep=3pt] (root) at (9.2,0) {root $y$};
\draw[->] (j)--(branch);
\draw[->] (branch.north) -- (2,1.25) -- node[above]{arbitrary} (response.west);
\draw[->] (branch.south) -- (2,-1.25) -- node[below=7pt]{recommended} (request.west);
\draw[->] (response)--(any);\draw[->] (request)--(rec);
\draw[->] (any)--(root);\draw[->] (rec)--(root);
\end{tikzpicture}
\caption{Deterministic replacement of one chance source $v$ in
Lemma~\ref{lem:r36-streett-mdp}. Circles belong to Max and squares to the
opponent; the dashed box returns to an original successor root. After Max chooses $j\in\{0,\ldots,d\}$, the branch determines
who chooses a successor $y$ from the full original support and which new
marker is emitted. Every local vertex copies $v$'s hard label. The proof
specifies the augmented pairs; the original soft condition is not added
as a separate conjunct.}\label{fig:streett-mdp-interface}
\end{figure}

\begin{proof}
Write the soft pairs as \((R_j,G_j)\), \(1\leq j\leq d\), and add a
dummy pair \((R_0,G_0)=(\varnothing,V)\). Retain each original \Max
vertex and call the copies of original vertices \emph{roots}. Replace
each chance vertex \(v\) by the following local game. First \Max
chooses \(j\in\{0,\ldots,d\}\). The opponent then chooses between:
\begin{enumerate}
\item an arbitrary branch, which emits a new response marker for pair
\(j\), after which the opponent selects any supported successor of \(v\);
\item a recommended branch, which emits a new request marker for pair
\(j\), after which \Max selects any supported successor of \(v\).
\end{enumerate}
The augmented request set \(R'_j\) consists of the original \(R_j\)
roots and the recommended markers of index \(j\); \(G'_j\) consists
of the original \(G_j\) roots and the arbitrary markers of index \(j\).
There is no separate conjunct imposing the unaugmented soft condition
on gadget paths. Every local vertex copies \(v\)'s hard label. An
infinite gadget path therefore projects to a supported original path
with exactly the same hard Inf-set. The dummy response recurs on every
infinite path because roots recur.

For the forward direction, fix a finite sure-\(A\)/AS-\(B\) controller
\(\sigma\). Its reachable product is a finite Markov chain \(P\).
Every bottom strongly connected component satisfies \(B\): a bottom
component meeting \(R_j\) must meet \(G_j\), since all its vertices
recur almost surely. The strategy in the local game maintains
\(\sigma\)'s memory and a phase. Let \(K\) be the SCC of the current
product state and put
\[
 I(K)=\{j\in\{1,\ldots,d\}:K\cap R_j\ne\varnothing\}.
\]
An SCC change resets the phase. Along every supported path only finitely
many SCC changes are possible. If \(I(K)=\varnothing\), choose the
dummy index. Otherwise cycle through \(I(K)\), keeping index \(j\)
until an original \(G_j\) root or an arbitrary-branch response marker
is seen. Advance at most once per root visit, so no zero-time loop is
introduced.

On a recommended branch in phase \(j\), follow a shortest product path
to \(G_j\cap K\), if this set is nonempty, and otherwise to an exit
from \(K\). The first target is reachable by strong connectivity.
In the second case \(K\) meets \(R_j\) but not \(G_j\), so it cannot
be bottom and has a reachable exit. Fixed \Max moves of \(\sigma\)
also decrease this distance outside the target, as each has only one
product successor.

Consider an infinite path and its eventual SCC \(K\). An infinite
unfinished phase would have no further arbitrary response marker and
no original response. Every later chance step would consequently be
recommended, forcing the chosen finite distance to decrease until a
response or an exit occurs, a contradiction. Hence all phases complete
infinitely often. Every index in \(I(K)\) receives infinitely many
responses; every other real index has neither original nor artificial
requests recurrent. If \(I(K)\) is empty the latter conclusion holds
for all real indices. Thus \(B'\) holds on every path. The projected
path follows \(\sigma\), and so is in \(A\), including on null paths.

Conversely, fix a finite pure winning strategy \(\eta\) in the
deterministic game; finite memory suffices for deterministic
omega-regular winning. At an original chance vertex with current
\(\eta\)-memory, compute its chosen index \(j\) and its recommended
successor \(r\). When chance produces \(w\), update memory through the
recommended branch if \(w=r\), and through the arbitrary branch
otherwise. This is a deterministic public-edge update, chosen after
the outcome is observed. Every supported path of the resulting finite
MDP controller has an \(\eta\)-consistent expansion, and therefore
satisfies \(A\).

If \(B\) were not satisfied almost surely, its failure would have
positive probability. The finite induced Markov chain would therefore
have a reachable bottom SCC \(C\) and a pair \(h\) with
\(C\cap R_h\ne\varnothing\) and \(C\cap G_h=\varnothing\).
Construct a gadget path over \(C\). Whenever the selected index is
\(h\), take the recommended branch and \(r\). It emits a request
for \(h\), and its updated product state remains in \(C\), because
this is exactly the update for a positive-probability outcome. For any
other selected index, outside \(R_h\) choose a product successor
decreasing the distance within \(C\) to \(C\cap R_h\). At a visit
to \(R_h\), take any supported product successor in \(C\) and resume
this descent. Expand each such chance edge by the same
recommended-if-equal rule. At fixed \Max roots retain \(\eta\)'s
unique move; outside the target, it also decreases the distance because
it is the only product successor. Bottom closure keeps all these
successors in \(C\).

There is no original response for \(h\) in \(C\), and no arbitrary
branch of index \(h\) was used. If index \(h\) is selected infinitely,
its artificial request markers recur. If it is selected only finitely,
all sufficiently late moves follow distance descent, renewed on each
target visit. Finitely many \(R_h\) visits would then force an integer
distance to decrease forever. Thus original \(R_h\) requests recur
infinitely in this case. Either way pair \(h\) is violated,
contradicting \(\eta\).

For Streett \(A\), copy its \(k\) pairs and append the \(d+1\)
augmented pairs. Deterministic Streett winning is in coNP. Its losing
certificate is a positional Rabin opponent: after fixing that policy,
absence of an accepting Streett path is checked by SCC pruning. In an
SCC missing a response, delete the corresponding request vertices and
repeat. This is polynomial. All gadget labels and policies have
polynomial written size.
\end{proof}

\section{General Boolean attainment at the unsupplied optimum}
\label{app:r47-general-boolean}

The designated-start problem \textsc{Own-Attain} asks whether the finite
supremum is attained by a pure finite controller that satisfies hard
Streett on every supported infinite path.  The opponent may use any pure
public-history strategy.  A start outside the original sure region $W$
is a no instance.  The construction imposes no additional structural restriction beyond
those stated in the theorem below.

\begin{theorem}\label{thm:r47-general-own}
For hard Streett, \textsc{Own-Attain} is $\mathrm P^{\mathrm{NP}}$-complete
for Boolean Streett reward.
Hardness already holds with Boolean B\"uchi reward, $W=V$, and every
chance vertex fair binary.  Every cyclic SCC is chance-free except one
two-vertex SCC consisting of a Max retry gate and a fair chance vertex;
all other chance vertices lie in acyclic lotteries.

For Boolean Streett reward there is also a lower-bound
family with just one hard coB\"uchi pair.  In that family the soft Streett
list is unbounded; the two objective allocations are separate restrictions.
\end{theorem}

The upper bounds are the previously proved supplied-threshold certificates,
polynomial rational reconstruction and final endpoint query
(Theorem~\ref{thm:tight-complexity}).  We give a uniform lower-bound
construction, including the source completeness argument.  No oracle
answer is used while constructing an arena.

\subsection{Attained entries and an adaptive numerical gate}

Table~\ref{tab:stage-interface} separates construction-time data from proof-time values.
\begin{table}[ht]\centering
\caption{The adaptive interface at each stage. Caps apply at designated entries, not arbitrary internal check states.}
\label{tab:stage-interface}
\begin{tabular}{@{}>{\raggedright\arraybackslash}p{.18\linewidth}>{\raggedright\arraybackslash}p{.23\linewidth}>{\raggedright\arraybackslash}p{.24\linewidth}>{\raggedright\arraybackslash}p{.23\linewidth}@{}}\toprule
Object & Constructor knows & Proof establishes & Older modules\\\midrule
Old entries $p_i,q_i$ & Names and common gap $\delta>0$ & Gap sign encodes $b_i$ & Every inherited mode is capped\\
Fresh lotteries & Known rational weights & Weighted old values, attained & Calls are irreversible\\
New query and reference & Syntax, entry names and scaling & Query lies above or below its unknown reference & Old rows stay unchanged\\
Scaled entries & A common positive scale & All gaps agree at the next stage & New names retain old interfaces\\\bottomrule
\end{tabular}
\end{table}

\begin{samepage}
\begin{definition}[Entry contract]
\label{def:r52-entry-contract}
An entry contract at a construction stage consists of two guarantees
for its designated entry vertices:
\begin{enumerate}
\item Each entry has a stated value attained by a pure finite controller
that is sure for the original hard condition against every public-history
Min strategy.
\item One pure positional Min policy is common to all entries. For every
entry and every pure finite continuation initialized there in any
inherited mode, if that continuation is sure for the original hard condition against
every public-history Min strategy, its expected reward against this policy is at most the stated
value.
\end{enumerate}
\end{definition}
\end{samepage}

The absorbers satisfy Definition~\ref{def:r52-entry-contract}.
Lotteries inherit the contract by weighted averaging and initialization
at the observed entry. Lemma~\ref{lem:r47-numerical-gate} extends it to
new query entries, and scaling preserves it at the next stage.
Old edges stay unchanged and calls to older entries are irreversible,
so their contracts remain available. The constructor uses entry names
and known coefficients; values and oracle answers occur only in the
correctness proof.

Good and bad absorbers are forced Max vertices
with deterministic self-loops and values $1$ and $0$.
For the ordinary hard-true cores, randomized strategies are behavioral
maps on the countably many finite public histories. The guard and suffix
arguments below give one positional Min policy capping every pure-history
Max strategy at the designated entries: an infinite losing guard play has
reward zero, and irreversible calls preserve the inductive caps. Independent
presampling at every history gives a product distribution on pure history
strategies whose mixture, against any fixed possibly randomized opponent,
agrees on finite cylinders and hence on bounded Borel payoff expectations.
Thus the common upper cap extends by averaging, as do the finite pure lower
witnesses' guarantees against every pure Min strategy, proving the ordinary
entry values. More generally, both players have pure finite
optimal strategies in finite omega-regular stochastic games
\cite[Corollary~2.8]{UW11}; that fact supplies the usual-value membership
bound as well.

Modules call only older designated entries.  Later construction adds incoming
edges and fresh objective occurrences; it never changes an old edge.
There is no return from an older module to its caller.  A fresh guard's
Streett pairs have request and response states only in that guard.  Hence
they are finite or absent after a call to an older module.  Fresh
B\"uchi-good states are also unreachable from older modules.  Thus old
values, finite witnesses and positional caps remain valid.

An acyclic lottery among older entries with attained values has their probability-weighted
value, attained by a finite controller.  For a lower witness, initialize a tagged optimal
controller for the observed outcome.  For an upper bound, use the old
common Min policy, which caps every possible inherited mode.  All
supported lottery branches are retained.  Lottery vertices are hard
neutral and B\"uchi bad; they are visited only for bounded finite time.

\begin{figure}[ht]
\centering
\begin{tikzpicture}[>=Stealth,
gate/.style={draw,circle,minimum size=4mm,inner sep=0pt},
mn/.style={draw,rectangle,align=center,minimum height=7mm,inner sep=3pt},
mx/.style={draw,circle,align=center,inner sep=3pt},
call/.style={draw,dashed,rounded corners,inner sep=4pt}]
\node[gate] (g0) at (0,0) {};
\node[mn] (a) at (1.3,0) {assign $\mathbf y$};
\node[gate] (g1) at (2.6,0) {};
\node[mx] (cl) at (3.6,0) {$C$};
\node[gate] (g2) at (4.6,0) {};
\node[mn] (ck) at (5.8,0) {$\mathrm{check}(C)$};
\node[call] (rail) at (8.1,.6) {$r_\ell^{\rm test}$};
\node[call] (hi) at (8.1,-.6) {$\mathsf{hi}$};
\node[call] (lo) at (3.6,-1.4) {$\mathsf{lo}$};
\draw[->] (g0)--(a);\draw[->] (a)--(g1);
\draw[->] (g1)--(cl);\draw[->] (cl)--(g2);\draw[->] (g2)--(ck);
\draw[->] (ck)--node[above,sloped]{$\ell$}(rail);
\draw[->] (ck)--node[pos=.52,below=3pt]{dummy}(hi);
\draw[->] (ck.north) -- (5.8,1.05) -- node[above]{continue} (0,1.05) -- (g0.north);
\draw[->] (g0.south) to[bend right=15] (lo.west);
\draw[->] (g1.south) -- (lo.north west);
\draw[->] (g2.south) -- (lo.north east);
\node[below=1mm of lo] {fallback};
\end{tikzpicture}
\caption{Schematic source round of Lemma~\ref{lem:r47-numerical-gate},
not the whole game graph. Squares are Min and circles Max; the small
circles offer fallback $\mathsf{lo}$. Each part abbreviates source vertices and
their copied markers: every actual source vertex has such a gate.
The check vertex and its exit set are specific to the selected clause $C$. Dashed boxes call lotteries with attained values; the test lottery for external literal
$\ell$ is the appropriate $r_i^+$ or $r_i^-$.}
\label{fig:numerical-guard}
\end{figure}
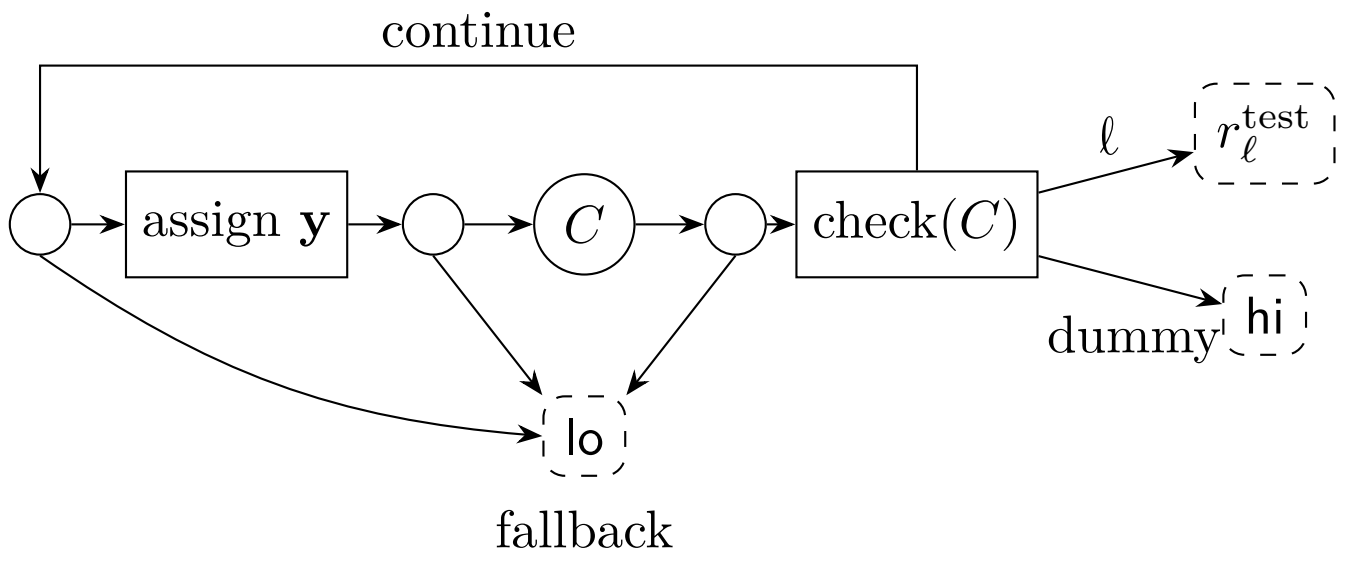

\begin{lemma}[Adaptive numerical gate]\label{lem:r47-numerical-gate}
Suppose a stage satisfies Definition~\ref{def:r52-entry-contract} and
represents $t$ Boolean parameters by paired entries $(p_i,q_i)$ whose
attained values lie in $[0,1]$, with known common rational gap
$0<\delta\leq1/4$:
\[
 p_i-q_i=\begin{cases}\delta&b_i=1,\\-\delta&b_i=0.\end{cases}
\]
For an explicit CNF $\varphi(b_1,\ldots,b_t,y_1,\ldots,y_{k_y})$, a polynomial
construction uses only known coefficients and gives entries $a$,
$\mathsf{ref}$, $\mathsf{lo}$ and $\mathsf{hi}$. Their attained values satisfy $0<\mathsf{lo}<\mathsf{ref}<\mathsf{hi}<1$ and
\[
 a=\begin{cases}
 \mathsf{hi}&\varphi(b_1,\ldots,b_t,\cdot)\text{ is unsatisfiable},\\
 \mathsf{lo}&\varphi(b_1,\ldots,b_t,\cdot)\text{ is satisfiable},
 \end{cases}\qquad \mathsf{hi}-\mathsf{ref}=\mathsf{ref}-\mathsf{lo}=\lambda\delta,
\]
where $\lambda=2^{-\lceil\log_2(2t+4)\rceil}$.
Every state remains in the original sure region.  The new source guard is
chance-free.  One may use either hard Streett and Boolean B\"uchi soft,
or hard true and Boolean Streett soft.  In both variants the common
positional Min cap extends to all new designated entries.
\end{lemma}

\begin{proof}
\paragraph*{Write lotteries from known coefficients.}
Write $S=\sum_{i=1}^t(p_i+q_i)$, using entry names also for their numerical
values.  This is algebraic notation: the constructor does not evaluate $S$.
Build the lotteries
\[
 \mathsf{lo}=\lambda(\delta+S),\qquad
 \mathsf{ref}=\lambda(2\delta+S),\qquad
 \mathsf{hi}=\lambda(3\delta+S).
\]
For $\mathsf{lo},\mathsf{ref},\mathsf{hi}$, give probability $\lambda$ to each old $p_i,q_i$, probability
$\lambda\delta,2\lambda\delta,3\lambda\delta$, respectively, to good,
and all remaining probability to bad.  Since $S\leq2t$ and
$\delta\leq1/4$,
\[
       0<\mathsf{lo}<\mathsf{ref}<\mathsf{hi}\leq\lambda(2t+3/4)<1.
\]
Every remaining probability is strictly positive, including when $t=0$.
For each parameter build
\begin{align*}
 r_i^+&=\lambda\left(2\delta+2p_i+
                           \sum_{j\ne i}(p_j+q_j)\right),\\
 r_i^-&=\lambda\left(2\delta+2q_i+
                           \sum_{j\ne i}(p_j+q_j)\right).
\end{align*}
Their nonbad masses sum to $\lambda(2t+2\delta)<1$.  Their values are attained,
and
\[
 r_i^+=\mathsf{ref}+\lambda(p_i-q_i),\qquad
 r_i^-=\mathsf{ref}-\lambda(p_i-q_i).
\]
Thus $r_i^+$ is $\mathsf{hi}$ when $b_i$ is true and $\mathsf{lo}$ otherwise; $r_i^-$ has the
opposite values.  The test lottery for literal $b_i$ is $r_i^-$,
and for literal $\neg b_i$ it is $r_i^+$.  A literal's test lottery has value
$\mathsf{hi}$ exactly when that literal is false.  Both test lotteries are written without
knowing the parameter's truth value.

\paragraph*{Build the clause-specific guard.}
The deterministic source guard repeats the following round.  Min assigns
the free variables $y_1,\ldots,y_{k_y}$ in order, visiting distinct markers
$A_y^0,A_y^1$ for the selected values.  Max chooses a clause $C$, visits
its marker $L_C$, and reaches a clause-specific Min state $\mathrm{check}(C)$.

Its outgoing edges are exactly: continuation to the next assignment round, the test lottery of each external literal occurring in $C$, and the dummy $\mathsf{hi}$ lottery. Thus different clauses have different check states and their own exit sets.
Put a Max gate
before every source state, offering normal continuation or fallback to $\mathsf{lo}$.
Gates copy source labels.  Each free-literal occurrence $\ell$ of variable
$y$ in $C$ contributes the Streett pair
\[
                (\{L_C\},\{A_y^{\mathrm{fals}(\ell)}\}),
\]
where $\mathrm{fals}(\ell)$ falsifies $\ell$; sets include the corresponding gate
copies.  No pair is added for external literals.  Maintain one global B\"uchi-good set $\mathit{Good}$, initially containing the good absorber.
Add all new source states and their gates to $\mathit{Good}$, retaining all older good
states; the bad absorber and lottery vertices stay outside $\mathit{Good}$.  If necessary add a
tautological clause on a fresh free variable so that the clause set is
nonempty.  Empty clauses, repeated literals and tautological clauses
otherwise need no special treatment.

In the first variant these guard pairs are hard, and soft is
B\"uchi$(\mathit{Good})$.  In the second variant hard is true, and soft is the
conjunction of all guard pairs and the B\"uchi pair $(V,\mathit{Good})$, where $V$
denotes the whole current arena.  The set $\mathit{Good}$ is shared by all guards; only the request/response marker sets
of different guards are local and disjoint.

\paragraph*{UNSAT: attain and cap the high value.}
Suppose first that the fixed-parameter CNF is unsatisfiable.  A finite Max
controller stores the current free assignment and chooses a false clause.
It may hard-code the fixed external truth values; the proof of existence
does not give those values to the arena constructor.  Every external
literal in that clause is false, so every test exit has value $\mathsf{hi}$; the
dummy exit has value $\mathsf{hi}$ too.  On every infinite source play, if $L_C$
recurs then each free literal of $C$ was falsified in each round selecting
$C$.  Its response marker therefore recurs.  This proves the guard pairs
on every supported path, without discarding a null path.  The play also
visits $\mathit{Good}$ infinitely and has soft reward $1\geq \mathsf{hi}$.  On an exit, initialize
the appropriate finite optimal suffix controller.  Finitely many tags
suffice, giving expected reward at least $\mathsf{hi}$ in either variant.

For the matching cap, Min takes dummy $\mathsf{hi}$ at the first check state, and
uses the old common policy in every suffix.  Before that check there are
only boundedly many source steps; an earlier Max fallback has value
$\mathsf{lo}\leq \mathsf{hi}$.  This policy caps every inherited finite mode at the designated
assignment-start entry by $\mathsf{hi}$.  Thus $a=\mathsf{hi}$ and its endpoint is closed.

\paragraph*{SAT: cap and attain the low value.}
Suppose instead that the fixed-parameter CNF is satisfiable.  Fix one
satisfying free assignment.  Min chooses it in every round.  At a clause
having a true free literal, Min continues; if no free literal is true,
some external literal is true, and Min exits to its test lottery of value
$\mathsf{lo}$.  This is positional.  All finite exits and fallbacks have value $\mathsf{lo}$
under the old suffix cap.  An infinite source play repeats some continued
clause $C$ infinitely.  That clause has a free literal true in the fixed
assignment, so its falsifying response marker never occurs.  The
corresponding Streett pair fails.  In the first variant such a play is
forbidden by sure hard; in the second its soft reward is $0\leq \mathsf{lo}$.
Consequently the positional policy caps every admissible finite controller
by $\mathsf{lo}$.  Immediate fallback attains $\mathsf{lo}$, so $a=\mathsf{lo}$.

\paragraph*{Preserve entry contracts and sure feasibility.}
These policies are on fresh Min states and use the unchanged old policy
in every suffix.  They extend the entry contract of
Definition~\ref{def:r52-entry-contract}. All upper
arguments apply to any inherited finite memory mode, so no memory reset by
the opponent is assumed.

Finally, every fallback reaches a lottery with finite sure suffixes.  From
a raw source state the next edge reaches a fallback gate or a feasible
lottery in bounded time.  All new hard requests are finite after leaving
the source.  This proves $W=V$ inductively in the first variant; it is
immediate in the hard-true variant.  Every original chance support is
retained throughout.
\end{proof}

\paragraph*{Maintaining one known gap.}
Start with no pairs and $\delta_0=1/4$.  At stage $i$, use $t=i-1$ and
$\lambda_i=2^{-\lceil\log_2(2i+2)\rceil}$.  After building the new
$(a_i,\mathsf{ref}_i)$, replace every old stored pair by new scaled entries that go to its
old entry with probability $\lambda_i$ and otherwise to bad.  The existing
entries and all previously written edges remain unchanged.  Store these
scaled pairs together with $(a_i,\mathsf{ref}_i)$.  Every stored gap is now
\[
 \delta_i=\lambda_i\delta_{i-1}
 =2^{-2-\sum_{h=1}^i\lceil\log_2(2h+2)\rceil}.
\]
The sign of every earlier parameter is preserved.  The exponent is
$O(i\log(i+2))$, so all coefficients and residues have polynomial binary
length.  This rescaling prevents repeated adaptive use from requiring
unknown numerical offsets or exponentially many written bits.

\subsection{A sequential CNF source without an oracle constructor}

A sequential CNF program is an explicitly written list
$\varphi_i(b_1,\ldots,b_{i-1},Y_i)$, $1\leq i\leq k$, evaluated by
\[
 b_i=1\quad\Longleftrightarrow\quad
       \varphi_i(b_1,\ldots,b_{i-1},\cdot)\text{ is unsatisfiable}.
\]
Its question is whether the final formula is satisfiable, equivalently
$b_k=0$.  All formula and variable occurrences count towards input length.

\begin{lemma}\label{lem:r47-adaptive-source}
The final-satisfiability question for sequential CNF programs is
$\mathrm P^{\mathrm{NP}}$-complete.
\end{lemma}

\begin{proof}
Membership evaluates the formulas in order with an NP oracle.  For hardness
let $M$ be any deterministic polynomial-time NP-oracle machine on $x$.
Bound its calls by $m$ and pad after halting with a fixed unsatisfiable
query.  Under proposed earlier answer bits, a deterministic polynomial
computation produces the next query $q_i$.  A polynomial circuit computes
$q_i(x,1-b_1,\ldots,1-b_{i-1})$ and verifies an NP witness for that query.
Cook--Levin/Tseitin encoding yields a polynomial CNF
$\varphi_i(b_1,\ldots,b_{i-1},Y_i)$ whose free-variable extension exists
exactly when that witness exists; all computation and witness variables
are free $Y_i$.  The construction unrolls the time-bounded computation
under a proposed transcript and does not decide an oracle query.

Inductively its evaluated $b_i$ is exactly the complement of $M$'s actual
answer bit.  Append a final CNF encoding the deterministic acceptance
circuit of $M$ under the full answer transcript.  For fixed external bits
that CNF has an extension exactly when $M$ accepts.  Thus the program's
final formula is satisfiable exactly when $M$ accepts $x$.  All circuits,
CNFs and padding have polynomial size and are produced without oracle
answers, giving the many-one reduction.
\end{proof}

Apply the numerical gate sequentially to this program, rescaling old
pairs after each stage.  It stores exactly its evaluated bits, so
\[
 a_k=\mathsf{lo}_k\quad\Longleftrightarrow\quad b_k=0
 \quad\Longleftrightarrow\quad\text{the final formula is satisfiable},
\]
and $a_k=\mathsf{hi}_k$ otherwise.  Both alternatives are realized by the same
syntactically constructed arena.

\paragraph*{Polynomial arena size.}
Stage $i$ has $O(i)$ lottery rows for the test lotteries, $\mathsf{lo},\mathsf{ref},\mathsf{hi}$ and scaled entries,
and $O(i^2)$ raw mixture edges.  Its guard and Streett list have polynomial
size in the written program and $i$, with one pair per free-literal occurrence.
Responses and requests are singleton markers with bounded gate copies.
Older modules are shared rather than recursively copied; the call graph
is acyclic.  The raw mixture edge total is $O(k^3)$, with coefficient bit
length $O(k\log(k+2))$, and every other written occurrence is polynomial.
Finite Max witnesses may store exponentially many free assignments and
suffix modes; this is permitted.  No polynomial controller-size bound is
asserted.

\subsection{Fair lotteries and supported paths}

\begin{lemma}\label{lem:r47-fair-lotteries}
The lower-bound cores can be written with every chance vertex fair binary
and every cyclic SCC chance-free, preserving their attained designated
values, finite witnesses, simultaneous Min caps and original $W=V$.
\end{lemma}

\begin{proof}
For a row with $r$ positive dyadic outcomes and denominator $2^B$, put
the outcomes in consecutive integer intervals partitioning $[0,2^B)$.
Sample fair binary prefixes.  Stop when the prefix interval lies wholly
in one outcome interval, and point to that outcome; otherwise split it
into its two equal halves.  At each depth, every retained internal prefix
contains an outcome boundary, so there are at most $r-1$ internal nodes
per depth and at most $(r-1)B$ in total.  Every branch terminates within
$B$ steps and has exactly the prescribed outcome law.  Combine duplicate
outcome pointers before constructing the tree.

Every generated nonabsorbing row has at least two positive distinct
outcomes.  Normalized $\mathsf{lo},\mathsf{ref},\mathsf{hi}$ and test-lottery rows have positive good and bad
masses; scaled entries have an old positive entry and bad.  Absorbers are
forced deterministic Max vertices, not probability-one chance vertices.
Thus all actual chance vertices can literally be binary fair.

Extra public prefix bits may reveal more than the original outcome. Here each
lottery is finite and calls an irreversible entry with attained value. Lower strategies ignore
the prefix bits and initialize its finite witness. After each positive-probability
leaf prefix reaching an older entry $y$, the inherited finite controller is sure
and its continuation is capped at that entry's value by the old common Min policy,
even if its memory depends on the whole prefix. Averaging these bounds with the
exact hitting law gives the same attained value. This does not assert a strategy
isomorphism for arbitrary games. Intermediate vertices are hard neutral and outside $\mathit{Good}$.
No supported infinite path can remain in a finite acyclic tree, and older
calls cannot return.  Thus no hard null path is removed and no new
infinite lottery path is created.

The expansion uses $O(B_{\max})$ times the raw edge total, with
$B_{\max}=O(k\log(k+2))$, hence polynomially many states.  Every chance
tree is acyclic and has no return from an outcome, while source SCCs are
chance-free.  All new states are sure feasible by their bounded exit to
feasible suffixes.
\end{proof}

\subsection{Turning the last comparison into own attainment}

\begin{proof}[Lower bound in Theorem~\ref{thm:r47-general-own}]
Take the last entries with attained values $a_k,\mathsf{ref}_k$, with
$\mathsf{lo}_k<\mathsf{ref}_k<\mathsf{hi}_k$ and gap $\delta_k$.  Open only the reference $\mathsf{ref}=\mathsf{ref}_k>0$.
A new Max gate $g$ offers STOP to bad or TRY to a fair chance vertex $c$;
success enters $\mathsf{ref}$, and failure returns to $g$.  Add the hard coB\"uchi
pair $(\{g,c\},\varnothing)$.  These vertices are outside $\mathit{Good}$ and all
other objective sets.  The successful and stopped suffixes cannot return.
Stopping at the next gate proves feasibility from either retry state.

Trying at most $N$ times and then stopping is pure finite and sure-hard,
with guarantee $(1-2^{-N})\mathsf{ref}$.  Hence the retry supremum is at least $\mathsf{ref}$.
For any pure finite sure controller, follow its supported all-failure
path.  It must STOP after finitely many failures, or violate the new
hard pair.  That prefix has positive probability.  The common old Min
cap applies to every successful inherited $\mathsf{ref}$ mode, so this controller
has worst expectation at most $(1-2^{-N})\mathsf{ref}<\mathsf{ref}$ for its finite stopping
time $N$.  STOP is included in its finite memory.  Thus the reference
value is $\mathsf{ref}$ and its endpoint is open.

A fresh Min entry chooses $a_k$ or this open reference.  Its supremum is
$\min(a_k,\mathsf{ref})$.  If $a_k=\mathsf{hi}_k>\mathsf{ref}$, Min selects the reference and no finite
sure controller attains $\mathsf{ref}$.  If $a_k=\mathsf{lo}_k<\mathsf{ref}$, use its finite optimal
controller on that branch, and choose $N$ on the reference branch with
$2^{-N}\leq\delta_k$.  Since $\mathsf{ref}\leq1$,
\[
                  (1-2^{-N})\mathsf{ref}\geq \mathsf{ref}-\delta_k=\mathsf{lo}_k.
\]
The tagged controller attains $\mathsf{lo}_k$ against both Min choices.  Therefore
own attainment holds exactly when the final source formula is satisfiable,
proving the many-one lower bound from Lemma~\ref{lem:r47-adaptive-source}.

In the hard-Streett/soft-B\"uchi variant, the reward is Boolean Streett.
In the alternative variant, all source pairs are soft Streett and the
cores have hard true; only the final retry contributes a hard coB\"uchi
pair. Thus Boolean reward supports both separately stated allocations.

By Lemma~\ref{lem:r47-fair-lotteries}, all core chance vertices are fair
and acyclic.  The retry chance is already fair.  The sole SCC containing
both chance and Max consists exactly of $g,c$; it has no Min vertex.
All other cyclic SCCs are chance-free.  The final Min entry has no return
path, and $W=V$ still holds.  This proves all stated restrictions.
\end{proof}

\subsection{A two-query illustration}
Here is a two-query instance of the construction.
Start with gap $\delta_0=1/4$, and take the first formula to be either $y$
or $y\wedge\neg y$. The second formula is the previous UNSAT bit $b_1$.
Writing its query values as $a_1,a_2$, the first scale is $\lambda_1=1/4$
and the second is $\lambda_2=1/8$.
Table~\ref{tab:two-queries} displays these two cases.

\begin{table}[ht]\centering
\caption{Two queries with the same final query value and different
references. The first formula sets the stored UNSAT bit $b_1$; the second formula
is the literal $b_1$.
Opening only its reference turns the last comparison into an attainment
flag.}\label{tab:two-queries}
\begin{tabular}{@{}lcccc@{}}\toprule
First formula & $b_1$ & Reference $\mathsf{ref}_2$ & Query value $a_2$ & Endpoint\\\midrule
$y$ & $0$ & $5/128$ & $3/64$ & open\\
$y\wedge\neg y$ & $1$ & $7/128$ & $3/64$ & closed\\\bottomrule
\end{tabular}
\end{table}

The arithmetic is as follows.

For the satisfiable first formula, $a_1=\mathsf{lo}_1=1/16$ and
$\mathsf{ref}_1=1/8$; for the unsatisfiable one, $a_1=\mathsf{hi}_1=3/16$ and
$\mathsf{ref}_1=1/8$. In both cases $\delta_1=1/16$.
The second reference is $\mathsf{ref}_2=(a_1+\mathsf{ref}_1+2\delta_1)/8$.
Thus the second query has the same value $a_2=3/64$, but that value lies
above or below its own reference. The final retry therefore produces
different attainment flags. The comparison baseline, not the query value
alone, carries the adaptive information.

\subsection{Relative values and one exact value}

\begin{corollary}\label{cor:r47-value-comparison}
Given an ordinary stochastic Streett game $G$ and states $s,t$, deciding
whether $\val_G(s)\ge\val_G(t)$ is $\mathrm P^{\mathrm{NP}}$-complete.
Hardness holds with attained designated entries, fair binary chance and
chance-free cyclic SCCs.
\end{corollary}

\begin{proof}
Use the alternative cores with attained entry values before adding retry.  We have
$a_k<\mathsf{ref}_k$ exactly when the final source formula is satisfiable, and
$\mathsf{ref}_k\leq a_k$ exactly when it is unsatisfiable.  The two inequalities are strict
alternatives.  Complement closure of $\mathrm P^{\mathrm{NP}}$ gives
hardness for either orientation of the non-strict comparison language.
For membership, the usual possibly randomized history-dependent values
equal pure finite values, by the optimal strategies of both players in
\cite[Corollary~2.8]{UW11}. Apply Theorem~\ref{thm:tight-complexity}
with hard true. The common-denominator bound of
Theorem~\ref{thm:r48-general-denominators}, with Boolean reward $R=1$,
gives $h\le Q$; the threshold oracle and rational reconstruction of
the function part of Theorem~\ref{thm:tight-complexity} compute each value in
$\mathrm{FP}^{\mathrm{NP}}$. Comparing the two fractions is in
$\mathrm P^{\mathrm{NP}}$. This ordinary denominator consequence also
follows from the prior original-state chain of \cite[Lemma~9]{CH07}.
The reduction compares usual values and does not ask an endpoint question.
\end{proof}

\begin{theorem}\label{thm:r47-single-value}
Computing one exact designated value, as a reduced fraction with binary
numerator and denominator, is $\mathrm{FP}^{\mathrm{NP}}$-complete under
polynomial metric reductions. Hardness holds for ordinary stochastic
Streett games with fair binary chance, attained designated entries and
chance-free cyclic SCCs.
\end{theorem}

\begin{proof}
Membership in the constrained model follows from
Theorem~\ref{thm:tight-complexity}; for ordinary games, the preceding
proof identifies the usual value with its hard-true case. The reduction
uses that hard-true variant, so every generated target has $W=V$ and
lies in the exact-value domain of Section~\ref{sec:exact-output}.
For hardness write
$S_t$ for the sum of all stored pair values after $t$ stages, with common
gap $\delta_t$.  For the next UNSAT bit $b_{t+1}$ the new pair has
\[
 a=\lambda\bigl(S_t+(1+2b_{t+1})\delta_t\bigr),\qquad
 \mathsf{ref}=\lambda(S_t+2\delta_t).
\]
Old pairs contribute $\lambda S_t$ after rescaling.  Thus
\[
 S_{t+1}=\lambda\bigl(3S_t+(3+2b_{t+1})\delta_t\bigr),\qquad
 \delta_{t+1}=\lambda\delta_t.
\]
Consequently $\Xi_t=S_t/\delta_t$ satisfies $\Xi_0=0$ and
$\Xi_{t+1}=3\Xi_t+3+2b_{t+1}$, giving
\[
 \Xi_m=\frac{3(3^m-1)}2+
                     2\sum_{i=1}^m b_i3^{m-i}.
\]
Add a final lottery $Z$ giving probability $1/D_0$ to each of the $2m$
stored entries and otherwise bad, where $D_0$ is a known power of two strictly
larger than $2m$.  It has attained value $Z=S_m/D_0$.  Its exact value
therefore determines the integer
\[
 J=\frac12\left(\frac{D_0Z}{\delta_m}-\frac{3(3^m-1)}2\right),
\]
whose base-three digits are precisely $b_1,\ldots,b_m$, including leading
zeroes.  The exponent of $\delta_m$ is $O(m\log(m+2))$, and all other
displayed integers have polynomial bit length.  Exact rational arithmetic
and repeated division by three decode the entire string in polynomial
time.  The final lottery has at least two positive distinct outcomes and
can be expanded as in Lemma~\ref{lem:r47-fair-lotteries}.

To reduce an arbitrary $\mathrm{FP}^{\mathrm{NP}}$ transducer directly,
unroll its polynomial query-generation computation on proposed previous
answer bits.  For query $i$, Cook--Levin/Tseitin encoding produces a CNF
\[
                 \varphi_i(b_1,\ldots,b_{i-1},Y_i)
\]
that has a free-variable extension iff the $i$th NP query has an accepting
witness under earlier answers $\alpha_j=1-b_j$.  Computation and witness
variables are all free $Y_i$.  Pad after halting to a known polynomial
number $m$ of calls.  The adaptive gate then stores exactly the actual
oracle no bits.  The reduction builds these formulas and modules
syntactically, without deciding any query.  Given the single exact value
$Z$, its polynomial decoder recovers all oracle answers and runs the
transducer with that transcript to obtain its output.  This is a metric
reduction, not a decision-query simulation.  Only the $m$ query gates are
needed; no sequence of gates for all Tseitin auxiliary variables is used.

This theorem uses ordinary soft Streett, not ordinary B\"uchi. Likewise
the one-hard-coB\"uchi variant uses an unbounded soft Streett list. Neither
claim specializes to a single soft B\"uchi condition.
\end{proof}

\subsection{The strict SSG query is already present in own attainment}
\label{app:fixed-index-ssg-own}

Take a simple stochastic reachability game with entry $r$, absorbing good
target $g$, fair chance rows, and usual value $a_0$.
The source question is $a_0>1/2$~\cite{Condon89TR863}.
Take $r$ distinct from the good and bad absorbers, inserting a forced
predecessor as the entry if necessary; this preserves its value and makes
the lottery below have distinct endpoints.
There are pure positional original optima, valid against arbitrary
history opponents~\cite[Corollary~2.7]{UW11}. In particular one Min
policy caps every inherited continuation at $r$ by $a_0$. It can also
be seen directly by choosing value-preserving Min edges: the usual
nonnegative value process is a supermartingale against every Max history
strategy and bounds the probability of reaching $g$ at every finite
time, hence also in the limit.

Add a fair lottery $\ell$ choosing $g$ or $r$, with attained value
$a=(1+a_0)/2\ge1/2$. Add a Max retry state $t$ with STOP to a forced
bad absorber and TRY to a fair coin $c$ choosing $\ell$ or returning
to $t$. Add a two-fair-coin lottery attaining value $3/4$, and a new
Min entry $s$ choosing that lottery or $t$. Old states never return to
the new gadget. Set hard to coB\"uchi avoidance of $\{t,c\}$ and soft
to B\"uchi $g$. Both use two priorities; all chance rows are fair and
$W=V$, since STOP gives sure fallback from the retry region and old
play never revisits that region.

$N$ trials followed by STOP, with an original positional Max optimum
after a call of $r$, guarantee $(1-2^{-N})a$. Conversely every pure
finite sure controller stops after finitely many trials on the unique
all-return history. Repeating a TRY memory mode would repeat that
supported failure cycle forever and violate hard, even on a null path.
The same original positional Min cap bounds every continuation after
each observed call of $r$, regardless of inherited Max memory. Thus its
retry payoff is at most $(1-2^{-N})a<a$ for some finite $N$. The offset
$a\ge1/2$ ensures strictness even when $a_0=0$.

The entry value is $\min(a,3/4)$. If $a_0\le1/2$, then $a\le3/4$;
finite counters approach $a$, but Min can select the open branch and
prevent its attainment, including at equality $a_0=1/2$. If
$a_0>1/2$, then $a>3/4$, and some finite $N$ satisfies
$(1-2^{-N})a\ge3/4$. That controller attains the entry value $3/4$.
Its $N$ trial modes and one STOP mode are counted; all old Max actions
are positional in every mode, and old play cannot return to $t$, so no
extra phase or initialization mode is required. The arena constructor
uses neither $a_0$ nor $N$ as a numerical coefficient. Therefore
\[
 a_0>1/2\quad\Longleftrightarrow\quad
             \textsc{Own-Attain}(G',s).
\]
This is a polynomial many-one reduction with the stated fixed-index
restrictions. It is an SSG lower bound, not a completeness classification
for a named class or a fixed-size memory lower bound. Ordinary supplied
SSG thresholds are already the hard-true/soft-B\"uchi special case.
Accordingly no polynomial-time conclusion, unrestricted-Max result, or
memory bound is supplied by this reduction. The Boolean memory upper bound follows separately by taking $R=1$
in Theorem~\ref{thm:r51-rational-exact-memory}.

\subsection{Collected secondary consequences}
\begin{corollary}[Parity specialization and ordinary-game consequences]
\label{thm:fixed-index-parity}\label{thm:main-ordinary}
For hard and soft parity with at least one fixed index, supplied
thresholds and \textsc{Own-Attain} belong to
$\mathrm{NP}\cap\mathrm{coNP}$ when Max uses pure finite sure controllers
against arbitrary pure public-history Min.
For ordinary stochastic Streett games with arbitrary possibly randomized
history strategies for both players, deciding
$\val_G(s)\ge\val_G(t)$ for two specified states is
$\mathrm P^{\mathrm{NP}}$-complete, and one exact binary reduced-fraction
value is $\mathrm{FP}^{\mathrm{NP}}$-complete under polynomial metric
reductions.
\end{corollary}
\begin{proof}
The fixed-index claim is proved in
Appendix~\ref{app:fixed-index-certificates}.
The ordinary comparison and exact-output claims are
Corollary~\ref{cor:r47-value-comparison} and
Theorem~\ref{thm:r47-single-value}, respectively.
\end{proof}

\section{Closest models and expectation transfer}\label{app:transfer-comparison}
The split/merge construction for ordinary expected payoffs is due to
Gimbert and Kelmendi~\cite[Theorems~1.1, 1.2, 5.3 and~6.1]{GK23};
the qualitative split/pause ancestor is also prior~\cite{CP18full,CP19}.
The distinction here is an additional admissibility obligation, not an
impossibility of adapting those frameworks.

Table~\ref{tab:r52-model-comparison} compares the closest models and
outputs. The additional results concern the particular strategy class,
sure semantics and numerical request in each row; they do not subsume
every result of the cited work.

\begin{table}[ht]\centering
\caption{Closest settings and the remaining distinction. The memory theorem
uses original-row and reward denominators, rather than the parameters of
a target-specific MDP construction or a fixed memory skeleton.}
\label{tab:r52-model-comparison}
\begin{tabular}{@{}>{\raggedright\arraybackslash}p{.20\linewidth}
                  >{\raggedright\arraybackslash}p{.41\linewidth}
                  >{\raggedright\arraybackslash}p{.33\linewidth}@{}}
\toprule
Source & Model and requested output & Additional result here\\\midrule
\cite[Theorems~4 and~23]{BRR17}
& Sure parity in MDPs: reachability thresholds have finite witnesses;
parity-probability thresholds may require infinite memory.
& Adversarial Min and pure finite Max; classifications of three
numerical requests.\\
\cite[arXiv v1, Theorem~3]{BKW24}
& Clean sure-parity MDPs: existence of a lexicographic optimum for
sink reachability, with finite witnesses when it exists; randomization is allowed.
& Finite attainment and exact-output classifications in adversarial
games; arithmetic implementation bounds.\\
\cite[Theorem~8]{CP19}; \cite[Sections~3 and~6.1]{DG26}
& Qualitative sure/almost-sure parity: finite-memory equivalence,
conjunction monitors and fixed-index bounds.
& Arbitrary rational targets and finite quantitative attainment.
The qualitative constructions are reused.\\
\cite[Theorem~7]{DGG25}; \cite[Theorems~1.1, 1.2, 6.1]{GK23}
& Ordinary expected payoff without a separate sure requirement:
value classes, local thresholds and strategy transfer.
& Preserve the sure domain and open or closed finite endpoints,
including newly admissible controllers.\\
\cite[Theorem~7]{CDH04full}; \cite{CH06}; \cite[Corollary~2.8]{UW11}
& Ordinary Streett supplied thresholds, value-vector methods and pure
finite optima for usual arbitrary-history values.
& Adaptive lower bounds for comparing unsupplied values and decoding
one exact output; the upper bounds are prior.\\
\bottomrule
\end{tabular}
\end{table}

\begin{samepage}
The fixed-index finite-controller case at non-strict threshold one is
already qualitative: CP19's finite-memory equivalence and DG26's
parity-conjunction monitor give an $\mathrm{NP}\cap\mathrm{coNP}$ upper
bound~\cite[Theorem~8]{CP19}, \cite[Section~3, Theorem~2]{DG26}.
DG26 also treats fixed-index sure/almost-sure synthesis with unrestricted
memory~\cite[Section~6.1 and Table~1]{DG26}. Our parity specialization
extends the finite-controller upper bound to arbitrary supplied rational
targets and to attainment without supplying the optimum; the matching
own-attainment and exact-output lower bounds here use Streett objectives.\par
\end{samepage}

\begin{samepage}
Other combinations of stochastic and worst-case requirements have
different tasks. Beyond-worst-case synthesis keeps an adversarial
worst-case constraint, but evaluates expectation against a supplied
stochastic environment~\cite{BFRR14}; here the reward expectation is
minimized over Min histories as well. The parity-MDP equality test of
\cite[Theorem~4]{BRR17} deletes target-dependent edges and normalizes
probabilities; the equality arena here keeps every original chance row.
Qualitative Boolean combinations in MDPs~\cite{BGR20} also have a
different task and strategy scope. These distinctions explain why the
present bounds are not direct improvements of one common prior measure.
\par
\end{samepage}

\paragraph*{When value preservation already suffices.}
If $A=\mathrm{true}$ and
$f(S\cup T)\ge\min\{f(S),f(T)\}$ for all nonempty colour sets,
including disjoint ones, the payoff is shift invariant and
inverse-submixing. An infinite shuffle has recurrent set the union of
the two recurrent sets. Swapping players in the ordinary
half-positional theorem gives a common positional minimizing value
policy. Finite-valued Inf rewards have finite pure one-player optima:
stop at an optimal end component and tour that component. The exact
finite-memory transfer therefore supplies a finite maximizing optimum.
All endpoints are closed before and after fixing Min. In this stronger
union, unconstrained special case, value preservation already preserves
the lower-guarantee intervals. This is an included prior case, not an
additional endpoint result here.

\paragraph*{What the sure constraint changes.}
Ordinary local threshold tests and value classes are
prior~\cite[Theorem~7, Lemmas~8--9 and Corollary~10]{DGG25}.
A bounded penalty on hard-rejecting paths does not enforce a sure
constraint: in a retry example, retrying forever can succeed almost
surely while its supported all-failure path violates hard. Likewise,
almost-sure finiteness of resets, as used in the ordinary transfer proof,
does not give finiteness on every supported path. Neither observation
contradicts the ordinary result or rules out other transformations.
Our proof instead keeps finite child memories while pausing them.

The extra steps are the graded all-path attainment test, compatibility
of its weak and strict local tests in one harmonic arena, and a bound
for controllers that become sure only after Min is fixed. Figure~\ref{fig:endpoint}
shows why an arbitrary value policy need not preserve the endpoint.
The equality arena is an attainment test; deleting non-value-preserving
Max edges need not preserve the original admissible controllers.

\paragraph*{Related memory results.}
Fixed-skeleton memory lifts assume one-player sufficiency for both
players over a class closed under subarenas, splits and memory
products~\cite[Theorem~4.1]{BORV23}. Attainment at selected states of one
arena is a different premise. Distributional preferences in that theory
need not discard support information. Finite-choice subgame-optimality
and reset constructions for ordinary prefix-independent winning
probabilities are also prior~\cite[Theorems~1 and~4]{BBL23}.
The memory bound here caps completed positive-potential events,
including player gains and full-row chance slack, rather than using a
reset known only to stop almost surely.
No exhaustive novelty claim against every adaptation of these theories
is made.

\section{Finite checks and research methods}
\label{app:implementation}

\subsection{Construction and certificate checks}
Finite exact checks audit construction and certificate interfaces.
Frozen constructors write CNF syntax and rational rows without a SAT
oracle; separate truth enumeration checks their outputs. Whole controller
products retain every original chance successor and relevant opponent
choice, and graph certificates check hard acceptance on all supported
paths. Separately written rational implementations compare reward
potentials with event ranks and check STOP-inclusive retry frontiers and
all-state-attainment caps. Negative controls change these interfaces or
permit almost-sure resets with a supported hard-rejecting cycle.
Inputs, protocols, source hashes, products, certificates and outputs are
retained. These checks do not prove the general theorems, asymptotic
bounds or scalability, and do not instantiate a worst-case controller
bound or an efficient general compiler.

\subsection{Author contributions and use of AI tools}
We began with the following question about finite controllers satisfying
a specification on every execution: can one distinguish the supremal
guarantee from its attainment by a finite controller, and characterize
the complexity of the corresponding problems? During the research, we
considered proof strategies, scrutinized each argument proposed by AI
tools, and revised subsequent proof directions according to our
mathematical judgments.

We used Codex to support planning and the development of mathematical
ideas and proof proposals, including numerical guards, original-state
subpotentials, quantitative certificates, finite event caps,
attained-state anchoring and rational-reward memory bounds, including
serial-trial lower bounds. Codex also assisted us with critique,
primary-source inspection, experimental design, coding, finite data
generation, exact calculations and testing, writing and evidence
curation. We used separate AI agents to check proofs and code, sometimes
with separately written graph or arithmetic programs. These checks can
share assumptions and do not constitute independent human peer review.

We have completed human checking of the current manuscript, including
the changes since the previously checked version and the serial-trial
extension in Corollary~\ref{cor:r55-serial-product-memory}, and found no
issues. We have not performed proof-assistant certification.
We are responsible for the final claims, proofs and interpretations.

\bibliographystyle{plainurl}
\bibliography{references}
\end{document}